\pdfoutput=1  

\documentclass[aps,twocolumn,superscriptaddress,floatfix,preprintnumbers,nofootinbib,eqsecnum]{revtex4-2}

\usepackage{amsmath, float, graphicx,placeins,amssymb,multirow,pgfplots,pgfplotstable}
\usepackage{tikz,hyperref}
\usepackage{tabularx}
\usepackage{CJKutf8}
\usetikzlibrary{shapes.geometric,arrows}

\begin{document}

\preprint{MI-HET-759} 

\title{Sequential Displaced Vertices (``Tumblers''):\\ A Novel Collider Signature for Long-Lived Particles}

\def\andname{\hspace*{-0.5em}} 
\author{Keith R. Dienes}
\email{dienes@arizona.edu}
\affiliation{Department of Physics, University of Arizona, Tucson, AZ 85721, USA}
\affiliation{Department of Physics, University of Maryland, College Park, MD 20742, USA}
\author{Doojin Kim}
\email{doojin.kim@tamu.edu}
\affiliation{Mitchell Institute for Fundamental Physics and Astronomy, Department of Physics 
       and Astronomy, Texas A{\&}M University, College Station, TX  77843  USA}
\author{Tara T. Leininger}
\email{leiningt@lafayette.edu}
\affiliation{Department of Physics, Lafayette College, Easton, PA 18042, USA}
\author{Brooks Thomas}
\email{thomasbd@lafayette.edu}
\affiliation{Department of Physics, Lafayette College, Easton, PA 18042, USA}

\begin{abstract}
In this paper, we point out a novel signature of physics beyond the Standard Model 
which could potentially be observed both at the Large Hadron Collider (LHC) and at 
future colliders.  This signature, which emerges naturally within many proposed 
extensions of the Standard Model, results from the multiple displaced vertices associated 
with the successive decays of unstable, long-lived 
particles along the same decay chain.  We call such a sequence of displaced vertices a
``tumbler.''  We examine the prospects for observing tumblers at the LHC and assess the 
extent to which tumbler signatures can be distinguished from other signatures of 
new physics which also involve multiple displaced vertices within the same collider event.
As part of this analysis, we also develop a procedure for reconstructing the masses 
and lifetimes of the particles involved in the corresponding decay chains.
We find that the prospects for discovering and distinguishing tumblers can be greatly 
enhanced by exploiting precision timing information such as would 
be provided by the CMS timing layer at the high-luminosity LHC.~  Our analysis therefore provides 
strong additional motivation for continued efforts to improve the timing capabilities of collider 
detectors at the LHC and beyond.
\end{abstract}

\maketitle


\newcommand{\PRE}[1]{{#1}} 
\newcommand{\ul}{\underline}
\newcommand{\del}{\partial}
\newcommand{\nbox}{{\,\lower0.9pt\vbox{\hrule \hbox{\vrule height 0.2 cm
\hskip 0.2 cm \vrule height 0.2 cm}\hrule}\,}}

\newcommand{\postscript}[2]{\setlength{\epsfxsize}{#2\hsize}
   \centerline{\epsfbox{#1}}}
\newcommand{\gweak}{g_{\text{weak}}}
\newcommand{\mweak}{m_{\text{weak}}}
\newcommand{\mplanck}{M_{\text{Pl}}}
\newcommand{\mstar}{M_{*}}
\newcommand{\sigmaan}{\sigma_{\text{an}}}
\newcommand{\sigmatot}{\sigma_{\text{tot}}}
\newcommand{\sigmaSI}{\sigma_{\rm SI}}
\newcommand{\sigmaSD}{\sigma_{\rm SD}}
\newcommand{\OmegaM}{\Omega_{\text{M}}}
\newcommand{\OmegaDM}{\Omega_{\text{DM}}}
\newcommand{\ipb}{\text{pb}^{-1}}
\newcommand{\ifb}{\text{fb}^{-1}}
\newcommand{\iab}{\text{ab}^{-1}}
\newcommand{\ev}{\text{eV}}
\newcommand{\kev}{\text{keV}}
\newcommand{\mev}{\text{MeV}}
\newcommand{\gev}{\text{GeV}}
\newcommand{\tev}{\text{TeV}}
\newcommand{\pb}{\text{pb}}
\newcommand{\mb}{\text{mb}}
\newcommand{\cm}{\text{cm}}
\newcommand{\m}{\text{m}}
\newcommand{\km}{\text{km}}
\newcommand{\kg}{\text{kg}}
\newcommand{\g}{\text{g}}
\newcommand{\s}{\text{s}}
\newcommand{\yr}{\text{yr}}
\newcommand{\Mpc}{\text{Mpc}}
\newcommand{\etal}{{\em et al.}}
\newcommand{\eg}{{\em e.g.}}
\newcommand{\ie}{{\em i.e.}}
\newcommand{\ibid}{{\em ibid.}}
\newcommand{\Eqref}[1]{Equation~(\ref{#1})}
\newcommand{\secref}[1]{Sec.~\ref{sec:#1}}
\newcommand{\secsref}[2]{Secs.~\ref{sec:#1} and \ref{sec:#2}}
\newcommand{\Secref}[1]{Section~\ref{sec:#1}}
\newcommand{\appref}[1]{App.~\ref{sec:#1}}
\newcommand{\figref}[1]{Fig.~\ref{fig:#1}}
\newcommand{\figsref}[2]{Figs.~\ref{fig:#1} and \ref{fig:#2}}
\newcommand{\Figref}[1]{Figure~\ref{fig:#1}}
\newcommand{\tableref}[1]{Table~\ref{table:#1}}
\newcommand{\tablesref}[2]{Tables~\ref{table:#1} and \ref{table:#2}}
\newcommand{\Dsle}[1]{\slash\hskip -0.23 cm #1}
\newcommand{\met}{{\Dsle E_T}}
\newcommand{\mpt}{\not{\! p_T}}
\newcommand{\Dslp}[1]{\slash\hskip -0.23 cm #1}
\newcommand{\Dsl}[1]{\slash\hskip -0.20 cm #1}

\newcommand{\mB}{m_{B^1}}
\newcommand{\mq}{m_{q^1}}
\newcommand{\mf}{m_{f^1}}
\newcommand{\mKK}{m_{KK}}
\newcommand{\WIMP}{\text{WIMP}}
\newcommand{\SWIMP}{\text{SWIMP}}
\newcommand{\NLSP}{\text{NLSP}}
\newcommand{\LSP}{\text{LSP}}
\newcommand{\mWIMP}{m_{\WIMP}}
\newcommand{\mSWIMP}{m_{\SWIMP}}
\newcommand{\mNLSP}{m_{\NLSP}}
\newcommand{\mchi}{m_{\chi}}
\newcommand{\mgravitino}{m_{\gravitino}}
\newcommand{\mmed}{M_{\text{med}}}
\newcommand{\gravitino}{\tilde{G}}
\newcommand{\Bino}{\tilde{B}}
\newcommand{\photino}{\tilde{\gamma}}
\newcommand{\stau}{\tilde{\tau}}
\newcommand{\slepton}{\tilde{l}}
\newcommand{\snu}{\tilde{\nu}}
\newcommand{\squark}{\tilde{q}}
\newcommand{\mgaugino}{M_{1/2}}
\newcommand{\epsEM}{\varepsilon_{\text{EM}}}
\newcommand{\mmess}{M_{\text{mess}}}
\newcommand{\lmess}{\Lambda}
\newcommand{\nmess}{N_{\text{m}}}
\newcommand{\signmu}{\text{sign}(\mu)}
\newcommand{\Omegachi}{\Omega_{\chi}}
\newcommand{\lambdafs}{\lambda_{\text{FS}}}
\newcommand{\be}{\begin{equation}}
\newcommand{\ee}{\end{equation}}
\newcommand{\bea}{\begin{eqnarray}}
\newcommand{\eea}{\end{eqnarray}}
\newcommand{\beq}{\begin{equation}}
\newcommand{\eeq}{\end{equation}}
\newcommand{\beqn}{\begin{eqnarray}}
\newcommand{\eeqn}{\end{eqnarray}}
\newcommand{\baln}{\begin{align}}
\newcommand{\ealn}{\end{align}}
\newcommand{\lsim}{\lower.7ex\hbox{$\;\stackrel{\textstyle<}{\sim}\;$}}
\newcommand{\gsim}{\lower.7ex\hbox{$\;\stackrel{\textstyle>}{\sim}\;$}}

\newcolumntype{C}{>{\centering\arraybackslash}X}

\newcommand{\ssection}[1]{{\em #1.\ }}
\newcommand{\rem}[1]{\textbf{#1}}

\def\ie{{\it i.e.}\/}
\def\eg{{\it e.g.}\/}
\def\etc{{\it etc}.\/}
\def\calN{{\cal N}}

\def\mptwo{{m_{\pi^0}^2}}
\def\mp{{m_{\pi^0}}}
\def\sqtsn{\sqrt{s_n}}
\def\sqtsn{\sqrt{s_n}}
\def\sqtsn{\sqrt{s_n}}
\def\sqts0{\sqrt{s_0}}
\def\Dsqts{\Delta(\sqrt{s})}
\def\Omegatot{\Omega_{\mathrm{tot}}}
\def\rhotot{\rho_{\mathrm{tot}}}
\def\rhocrit{\rho_{\mathrm{crit}}}
\def\OmegaDM{\Omega_{\mathrm{DM}}}
\def\OmegaDMbar{\overline{\Omega}_{\mathrm{DM}}}
\def\tLS{t_{\mathrm{LS}}}
\def\aLS{a_{\mathrm{LS}}}
\def\zLS{z_{\mathrm{LS}}}
\def\tnow{t_{\mathrm{now}}}
\def\znow{z_{\mathrm{now}}}
\def\Ndof{N_{\mathrm{d.o.f.}}\/}
\def\ra{\rightarrow}


\section{Introduction\label{sec:Intro}}


 Ever since the seminal work of Glashow, Weinberg, and Salam in the 1970s that 
 gave birth to modern particle physics, the Standard Model (SM) has reigned supreme.   
 Although the discovery of neutrino oscillations and the preponderance of
 observational evidence for dark matter and dark energy have 
 indicated the need to extend the SM into new domains, the core of the SM has 
 remained intact and continues to accurately describe all existing collider data 
 despite decades of intense experimental research.  Indeed, unambiguous evidence for 
 possible SM extensions such as weak-scale supersymmetry or large extra dimensions has 
 not yet been found.  

There are, in principle, two possible reasons for this state of affairs.  On the one 
hand, the energy scale associated with the new physics may be sufficiently high that 
this physics lies beyond the reach of current experiments.   However, on the other hand, 
it is possible that the new physics resides at energy scales which are potentially 
accessible at current or imminent collider experiments, but that this physics is 
manifested through collider signatures that have not yet received much attention 
within the community (for reviews, see, \eg, 
Refs.~\cite{Alimena:2019zri,Fischer:2021sqw,Alimena:2021mdu}). 

In this paper we point out a novel collider signature which arises in a variety of 
scenarios for new physics.  This signature rests on the possible existence of long-lived 
particles (LLPs).  As discussed in Ref.~\cite{Curtin:2018mvb}, LLPs can arise in many 
proposed extensions of the Standard Model.  These include models which attempt to 
address the gauge hierarchy problem, models which provide new approaches to dark-matter 
physics, models which describe different scenarios for baryogenesis and leptogenesis, 
and even non-minimal models of neutrino physics.  Because of their relative long 
lifetimes, LLPs, once produced, can propagate across macroscopic distances before they 
decay.  For LLPs with proper decay lengths $c\tau$ ranging from millimeters 
to hundreds of meters, these decays can give rise to a number of distinctive signatures 
at colliders, including emerging jets~\cite{Schwaller:2015gea}, disappearing tracks, and
macroscopically displaced vertices (DVs). 
While searches for DVs are part of the standard experimental program at colliders, 
the signature on which we shall focus our attention involves the presence of multiple 
displaced vertices which result from the successive decays of multiple unstable LLPs within 
the same decay chain.  In such cases, the event unfolds by ``tumbling'' down the steps of 
the decay chain, terminating only once a collider-stable particle is reached.    

Given this decay topology, we shall refer to such a sequence of DVs as a 
``tumbler.''  In this work, we shall consider the special case of tumblers in which each such 
LLP decay yields a single, lighter LLP as well as one or more SM particles which can be 
detected directly by a collider detector.  The signatures of such tumblers are quite striking 
as they have very low SM backgrounds.  We shall examine the prospects for 
observing such tumblers at the LHC, and we shall assess the extent to which such tumbler 
signatures can be distinguished from other signatures of new physics which 
also involve multiple 
DVs within the same collider event.  We shall also develop a procedure for
reconstructing the masses and lifetimes of the particles involved in the corresponding 
decay chains. 

One important theme running through this work will be the observation that 
the prospects for discovering and distinguishing tumblers can be greatly enhanced by 
exploiting precision timing information.   Fortunately, this sort of information can be 
provided by a precision timing layer of the sort that will be installed within the CMS 
detector during the forthcoming high-luminosity upgrade of the Large Hadron 
Collider (LHC)~\cite{Gray:2017,Butler:2019rpu}.   As we shall see, this timing 
information can significantly improve the precision with which the masses and lifetimes 
of the particles within a tumbler can be measured. 

This paper is organized as follows.  In Sect.~\ref{sec:TumblersTiming}, we describe the 
basic properties of tumblers and discuss the role that timing information 
can play in characterizing them.
In Sect.~\ref{sec:ToyModel}, we introduce a concrete example model which 
can give rise to tumblers.  In Sect.~\ref{sec:SurveyParamSpace}, we 
survey the parameter space of this model and identify regions of this parameter
space wherein the prospects for identifying tumblers are particularly auspicious.
In Sect.~\ref{sec:Discovery}, we investigate the extent to which current LHC data
constrains this parameter space and assess the prospects for observing a significant 
number of tumbler events both before and after the high-luminosity LHC (HL-LHC) upgrade.
In Sect.~\ref{sec:ResultsRecon}, we develop an event-selection procedure which provides an
efficient way of distinguishing between events which involve tumblers and events 
which involve multiple DVs which were not in fact produced by the successive decays of
unstable particles within the same decay chain.  We also investigate the degree to which the 
masses and lifetimes of the dark-sector states can be measured from tumbler events.  
In Sect.~\ref{sec:Conclusion}, we 
conclude with a summary of our results and a discussion of the ways in which 
improvements in energy and timing resolution could enhance our ability to distinguish
tumbler signatures at the HL-LHC or at future colliders.


\FloatBarrier
\section{Tumblers at the LHC\label{sec:TumblersTiming}}


Macroscopically displaced vertices can result from the decays of long-lived particles 
(LLPs) that decay on distance scales 
$\mathcal{O}(1\mathrm{~mm}) \lesssim c\tau \lesssim \mathcal{O}(100\mathrm{~m})$
inside a collider detector.  Such vertices represent a striking potential signal of new 
physics~\cite{Curtin:2018mvb,Alimena:2019zri}.
While the DV signatures associated with the decays of even a single LLP species 
can yield a wealth of information about physics beyond the SM, the phenomenology associated 
with DVs can be far richer in extensions of the SM which involve multiple of LLP species.  
One intriguing possibility arises in scenarios in which one species of LLP can decay into a 
final state which includes both SM particles and a lighter LLP of a different 
species.  If this lighter LLP also decays within the detector, the result is a sequence 
of two or more DVs 
which result from the successive decays of unstable particles within the same decay chain.
Like DVs themselves, such {\it sequences}\/ of DVs --- \ie, such ``tumblers'' --- can 
arise naturally in many extensions of the SM.~  
These include models involving compressed supersymmetry~\cite{Martin:2007gf}; 
hidden-valley models~\cite{Strassler:2006im} and other, similar theories 
which give rise to phenomena referred to as emerging jets~\cite{Schwaller:2015gea},  
semi-visible jets~\cite{Cohen:2015toa}, dark jets~\cite{Park:2017rfb}, and/or 
soft bombs~\cite{Knapen:2016hky}; theories 
involving large numbers of additional degrees of freedom with a significant degree 
of disorder in their mass matrix~\cite{DAgnolo:2019cio}; and scenarios involving 
non-minimal dark sectors~\cite{Dienes:2019krh}.  Indeed, tumbler-like events of this
sort, under the name ``micro-cascades,'' were explicitly invoked more than a decade ago 
to explain possible anomalies in CDF data involving muons produced with large 
impact parameters~\cite{Giromini:2008xh,Strassler:2008jq}.  The possibility of 
tumbler-like events arising in a variety of hidden-valley models was also discussed 
in Refs.~\cite{Strassler:2006qa,Juknevich:2009gg,Juknevich:2009ji,Craig:2015pha}.

\begin{figure*}
  \centering
  \includegraphics[clip, width=0.75\textwidth]{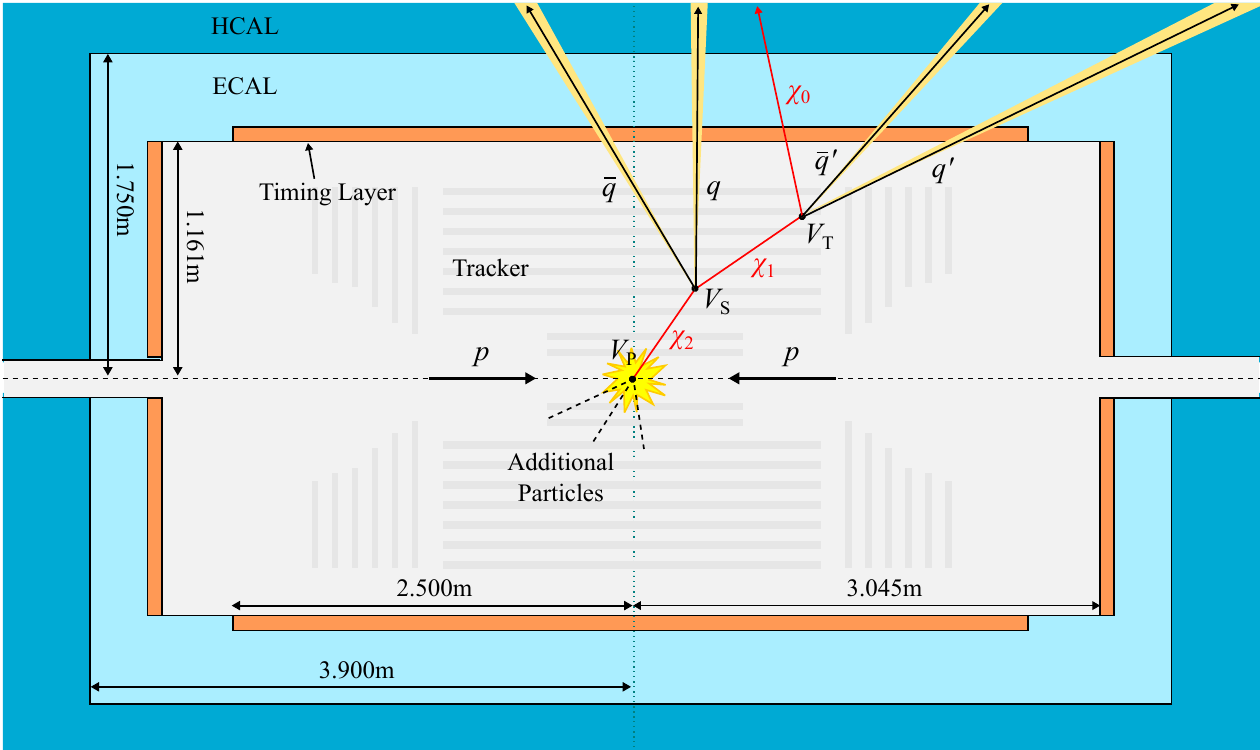}
  \caption{Schematic of a tumbler event within a collider detector
    modeled after the CMS detector at the HL-LHC.~  
    In this event, a heavy LLP $\chi_2$ is initially produced at the primary vertex $V_P$,
    along with some additional particles.  The $\chi_2$ particle then travels a 
    measurable distance before decaying into a lighter LLP $\chi_1$, a 
    quark $q$, and an anti-quark $\bar{q}$ at the secondary vertex $V_S$.  This $\chi_1$ 
    particle then travels a measurable distance away from $V_S$ and subsequently decays into 
    an even lighter LLP $\chi_0$, another quark $q'$, and another 
    anti-quark $\bar{q}'$ at the tertiary vertex $V_T$.  The $\chi_0$ particle manifests
    itself as missing energy $\met$, while the quarks and anti-quarks manifest themselves
    as hadronic jets.  Information about when each jet interacts with the timing layer, in 
    conjunction with additional information about the momentum of the jet from the tracker 
    and calorimeters, can be used to reconstruct the locations and times      
    at which $V_S$ and $V_T$ occurred.
  \label{fig:decayschematic}}
\end{figure*}

An example of a tumbler is illustrated in Fig.~\ref{fig:decayschematic}.  In this
example, an LLP $\chi_2$ is produced within a collider detector at the primary 
interaction vertex $V_P$, along with one or more additional SM particles.  This $\chi_2$ 
particle travels a measurable 
distance away from $V_P$ before it decays into a pair of SM particles (which for concreteness
we take to be a quark $q$ and an anti-quark $\bar{q}$), along with another, lighter LLP 
$\chi_1$ at the secondary vertex $V_S$.  This $\chi_1$ particle, in turn, travels a
measurable distance away from $V_S$ before it likewise decays at a tertiary vertex $V_T$ 
into a quark $q'$, an anti-quark $\bar{q}'$, and another, even lighter LLP $\chi_0$, which 
escapes the detector and manifests itself as missing transverse energy $\met$.

Fig.~\ref{fig:decayschematic} illustrates the topology of a tumbler involving only two DVs, 
as appropriate for a decay chain involving three LLPs ($\chi_2$, $\chi_1$, and $\chi_0$).   
In some sense, this is the minimal possible tumbler, and this case will be the focus of 
this paper.  However, there is nothing that requires tumblers to be limited to only two 
DVs or three LLPs, and indeed longer decay chains leading to more DVs are 
possible.  Indeed, in many SM extensions, entire ensembles of LLPs $\chi_n$ can arise.  
Such ensembles can then give rise to potentially long decay chains with many sequential 
DVs.  However, all such tumbler events share the same basic event topology, 
with sequential decays proceeding in linear fashion down the decay chain.

How might such a tumbler be detected and distinguished?  Since the SM backgrounds for 
processes involving DVs are quite low, signals involving DVs provide particularly 
striking indications of new physics.
A variety of LLP searches involving DVs have already been performed by the ATLAS
and CMS Collaborations.  Moreover, the sensitivity of the ATLAS and CMS detectors to DV
signatures will be significantly enhanced during the forthcoming HL-LHC 
upgrade, in part as a result of the installation of additional apparatus within both 
of these detectors which provides precision timing information about the particles 
produced in a collider event.  In particular, 
the upgraded ATLAS detector will include a high-granularity timing detector in front of 
each of the end-cap calorimeters in order to provide timing information for particles 
emitted in the forward direction~\cite{Lanni:2020}.  The upgraded CMS 
detector, by contrast, will include not only a pair of timing detectors located in 
front of the end-caps, but also a thin cylindrical timing layer situated between the 
tracker and the electromagnetic calorimeter (ECAL) which provides coverage within 
the barrel region of the detector~\cite{Gray:2017,Butler:2019rpu}.  This timing layer,
which is included in the illustration in Fig.~\ref{fig:decayschematic},
will provide a timing resolution of $\sigma_t \approx 30$~ps --- a vast improvement 
over the timing resolution $\sigma_t \approx 150$~ps currently afforded by the ECAL 
itself~\cite{delRe:2015hla}.  Such a significant enhancement in timing precision will 
significantly improve the sensitivity of LLP searches at the HL-LHC.~  Indeed, 
not only can information from the timing layer be used to reduce SM backgrounds for such 
searches~\cite{Liu:2018wte,Liu:2020vur}, but it can also aid in the reconstruction of the 
LLP masses~\cite{Kang:2019ukr}, strategies for which have been developed
for a number of LLP-decay scenarios~\cite{Cottin:2018hyf,Bae:2020dwf}.
In particular, the momenta 
$\vec{\mathbf{p}}_q$ and $\vec{\mathbf{p}}_{\bar{q}}$ of the hadronic
jets associated with $q$ and $\bar{q}$, in conjunction with timing the information for these
jets provided by either the timing layer or the ECAL, can be used to identify both the time 
$t_S$ and spatial location $\vec{\mathbf{x}}_S$ of $V_S$.  Similarly, the momenta
$\vec{\mathbf{p}}_{q'}$ and $\vec{\mathbf{p}}_{\bar{q}'}$ of the jets associated 
with $q'$ and $\bar{q}'$, in conjunction with the corresponding timing information, can be 
used to identify the time $t_T$ and spatial location $\vec{\mathbf{x}}_T$ of $V_T$.  
Information about the momenta of the additional SM particles produced at $V_P$, in conjunction
with the corresponding timing information, can be used to identify the time $t_P$ and 
spatial position $\vec{\mathbf{x}}_P$ of this vertex.


\FloatBarrier
\section{A Concrete Example Model\label{sec:ToyModel}}


In order to perform a more quantitative assessment of the prospects for detecting 
tumbler signatures at the LHC and beyond, it is necessary to work within the context of a 
concrete model.  Such a model can therefore also serve as an existence proof that tumblers 
may indeed arise at colliders such as the LHC, and yet be consistent with current experimental 
results.  The model that we adopt for this purpose is drawn from a general class of 
non-minimal dark-sector scenarios in which there exist multiple dark-sector states $\chi_n$ 
with similar quantum numbers, all of which can interact with the fields of the visible sector 
via a common mediator particle $\phi$.  Not only do these 
interactions provide a portal through which the $\chi_n$ can be produced, but they also render 
the heavier $\chi_n$ unstable.  Since the final states into which these $\chi_n$ 
decay in such scenarios generically involve both SM particles and other, lighter dark-sector 
states $\chi_m$, extended decay chains can develop.    

In Ref.~\cite{Dienes:2019krh} we constructed such a model within this class and focused on 
a region of parameter space in which the $\chi_n$ particles involved in these decay chains 
had lifetimes leading to prompt decays rather than macroscopically displaced vertices.  
We then discovered that such decay chains can lead to striking signatures involving large
multiplicities of produced SM states.   

In this paper, by contrast, we shall focus on a different region within the parameter space of this model, one in which the $\chi_n$ have lifetimes within the range 
$\mathcal{O}(1\mathrm{~mm}) \lesssim c\tau_n \lesssim \mathcal{O}(100\mathrm{~m})$.   
As we shall explain further below, we thus obtain decay chains involving 
DVs --- \ie, tumblers.  Moreover, although our analysis in 
Ref.~\cite{Dienes:2019krh} considered arbitrary numbers of $\chi_n$ states within the 
associated decay chains, we shall here restrict our attention to cases with only 
three $\chi_n$ particles, with $n=0,1,2$ labeling these states in order of increasing mass.

More specifically, this model is defined as follows.  We shall take the $\chi_n$ to be 
Dirac fermions and to be singlets under the SM gauge group.  We take the masses $m_n$ of 
the $\chi_n$ to be free parameters, subject to the condition $m_2 > m_1 > m_0$.
The particle $\phi$ which mediates the interactions between the $\chi_n$ and the fields of 
the SM in our model is taken to be a complex scalar which transforms as a triplet both under 
the SM $SU(3)_c$ gauge group and under the approximate $U(3)_u$ flavor symmetry of the 
right-handed up-type quarks.  In order to alleviate issues involving flavor-changing effects, 
we shall assume that the up-type quarks $q \in \{u,c,t\}$  and the component fields $\phi_q$ 
within $\phi$ share a common mass eigenbasis.  Expressed in this eigenbasis, the interaction
Lagrangian which couples the dark and visible sectors then takes the form
\begin{equation}
  \mathcal{L}_{\mathrm{int}} ~=~ \sum_q \sum_{n=0}^{2} \big[ c_{n q} \phi_q^\dagger
     \bar{\chi}_n P_R q  + \mathrm{h.c.}\big]~,
\label{eq:Lint}
\end{equation}
where $P_R \equiv \frac{1}{2}(1 + \gamma^5)$ is the usual right-handed projection operator
and where $c_{n q}$ is a dimensionless coupling constant which in principle depends both 
on the value of the index $n$ for the dark-sector field and on the flavor of the quark.
Such a coupling structure implies that each of the $\phi_q$ couples only to a single 
quark flavor $q$.

For simplicity, we shall focus on the case in which the masses of the 
mediators $\phi_c$ and $\phi_t$ which couple to the charm and top quarks are sufficiently
large that that they greatly exceed the mass of the mediator $\phi_u$ 
(\ie, $m_{\phi_c}, m_{\phi_t}\gg m_{\phi_u}$) and also have no appreciable impact on the 
collider phenomenology of the model.  From a low-energy perspective, this is equivalent to 
adopting a coupling structure in which $c_{n c} \approx 0$ and $c_{n t} \approx 0$ for all $n$, 
while the $c_n \equiv c_{n u}$ are in general non-vanishing.  Moreover, for concreteness, 
we shall assume that 
the $c_n$ scale according to the power-law relation
\begin{equation}
  c_{n} ~=~ c_{0} \left(\frac{m_n}{m_0}\right)^\gamma~,
\label{eq:CouplingScaling}
\end{equation}
where $c_0$ is the coupling associated with the lightest ensemble constituent
$\chi_0$ and where $\gamma$ is a dimensionless scaling exponent.

In summary, our model is characterized by six free parameters.  These are the 
masses $m_n$ of the three $\chi_n$, the parameters $c_0$ and $\gamma$ which specify the 
couplings between these fields and the mediator $\phi_u$, and the mass $m_{\phi_u}$ of 
the mediator itself.  For ease of notation, since we are assuming that $\phi_c$ and $\phi_t$ 
are sufficiently heavy that they play no role in the collider phenomenology of the model, 
we shall henceforth simply refer to $\phi_u$ and $m_{\phi_u}$ as $\phi$ and $m_\phi$, 
respectively.

\begin{figure}
  \centering
  \includegraphics[clip, width=0.45\textwidth]{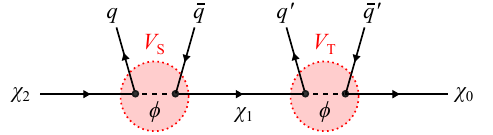}
  \caption{Realization of the tumbler event topology shown in 
  Fig.~\protect\ref{fig:decayschematic} within the context of our model.
  In particular, within our model, the secondary and tertiary vertices 
  $V_S$ and $V_T$ in Fig.~\protect\ref{fig:decayschematic} 
  are each now effectively realized as a pair of three-point vertices
  mediated by $\phi$.
  \label{fig:chainschematic}}
\end{figure}

Our interest in this model is primarily due to the tumbler signatures which result from
successive decays of the dark-sector states.  Indeed, the interaction Lagrangian in 
Eq.~(\ref{eq:Lint}) renders $\chi_1$ and $\chi_2$ unstable.  We shall primarily be 
interested in the regime within which the mediator is heavy, with 
$m_\phi > m_n$ for all $n$.  Within this regime, the leading contribution to the 
decay width $\Gamma_\phi$ of the mediator arises from to two-body decay processes of the 
form $\phi \ra q \bar{\chi}_n$.  By contrast, the leading contribution to the decay
widths of each $\chi_1$ and $\chi_2$ arise from three-body decay processes of the form 
$\chi_n \ra q\bar{q} \chi_m$ with $m < n$, each of which involves an off-shell 
mediator.  Thus, when a $\chi_2$ particle is produced at the primary 
interaction vertex $V_P$, there is a non-vanishing probability that it will decay via 
the process $\chi_2 \ra q\bar{q} \chi_1$, with $\chi_1$ in turn decaying via the process 
$\chi_1 \ra q\bar{q} \chi_0$.  The resulting decay chain is illustrated in
Fig.~\ref{fig:chainschematic}, where each black dot represents an interaction vertex 
associated with one of the Lagrangian terms in Eq.~(\ref{eq:Lint}).  Since the $\phi$
particles involved in the decay processes are both off shell, the red circles indicated
in the diagram, each of which encompasses two such interaction vertices, represent 
localized spacetime events.  If the $\chi_1$ and $\chi_2$ particles are both long-lived and 
each travel a macroscopic distance before they decay, the result is a tumbler, with these 
spacetime events corresponding to the secondary and tertiary decay vertices $V_S$ and $V_T$ 
indicated in Fig.~\ref{fig:decayschematic}. 

Although $\chi_1$ and $\chi_2$ are unstable, the lightest dark-sector state 
$\chi_0$ in our model is stabilized by an accidental $\mathbb{Z}_2$ symmetry
of the model under which $\phi$ and the $\chi_n$ are odd, whereas the fields of the SM are even.  
This symmetry, if unbroken, would render this particle absolutely stable --- and 
a potential dark-matter candidate~\cite{Dienes:2019krh}.  Alternatively, this 
symmetry could be broken by additional, highly suppressed interactions which permit $\chi_0$ 
to decay into final states involving SM particles alone.  However, as long as $\chi_0$ is
collider-stable --- \ie, sufficiently long-lived that virtually every $\chi_0$ particle 
produced within a collider detector escapes the detector well before it decays --- 
we shall not need to specify whether this $\mathbb{Z}_2$ symmetry is exact or approximate
for the purposes of understanding the collider phenomenology of the model.  In what follows, 
we shall therefore simply assume that $\chi_0$ is indeed collider-stable and consequently 
manifests itself as $\met$.


\FloatBarrier
\section{Surveying the Parameter Space\label{sec:SurveyParamSpace}}


Our first step is to identify regions of the parameter space of our model within which 
the prospects for observing a tumbler signature, either at the LHC or at a future hadron 
collider, are particularly auspicious.  The event rate for collider processes involving 
tumblers depends on several factors.  These include the cross-sections for the relevant 
production processes; the lifetimes of $\chi_1$, $\chi_2$, and $\phi$; and the probability 
that an on-shell $\phi$ or $\chi_2$ particle initially produced via one of these production 
processes will decay via an appropriate decay chain.

We begin by evaluating the total decay widths $\Gamma_\phi$ and $\Gamma_n$ and the 
branching fractions ${\rm BR}_{\phi n}$ and ${\rm BR}_{n \ell}$ for decay processes of the 
form $\phi^\dagger \ra \bar{q}\chi_n$ and $\chi_n \ra \bar{q}q\chi_\ell$, respectively.
In order to calculate these branching fractions, we must first evaluate the partial widths 
for all kinematically accessible decays of $\phi$, $\chi_2$, and $\chi_1$.  The partial 
width $\Gamma_{\phi n}\equiv \Gamma(\phi^\dagger \rightarrow \bar{q}\chi_n)$ for 
the decay process in which an on-shell mediator decays into a quark and an ensemble
constituent $\chi_n$ is~\cite{Dienes:2019krh}  
\begin{equation} 
  \Gamma_{\phi n} ~=~ \frac{c_{n}^2}{16 \pi}  
    \frac{( m_\phi^2 - m_n^2)^2}{m_{\phi}^3}~.
  \label{eq:Gammaphi}
\end{equation}
Likewise, the partial width $\Gamma_{n\ell} \equiv \Gamma(\chi_n \rightarrow \bar{q}q\chi_\ell)$ 
takes the form~\cite{Dienes:2019krh}   
\begin{eqnarray}
  \Gamma_{n \ell} &~=~& \frac{3c_{n}^2c_{\ell}^2}
    {256 \pi^2}\frac{m_\phi}{ r_{\phi n}^3} \Bigg[
    f_{\phi n \ell }^{(1)} -  f_{\phi n \ell }^{(2)} \ln (r_{n \ell}) \nonumber \\
    & &
    ~ + f_{\phi n \ell }^{(3)} \ln\left(\frac{1-r_{\phi n}^2}{1-r_{\phi n}^2r_{n \ell }^2}\right) 
    \Bigg]~,
  \label{eq:Gammanmeq}
\end{eqnarray}
where $r_{n\ell}\equiv m_\ell/m_n$, where $r_{\phi n}\equiv m_n/m_\phi$, and where 
\begin{eqnarray}
  f_{\phi n \ell }^{(1)} &~\equiv ~& 6r_{\phi n}^2(1-r_{n \ell}^2) 
      - 5r_{\phi n}^4(1-r_{n \ell}^4) \nonumber \\
     & & ~+ 2r_{\phi n}^6r_{n \ell}^2 (1-r_{n \ell}^2)\nonumber \\
  f_{\phi n \ell }^{(2)} &~\equiv~ & 4r_{\phi n}^8r_{n \ell}^4 \nonumber \\
  f_{\phi n \ell}^{(3)} &~\equiv ~&  6 - 8r_{\phi n }^2(1+r_{n \ell}^2) 
      - 2r_{\phi n}^8r_{n \ell}^4 \nonumber \\ 
    & & ~+ 2r_{\phi n}^4(1+4r_{n \ell}^2+r_{n \ell}^4)~.
\end{eqnarray}
The branching fractions of interest are then given by
\begin{equation}
  {\rm BR}_{\phi n} ~=~ \frac{\Gamma_{\phi n}}{\Gamma_\phi}~,~~~~~
  {\rm BR}_{n \ell} ~=~ \frac{\Gamma_{n \ell}}{\Gamma_n}~,
  \label{eq:BRS}
\end{equation}
where the total widths of $\phi$ and $\chi_n$ are respectively given by
\begin{equation}
  \Gamma_\phi ~=~ \sum_{n=0}^2 \Gamma_{n\phi}~,~~~~~
  \Gamma_n ~=~ \sum_{\ell=0}^{n-1} \Gamma_{n \ell}~.
  \label{eq:TotalWidths}
\end{equation}

We observe from the partial-width expressions in Eqs.~(\ref{eq:Gammaphi}) and 
(\ref{eq:Gammanmeq}) that $\Gamma_\phi \propto c_0^2$, while $\Gamma_n\propto c_0^4$.  
For $m_n \sim \mathcal{O}(100\mathrm{~GeV})$ and 
$m_\phi \sim \mathcal{O}(\mathrm{TeV})$, these expressions also imply that we must take 
$c_0 \ll 1$ in order for $\chi_1$ and $\chi_2$ to be sufficiently long-lived that 
their decays give rise to DVs.  Together, these two considerations imply that 
$\Gamma_\phi \gg \Gamma_n$ within regions of parameter space which give rise 
to tumblers.  As a result, within these regions of interest, any on-shell $\phi$ 
particle produced at the primary interaction vertex typically decays promptly into
a quark and one of the $\chi_n$.  

From the branching fractions in Eq.~(\ref{eq:BRS}), we may in turn determine 
the probability that a particular decay chain will arise from the decay of a 
$\phi$ or $\chi_n$ particle.  We shall let $P_{a_1a_2\ldots a_f}$ denote the 
probability of a given decay chain, where the sequence of $a_i \in \{\phi,2,1,0\}$ in 
the subscript indicates the set of $\phi$ and $\chi_n$ particles produced along the decay 
chain.  For example, $P_{\phi 2 0}$ represents the probability that an on-shell $\phi$ 
particle, once produced, decays directly to $\chi_2$, which subsequently decays directly
to $\chi_0$.  These decay-chain probabilities are simply the products of the relevant 
branching fractions.  Since $\chi_1$ decays via the process $\chi_1\ra \bar{q}q\chi_0$ with 
branching fraction ${\rm BR}_{10} = 1$, we have $P_{10} =1$.  There are two possible decay 
chains which can arise from the decay of a $\chi_2$ particle, given that $\chi_2$ can decay 
either to $\chi_0$ directly, or to $\chi_1$ which then subsequently decays to $\chi_0$.
The respective decay-chain probabilities are therefore $P_{20} = {\rm BR}_{20}$ 
and $P_{210} = {\rm BR}_{21}$.  The probabilities associated with decay chains initiated
by the decays of $\phi$ and $\chi_2$ can be evaluated in a similar manner.  We 
emphasize that each of these decay-chain probabilities represents the {\it total}\/ 
probability associated with the corresponding sequence of decays {\it regardless}\/ 
of the likelihood that these decays would occur within a collider detector.

We now consider the production processes through which $\phi$ and $\chi_n$ particles
can be produced at a hadron collider.  The accidental $\mathbb{Z}_2$ symmetry of our 
model ensures that particles which are odd under this symmetry will always be produced
in pairs.  The dominant scattering processes which give rise to a signal in our toy model are 
therefore $pp\ra\phi^\dagger\phi$, $pp\ra\phi\chi_n$ (and its Hermitian-conjugate process), and 
$pp\ra\overline{\chi}_m\chi_n$.  The Feynman diagrams which provide the leading contributions
to the cross-sections for these processes are shown in Ref.~\cite{Dienes:2019krh}.

\begin{figure*}
  \includegraphics[clip, width=0.43\linewidth]{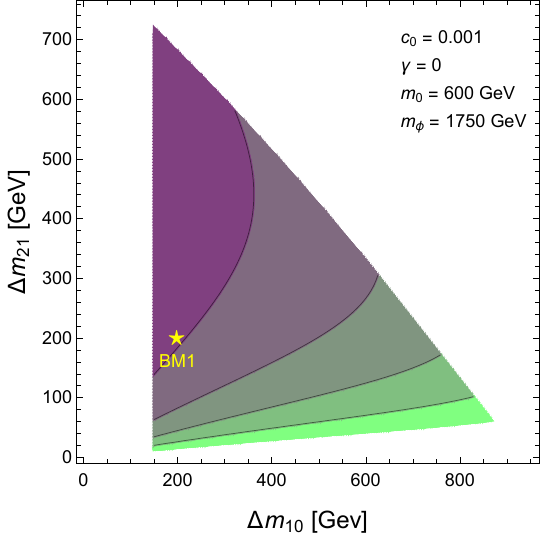}
  \includegraphics[clip, width=0.43\linewidth]{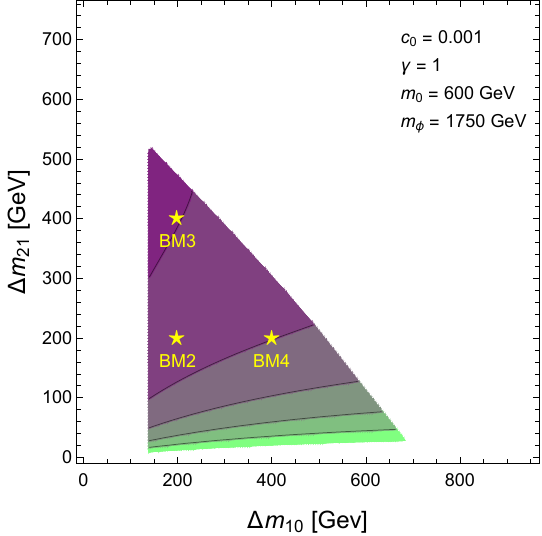}
  \includegraphics[clip, width=0.1\linewidth]{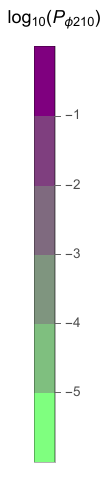}
\caption{Contours within the $(\Delta m_{10}, \Delta m_{21})$-plane of the overall probability 
  $P_{\phi 2 1 0} = \mathrm{BR}_{\phi 2}\mathrm{BR}_{21}$ 
  that an on-shell mediator $\phi$ will decay via the three-step decay chain which yields 
  a tumbler.  The results shown in the left panel correspond to the parameter assignments
  $m_\phi = 1750$~GeV, $m_0 = 600$~GeV, $c_0 = 0.001$, and $\gamma = 0$.  The results
  shown in the right panel correspond to the same assignments for $m_\phi$, $m_0$, and $c_0$, 
  but with $\gamma = 1$.  Regions of parameter space shown in white are
  not of interest from a tumbler perspective, either because one of the relevant decay 
  processes is kinematically forbidden, because one or both of the proper decay lengths 
  $c\tau_1$ and $c\tau_2$ of the unstable LLPs lies below $1$~mm or above 
  $10$~m, or because $P_{\phi 2 1 0} < 10^{-6}$.  The four stars which appear in the 
  panels of this figure indicate the parameter-space benchmarks defined in Table~\protect\ref{tab:benchmarks}. 
  \label{fig:BRPlots}}
\end{figure*}

Since $\phi$ carries color charge, the dominant contribution to the 
cross-section $\sigma_{\phi\phi}$ for the process $pp\ra\phi^\dagger\phi$ comes from 
diagrams which involve strong interactions alone.  By contrast, the diagrams which provide 
the dominant contribution to the cross-section $\sigma_{\phi n}$ for any process of the 
form $pp\ra\phi\chi_n$ each include one vertex which follows from the interaction Lagrangian 
in Eq.~(\ref{eq:Lint}).  Likewise, the diagrams which provide the dominant contribution to 
the cross-section $\sigma_{mn}$ for any process of the form $pp\ra\overline{\chi}_m\chi_n$ 
each include two such vertices.  These considerations imply that $\sigma_{\phi\phi}$ is 
independent of $c_0$, while $\sigma_{\phi n} \propto c_0^2$ and $\sigma_{m n}\propto c_0^4$.  
Thus, since $c_0 \ll 1$ within regions of parameter space which give rise to tumblers,
$pp\ra \phi^\dagger\phi$ typically dominates the production rate for tumbler 
events by several orders of magnitude within those regions.\footnote{In unusual circumstances 
wherein ${\rm BR}_{\phi 2}$ is suppressed by phase-space considerations and $\phi$ decays do not 
tend to produce tumblers, it is also possible that $pp\ra\phi\chi_2$ dominates this event 
rate.  However, since this possibility requires that the masses $m_2$ and $m_\phi$ be tuned 
such that they are nearly equal, we do not consider it further.}  As a result, while the 
branching fractions ${\rm BR}_{\phi n}$ and ${\rm BR}_{n \ell}$ depend on the values of 
$\gamma$, $c_0$, $m_0$, $m_1$, and $m_2$, the cross-section $\sigma_{\phi\phi}$ for the 
sole scattering process relevant for tumbler production at hadron colliders depends essentially 
on $m_\phi$ alone. 

Since $pp\ra \phi^\dagger\phi$ typically provides the dominant contribution to the 
tumbler event rate within our parameter-space region of interest, it is the decays of 
on-shell mediator particles which typically provide the dominant contribution to the 
tumbler-event rate.  The sole decay chain through which an on-shell $\phi$ particle, once 
produced by this process, can give rise to a tumbler is the chain in which this $\phi$ particle 
decays promptly to a $\chi_2$ particle, which then decays to a $\chi_1$ particle (which itself 
subsequently decays to a $\chi_0$ particle with ${\rm BR}_{10}=1$).  Thus, the decay-chain 
probability $P_{\phi 2 1 0} = {\rm BR}_{\phi 2}{\rm BR}_{21}$ for this sequence of decays is
a crucial figure of merit in assessing whether or not a given choice of our model parameters 
is likely to lead to a significant number of tumbler events at a hadron collider.

\begin{table*}
\begin{tabularx}{0.55\linewidth}{||C||c|c|c|c|c|c||}
\hline\hline  \multirow{3}{*}{~Benchmark~} &  
  \multicolumn{6}{c||}{~Input Parameters~}  \\
\cline{2-7} & \multirow{2}{*}{~$c_0$~} & \multirow{2}{*}{~~$\gamma$~~} 
  & ~$m_0$~ & ~$m_1$~ & ~$m_2$~~ 
  & ~$m_\phi$~  \\
  & & & ~(GeV)~ & ~(GeV)~ & ~(GeV)~ & ~(GeV)~ \\
  \hline 
BM1 & ~$0.001$~ & 0 & 600 & 800  & 1000 & 1750 \\
BM2 & ~$0.001$~ & 1 & 600 & 800  & 1000 & 1750 \\
BM3 & ~$0.001$~ & 1 & 600 & 800  & 1200 & 1750 \\
BM4 & ~$0.001$~ & 1 & 600 & 1000 & 1200 & 1750 \\
\hline \hline
\end{tabularx}
\caption{Definitions of our parameter-space benchmarks BM1 -- BM4.
\label{tab:benchmarks}}
\vskip 0.4cm
\begin{tabularx}{0.875\linewidth}{||C||c|c||c|c|c|c|c||c|c||}
\hline\hline  \multirow{3}{*}{~Benchmark~} & \multicolumn{2}{c||}{~Mass Splittings~} &
  \multicolumn{5}{c||}{~Branching Fractions~} & 
  \multicolumn{2}{c||}{~Proper Decay Lengths~} \\
\cline{2-10} & ~~$\Delta m_{10}$~~ & ~~$\Delta m_{21}$~~ & 
\multirow{2}{*}{~$\mathrm{BR}_{\phi 2}$~} & \multirow{2}{*}{~$\mathrm{BR}_{\phi 1}$~} & \multirow{2}{*}{~$\mathrm{BR}_{\phi 0}$~} & \multirow{2}{*}{~$\mathrm{BR}_{2 1}$~} & \multirow{2}{*}{~$\mathrm{BR}_{2 0}$~} & ~~$c\tau_2$~~ & ~~$c\tau_1$~~ \\ 
  & (GeV) & (GeV) & & & & & & (m) & (m)  \\ 
  \hline 
BM1 & 200 & 200 & 0.24 & 0.34 & 0.42 & 0.05 & 0.95 & ~$8.33\times 10^{-2}$~ & 2.42 \\
BM2 & 200 & 200 & 0.40 & 0.35 & 0.25 & 0.08 & 0.92 & ~$2.89\times 10^{-2}$~ & 1.36 \\
BM3 & 200 & 400 & 0.37 & 0.37 & 0.26 & 0.28 & 0.72 & ~$2.14\times 10^{-3}$~ & 1.36 \\
BM4 & 400 & 200 & 0.36 & 0.40 & 0.25 & 0.03 & 0.97 & ~$2.89\times 10^{-3}$~ & ~$3.15\times 10^{-2}$~ \\
\hline \hline
\end{tabularx}
\caption{Values for the mass splittings $\Delta m_{10} \equiv m_1 - m_0$ and 
  $\Delta m_{21} \equiv m_2 - m_1$, the branching fractions for all of the processes 
  via which $\phi$ and $\chi_2$ can decay, and the proper decay lengths $c\tau_1$ 
  and $c\tau_2$ of the unstable LLPs for each of the parameter-space benchmarks
  defined in Table~\protect\ref{tab:benchmarks}.
  \label{tab:benchquants}}
\end{table*}

\begin{figure*}
  \includegraphics[clip, width=0.43\linewidth]{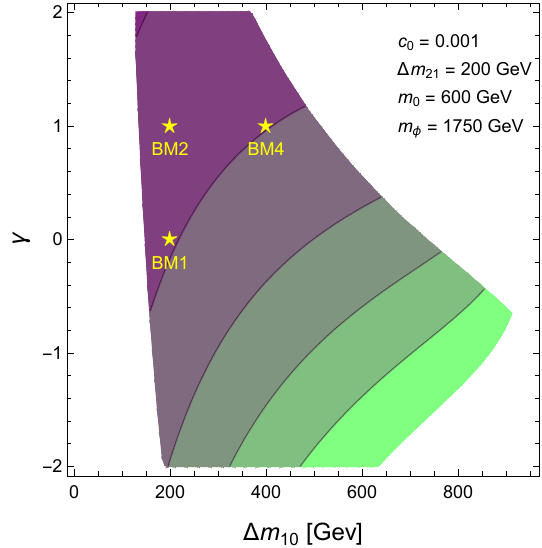}
  \includegraphics[clip, width=0.43\linewidth]{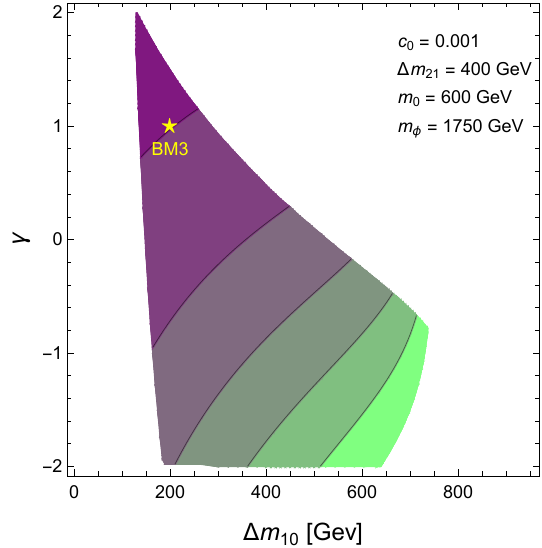}
  \includegraphics[clip, width=0.1\linewidth]{BRLegend.pdf}
\caption{Same as in Fig.~\protect\ref{fig:BRPlots}, except that the contours of $P_{\phi 2 1 0}$
  are shown within the $(\Delta m_{10},\gamma)$-plane for $\Delta m_{21} = 200$~GeV (left panel) and 
  $\Delta m_{21} = 400$~GeV (right panel).
  \label{fig:BRPlotsGamma}}
\end{figure*}


In order to assess which regions of the parameter space of our model are the
most promising for tumbler detection, we search for points at which the following criteria 
are satisfied.  First, the proper decay distances $c\tau_1$ and $c\tau_2$ of the unstable 
LLPs must each lie within the range $1~{\rm mm} < c\tau_n < 10~{\rm m}$.  These conditions 
ensure not only that a $\chi_1$ 
or $\chi_2$ particle has a significant probability of traveling an appreciable distance 
away from the location at which it was produced before it decays, but also that it has
a significant probability of decaying before it leaves the detector tracker.
Second, we require that $m_2 < m_\phi$ 
in order to ensure that the decay $\phi^\dagger \rightarrow \bar{q}\chi_2$ is kinematically 
allowed.  Third, we require that $P_{\phi 2 1 0}$  exceed a certain threshold.   
In general, P can be as high as $P_{\phi 2 1 0} \sim \mathcal{O}(0.1)$; indeed, 
this occurs despite the fact that $P_{\phi 2 1 0}$ is often suppressed by 
phase-space considerations which favor the decay of $\phi$, $\chi_2$, and 
$\chi_1$ directly to $\chi_0$.  That said, we shall nevertheless adopt the 
far more modest requirement $P_{\phi 2 1 0} \gtrsim 10^{-6}$ in our survey 
in order that we may better explore how this decay-chain probability varies 
across the parameter space as a whole.

In Fig.~\ref{fig:BRPlots}, we plot contours of $P_{\phi 2 1 0}$ in 
$(\Delta m_{10}, \Delta m_{21})$-space, where $\Delta m_{10} \equiv m_1 - m_0$ and
$\Delta m_{21} \equiv m_2 - m_1$.  Results are only shown for regions wherein all of the 
three criteria discussed above are satisfied; other regions appear in white.  
The results shown in the left panel correspond to the 
parameter assignments $m_\phi = 1750$~GeV, $m_0 = 600$~GeV, $c_0 = 0.001$, and 
$\gamma = 0$.  The results shown in the right panel correspond to the same assignments for 
$m_\phi$, $m_0$, and $c_0$, but with $\gamma = 1$.  

Broadly speaking, within these regions, the largest values of $P_{\phi 2 1 0}$ are obtained 
when $\Delta m_{10}$ is small and $\Delta m_{21}$ is large.  Moreover, we see that 
tumbler decay-chain probabilities as large as $P_{\phi 2 1 0} \sim \mathcal{O}(0.1)$ can 
arise within this region for $\gamma = 1$, whereas probabilities as large as 
$P_{\phi 2 1 0} \sim \mathcal{O}(0.01)$ can arise even for $\gamma = 0$.
Within the white region on the left side of each panel, the available phase space for the
decay $\chi_1\rightarrow \bar{q}q\chi_0$ is extremely small, and consequently 
$c\tau_1 > 10$~m.  By contrast, within the white region in the upper right corner 
of each panel, $m_2$ is quite large.  As a result, either the partial width for the 
decay $\chi_2\rightarrow \bar{q}q\chi_0$ becomes so large that $c\tau_2 < 1$~mm, or else 
$m_2 > m_\phi$ and the three-step decay chain which gives rise to tumblers is 
kinematically forbidden.  While the results shown in Fig.~\ref{fig:BRPlots} 
by no means represent an exhaustive survey of the parameter space of our model, they serve 
to highlight those regions which could potentially yield a significant 
number of tumbler events at the LHC or at future colliders.  

Guided by these results, then, we shall identify a set 
of four benchmark points within these regions for further study.  The parameter assignments 
which define these benchmark points are provided in Table~\ref{tab:benchmarks}.~  Each 
point is also labeled with a star in Fig.~\ref{fig:BRPlots}.  These benchmark points 
represent different combinations of the parameters $\gamma$, $m_1$, and $m_2$.
The mass splittings $\Delta m_{10}$ and $\Delta m_{21}$, the branching fractions 
for the different possible decay channels for $\phi$ and $\chi_2$, and the proper 
decay lengths of $\chi_1$ and $\chi_2$ for each of these benchmarks are provided 
in Table~\ref{tab:benchquants}.

It is also interesting to consider how our results for $P_{\phi 2 1 0}$ vary as a
function of the choice of the scaling exponent $\gamma$.
In Fig.~\ref{fig:BRPlotsGamma}, we plot contours of $P_{\phi 2 1 0}$ within the 
$(\Delta m_{10},\gamma)$-plane for $\Delta m_{21} = 200$~GeV (left panel) and 
$\Delta m_{21} = 400$~GeV (right panel).  The values we have adopted for $m_\phi$, 
$m_0$, and $c_0$ in both panels of the figure are the same as those adopted in
Fig.~\ref{fig:BRPlots}.  The locations of our parameter-space benchmarks are once
again indicated by the stars.  We see that increasing $\gamma$ with all other
parameters held fixed generally increases $P_{\phi 2 1 0}$.  Indeed, increasing 
this scaling exponent increases the ratios $c_2/c_1$ and $c_2/c_0$, and thereby increases
the branching fraction ${\rm BR}_{\phi 2}$ for the decay $\phi\ra q\overline{\chi}_2$ that
initiates the three-step decay chain which gives rise to tumblers.  By the same token,
however, increasing $\gamma$ also increases the total decay width of $\chi_2$.  For 
sufficiently large $\gamma$, the lifetime of this particle becomes such that 
$c\tau_2 < 1$~mm.  This is what occurs in the white region in the upper right corner of 
each panel.  On the other hand, when $\gamma < 0$, the decay $\phi\ra q\overline{\chi}_0$
dominates the width of $\phi$.  As a result, $P_{\phi 2 1 0}$ decreases rapidly with $\gamma$
until it drops below the threshold $P_{\phi 2 1 0} > 10^{-6}$, leading to the white 
region in the lower right of the plot. 
As in Fig.~\ref{fig:BRPlots}, the white region on the left side of each panel corresponds
to the region in which the available phase space for the decay 
$\chi_1\rightarrow \bar{q}q\chi_0$ is small and $c\tau_1 > 10$~m.

\FloatBarrier
\section{Constraints and Event Rates\label{sec:Discovery}}


In the previous section, we identified the parameter-space regions of our model 
which are particularly auspicious for producing tumblers.  In this section, we 
focus on these parameter-space regions of interest and 
assess whether a substantial population of tumbler events could yet await us at the 
LHC, given that no significant excess in discovery channels involving multiple DVs has 
been observed to date.

One important consideration is that 
our model not only gives rise to tumblers, but also yields contributions to the event rates 
in several additional detection channels for new physics.  These channels include the 
monojet~+~$\met$ channel, the multi-jet~+~$\met$ channel, and various channels involving 
displaced hadronic jets.  The results of new-physics searches which have been performed in 
these channels by the ATLAS and CMS Collaborations place additional constraints on the 
parameter space of our model.  Thus, we begin our analysis with a summary of the relevant
constraints from these searches.

\subsection{Displaced-Vertex Search Constraints\label{sec:DVsearch}}

A variety of searches for signatures of new physics involving displaced hadronic jets 
have been performed by both the CMS and ATLAS Collaborations.  The CMS Collaboration, for
example, has recently performed one search for displaced jets with 137~fb$^{-1}$ of integrated 
luminosity which incorporates timing information from the ECAL~\cite{Sirunyan:2019gut}, as
well as another, similar search with $132$~fb$^{-1}$ of integrated luminosity in which dedicated 
displaced-jet triggers and background-reduction techniques were applied~\cite{Sirunyan:2020cao}.  
A CMS search for displaced jets emanating from a pair of DVs resulting from the decays of 
pair-produced of LLPs was also recently performed with 140~fb$^{-1}$ of integrated 
luminosity~\cite{Sirunyan:2021kty}.  The results of these searches collectively supersede those 
from similar CMS searches for displaced jets performed at 36~fb$^{-1}$~\cite{Sirunyan:2018vlw} 
and $38.5$~fb$^{-1}$~\cite{Sirunyan:2018pwn} of integrated luminosity.
The extent to which machine-learning techniques could be used in order to further improve 
the reach of searches involving displaced jets was investigated in Ref.~\cite{Sirunyan:2019nfw}.

The ATLAS Collaboration has likewise performed a number of different searches for 
LLPs decaying into displaced jets.  These include searches for events in which the decay which 
produces the jets occurs within the tracker~\cite{Aaboud:2017iio}, within the
calorimeter~\cite{Aaboud:2019opc}, or in the muon chamber~\cite{Aaboud:2018aqj}.
An ATLAS search has also been performed for multiple LLPs decaying to
jets in the same event, where one LLP decays within the tracker and the other decays 
within the muon chamber~\cite{Aad:2019xav}.  All of these searches are performed with
roughly 35~fb$^{-1}$ of integrated luminosity, though the precise value of 
the integrated luminosity varies slightly among these searches.  Owing primarily to the
substantially lower integrated luminosity, these ATLAS searches are not as constraining as 
the CMS searches.  For this reason, we focus on the results of the CMS searches in what follows. 

The results in Refs.~\cite{Sirunyan:2019gut,Sirunyan:2020cao,Sirunyan:2021kty} collectively
constrain new-physics scenarios involving LLPs with lifetimes $\tau_\chi$ in the range 
$10^{-4}\mathrm{~m} \lesssim c\tau_\chi \lesssim 10\mathrm{~m}$ which decay into final states
involving hadronic jets.  In particular, they impose an upper bound on the product 
$\sigma_{\chi\chi}{\rm BR}_{\chi j}^2$ of the LLP pair-production cross-section and the 
square of the branching fraction of the LLP into such final states.  
While the precise numerical value of this upper bound depends on the production and decay 
kinematics of the LLP and on $\tau_\chi$, the bound falls within the range $0.05$ -- $0.5$~fb 
across almost this entire range of $\tau_\chi$.

\subsection{Multi-Jet Search Constraints\label{sec:Multijetsearch}}

Searches performed by both the CMS and ATLAS Collaborations also place constraints on 
beyond-the-Standard-Model (BSM) contributions to the event rate for processes 
involving multiple hadronic jets and $\met$.  The searches most relevant for 
constraining the parameter space of our model are those designed to uncover 
evidence of heavy decaying particles --- \eg,  squarks and gluinos in supersymmetry. 
The leading CMS constraints from multi-jet~+~$\met$ searches are those derived from 
searches~\cite{Sirunyan:2019ctn,Sirunyan:2019xwh} performed with 137~fb$^{-1}$ of integrated 
luminosity.  These include searches involving standard techniques developed in order to 
search for squarks and gluinos more generally, as well as searches which focus on 
specific scenarios 
for which the use of the $M_{T2}$ variable is particularly advantageous in terms of 
discovery potential.  The results of these analyses supersede those of a prior CMS
study~\cite{Sirunyan:2018vjp} performed with 36~fb$^{-1}$ of integrated luminosity.

The leading ATLAS constraints on excesses in the multi-jet~+~$\met$ channel of the 
sort obtained in our model are those derived from a search for squarks and gluinos
performed with 139~fb$^{-1}$ of integrated luminosity~\cite{Aad:2020aze}.
These results supersede those obtained from a prior ATLAS study~\cite{Aaboud:2017vwy}
performed with 36~fb$^{-1}$ of integrated luminosity.

In each of these ATLAS or CMS analyses, 95\%-C.L.\ exclusion limits on the product 
of the production cross-section $\sigma$, the signal acceptance $A$, and the detection 
efficiency $\epsilon$ are obtained for a variety of signal regions, which are defined 
differently in the different studies.  These limits are also interpreted in each case as 
constraints on the parameter space of a simplified supersymmetric model involving
a single flavor of squark $\tilde{q}$ which is pair-produced via the process 
$pp\ra \tilde{q}^\dagger \tilde{q}$ and subsequently decays directly to a light quark and the 
lightest neutralino $\widetilde{\chi}_1$.  All other sparticles are assumed to be extremely 
heavy in this scenario, and therefore to play no role in the pair-production process.  Since
$\tilde{q}$ and $\widetilde{\chi}_1$ in this supersymmetric model have the same quantum numbers 
as $\phi$ and $\chi_0$ in our model, respectively, these bounds may be applied to our model
directly.  The constraint contours within the $(m_{\tilde{q}}, m_{\widetilde{\chi}_1})$-plane 
obtained in Refs.~\cite{Sirunyan:2019xwh,Sirunyan:2019ctn,Aad:2020aze} are all roughly 
commensurate and, roughly speaking, exclude the region of this plane wherein 
$m_{\tilde{q}} \lesssim 1250$~GeV and $m_{\widetilde{\chi}_1} \lesssim 500$~GeV.  

Given that the values of the parameters $m_\phi$ and $m_0$ for all of our 
parameter-space benchmarks lie well outside the corresponding region in the 
$(m_\phi,m_0)$-plane, we may safely assume that our benchmarks are consistent 
with these constraints.  Moreover, in many of these searches, events
are vetoed in which a significant fraction of the jets are produced at locations 
other than the primary vertex.  

\subsection{Monojet Search Constraints\label{sec:Monojetsearch}}

The most stringent bound on excesses of events in the monojet~+~$\met$ channel is
that from an ATLAS study~\cite{Aad:2021egl} 
performed with 139~fb$^{-1}$ of integrated luminosity.  The results of this study 
supersede those from a similar ATLAS study~\cite{Aaboud:2017phn} performed with 36~fb$^{-1}$ 
of integrated luminosity.  Similar searches have been performed by the CMS Collaboration, 
but with far lower integrated luminosity.

The results in Ref.~\cite{Aad:2021egl} are quoted in a model-independent way for
several different signal regions corresponding with different threshold values taken 
for the magnitude $|\vec{\mathbf{p}}_{\rm T}^{~(\mathrm{rec})}|$ of the transverse 
momentum which recoils against the jet.  For each of these signal regions, a 
95\%-C.L.\ exclusion limit on the product of the production cross-section $\sigma$, 
signal acceptance $A$, and detection efficiency $\epsilon$ is obtained.  
These limits range from 
$\sigma \times A \times \epsilon < 736$~fb for a threshold of 
$|\vec{\mathbf{p}}_{\rm T}^{~(\mathrm{rec})}| > 200$~GeV to 
$\sigma \times A \times \epsilon < 0.3$~fb for
$|\vec{\mathbf{p}}_{\rm T}^{~(\mathrm{rec})}| > 1200$~GeV.  
Moreover, these limits are also interpreted as constraints
on the parameter space of the same simplified supersymmetric model
that was considered in the multi-jet analysis discussed above.
Once again, these constraints may be applied to our model directly.

The monojet constraints on this simplified supersymmetric model turn out
to be relevant within the same rough region of the 
$(m_{\tilde{q}}, m_{\widetilde{\chi}_1})$-plane as the multi-jet constraints
discussed above, but also are slightly less restrictive.  We
therefore expect that the same is true of the monojet constraints on our 
example model within the $(m_\phi,m_0)$-plane.  Thus, we may safely assume 
that our benchmarks are consistent with these constraints.  
In summary, then, it is clear that the dominant constraints on our model within our 
parameter-space region of interest are those from displaced-jet searches.  We shall
therefore focus primarily on these constraints in what follows.

\subsection{Effective Cross-Sections and Event Rates\label{sec:EffXSecs}}

In order to assess the impact of these experimental constraints on our model, we must  
evaluate the net contributions to the event rates for a
number of different detection channels.  In particular, we can identify four relevant 
channels, each of which is associated with a particular set of collider processes:
\begin{itemize}

  \item \underline{Tumbler class}: processes which involve at least one tumbler.  Processes 
    in this class are the primary focus of this paper.

  \item \underline{DV class}: processes which involve at least one DV, regardless of whether
    this DV is part of a tumbler.  The event rates associated with processes in this
    class are constrained by the results of displaced-jet searches. 
    
  \item \underline{Multi-jet class}: processes which do not give rise to any DVs, but 
    instead yield a pair of prompt hadronic jets and missing transverse energy.  Processes in 
    this class contribute to the event rate in the multi-jet~+~$\met$ channel.
    
  \item \underline{Monojet class}: processes which do not give rise to any DVs, 
    but instead yield a single prompt hadronic jet and missing transverse energy.  Processes in 
    this class contribute to the event rate in the monojet~+~$\met$ channel.
     
\end{itemize}
We emphasize that these classes are not mutually exclusive.  For example, all processes 
in the tumbler class necessarily include DVs and are therefore also part of the
DV class.  We also emphasize that all processes within a particular class are not 
completely equivalent.  One example of this is that the contributions from some DV-class processes 
may not be as stringently constrained by existing DV searches as the contributions from other
such processes as a consequence of differences in kinematics and the event-selection criteria 
involved.  Another example, as we shall discuss further in Sect.~\ref{sec:ResultsRecon}, is that
tumbler-class processes in which one or more additional hard jets are produced at the primary 
vertex are significantly more useful for reconstructing the masses and lifetimes of the $\chi_n$.
Nevertheless, as we shall see, this classification is useful in categorizing the contributions
from our model to the event rates in different detection channels.

\begin{table*}
\begin{tabular}{||c|c|c|c|c||}
\hline\hline  \multirow{2}{*}{~~~~~First Chain~~~~~} & 
  \multirow{2}{*}{~~~~~Second Chain~~~~~~} & 
  \multirow{2}{*}{~Tumblers~} & ~Displaced~ & ~Prompt~ \\
 & & & Vertices & Jets \\
\hline
\multicolumn{5}{||c||}{From $pp\rightarrow \phi\phi$ Production}\\
\hline 
$~\phi\ra \chi_2\ra \chi_1\ra \chi_0$~ & $\phi\ra \chi_2\ra \chi_1\ra \chi_0$ & 2T &     & $2j$ \\
$\phi\ra \chi_2\ra \chi_1\ra \chi_0$ & $\phi\ra \chi_2\ra \chi_0$           &  T &  DV & $2j$ \\
$\phi\ra \chi_2\ra \chi_1\ra \chi_0$ & $\phi\ra \chi_1 \ra\chi_0$           &  T &  DV & $2j$ \\
$\phi\ra \chi_2\ra \chi_1\ra \chi_0$ & $\phi\ra \chi_0$                     &  T &     & $2j$ \\
$\phi\ra \chi_2\ra \chi_0$           & $\phi\ra \chi_2\ra\chi_0$            &    & 2DV & $2j$ \\
$\phi\ra \chi_2\ra \chi_0$           & $\phi\ra \chi_1\ra\chi_0$            &    & 2DV & $2j$ \\
$\phi\ra \chi_2\ra \chi_0$           & $\phi\ra \chi_0$                     &    &  DV & $2j$ \\
$\phi\ra \chi_1\ra \chi_0$           & $\phi\ra \chi_2\ra\chi_0$            &    & 2DV & $2j$ \\
$\phi\ra \chi_1\ra \chi_0$           & $\phi\ra \chi_1\ra\chi_0$            &    & 2DV & $2j$ \\
$\phi\ra \chi_0$                     & $\phi\ra \chi_0$                     &    &     & $2j$ \\
\hline
\multicolumn{5}{||c||}{From $pp\rightarrow \phi\chi_n$ Production}\\
\hline
$\phi\ra \chi_2\ra \chi_1\ra \chi_0$ & $\chi_2\ra \chi_1\ra \chi_0$         & 2T &     &  $j$ \\
$\phi\ra \chi_2\ra \chi_1\ra \chi_0$ & $\chi_2\ra \chi_0$                   &  T &  DV &  $j$ \\
$\phi\ra \chi_2\ra \chi_1\ra \chi_0$ & $\chi_1 \ra\chi_0$                   &  T &  DV &  $j$ \\
$\phi\ra \chi_2\ra \chi_1\ra \chi_0$ & $\chi_0$                             &  T &     &  $j$ \\
$\phi\ra \chi_2\ra \chi_0$           & $\chi_2\ra \chi_1\ra \chi_0$         &  T &  DV &  $j$ \\
$\phi\ra \chi_2\ra \chi_0$           & $\chi_2\ra \chi_0$                   &    & 2DV &  $j$ \\
$\phi\ra \chi_2\ra \chi_0$           & $\chi_1 \ra\chi_0$                   &    & 2DV &  $j$ \\
$\phi\ra \chi_2\ra \chi_0$           & $\chi_0$                             &    &  DV &  $j$ \\
$\phi\ra \chi_1\ra \chi_0$           & $\chi_2\ra \chi_1\ra \chi_0$         &  T &  DV &  $j$ \\
$\phi\ra \chi_1\ra \chi_0$           & $\chi_2\ra \chi_0$                   &    & 2DV &  $j$ \\
$\phi\ra \chi_1\ra \chi_0$           & $\chi_1 \ra\chi_0$                   &    & 2DV &  $j$ \\
$\phi\ra \chi_1\ra \chi_0$           & $\chi_0$                             &    &  DV &  $j$ \\
$\phi\ra \chi_0$                     & $\chi_2\ra \chi_1\ra \chi_0$         &  T &     &  $j$ \\
$\phi\ra \chi_0$                     & $\chi_2\ra \chi_0$                   &    &  DV &  $j$ \\
$\phi\ra \chi_0$                     & $\chi_1 \ra\chi_0$                   &    &  DV &  $j$ \\
$\phi\ra \chi_0$                     & $\chi_0$                             &    &     &  $j$ \\
\hline
\multicolumn{5}{||c||}{From $pp\rightarrow \chi_m\chi_n$ Production}\\
\hline
$\chi_2\ra \chi_1\ra \chi_0$         & $\chi_2\ra \chi_1\ra \chi_0$         & 2T &     &      \\
$\chi_2\ra \chi_1\ra \chi_0$         & $\chi_2\ra \chi_0$                   &  T &  DV &      \\
$\chi_2\ra \chi_1\ra \chi_0$         & $\chi_1 \ra\chi_0$                   &  T &  DV &      \\
$\chi_2\ra \chi_1\ra \chi_0$         & $\chi_0$                             &  T &     &      \\
$\chi_2\ra \chi_0$                   & $\chi_2\ra \chi_0$                   &    & 2DV &      \\
$\chi_2\ra \chi_0$                   & $\chi_1\ra \chi_0$                   &    & 2DV &      \\
$\chi_2\ra \chi_0$                   & $\chi_0$                             &    &  DV &      \\
$\chi_1\ra \chi_0$                   & $\chi_1\ra \chi_0$                   &    & 2DV &      \\
$\chi_1\ra \chi_0$                   & $\chi_0$                             &    &  DV &      \\
$\chi_0$                             & $\chi_0$                             &    &     &      \\
\hline \hline
\end{tabular}
\caption{List of the possible event topologies which can arise within our model from 
  pair-production process of the form $pp\ra \phi\phi$, $pp\ra\phi\chi_n$, and 
  $pp\ra\chi_m\chi_n$.  The entries in each column describe the corresponding properties 
  of these topologies, with notation as described in the text.
  \label{tab:DecayChainList}}
\end{table*}

Contributions to the total event rate for each of these four classes of processes 
can in principle arise from a variety of different event topologies --- \ie, different 
combinations of
production processes.  In Table~\ref{tab:DecayChainList}, we list all possible such event 
topologies which can arise from pair-production processes of the forms $pp\ra \phi\phi$, 
$pp\ra\phi\chi_n$, and $pp\ra\chi_m\chi_n$.  The first column indicates the structure of 
the longer decay chain in the event, while the second column indicates the structure of 
the shorter decay chain.  An additional jet is produced by the decay of each 
mediator, while an additional pair of jets is produced by the decay of each LLP.~
However, for clarity, we have omitted mention of these particles in these columns
of the table.  Moreover, since there is no heuristic difference in terms of collider 
phenomenology between the decay chains precipitated by the decays of $\phi$ and $\chi_n$ and 
the decay chains precipitated by the decays of their anti-particles $\phi^\dagger$ 
and $\overline{\chi}_n$, we do not distinguish between particle and anti-particle decay chains.  
The third column of the table indicates whether the process gives rise to one or more tumblers 
at a collider.  An entry of ``T'' in this column indicates that the process gives rise to a 
single tumbler, while an entry of ``2T'' indicates that the process gives rise to two tumblers, 
one from each decay chain.  Likewise, the fourth column indicates whether or not the process 
gives rise to an isolated DV --- \ie, a DV which is {\it not}\/ part 
of a tumbler.  An entry of ``DV'' in this column indicates the presence of a single such vertex, 
while an entry of ``2DV'' indicates the presence of such vertices.  Finally, the fifth column 
indicates the presence of one or more prompt jets in the event.  An entry of ``$j$'' indicates 
the presence of one such jet, while an entry of ``2$j$'' indicates the presence of two such jets.  
We note that since every decay chain which occurs in our model terminates with $\chi_0$, 
every event which results from any of the processes listed in this table also includes $\met$.  

\begin{figure*}
  \includegraphics[clip, width=0.43\linewidth]{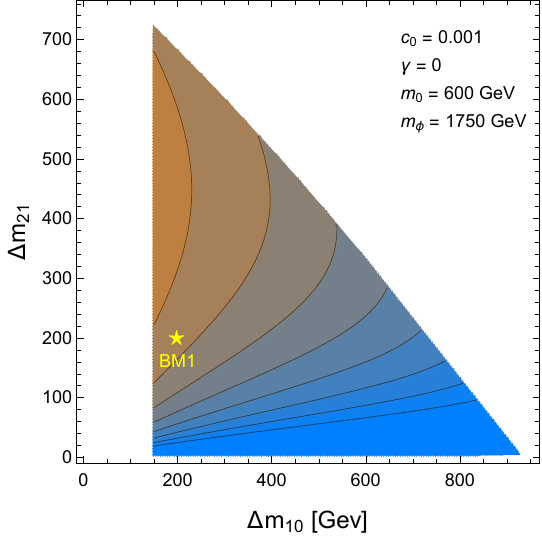}
  \includegraphics[clip, width=0.43\linewidth]{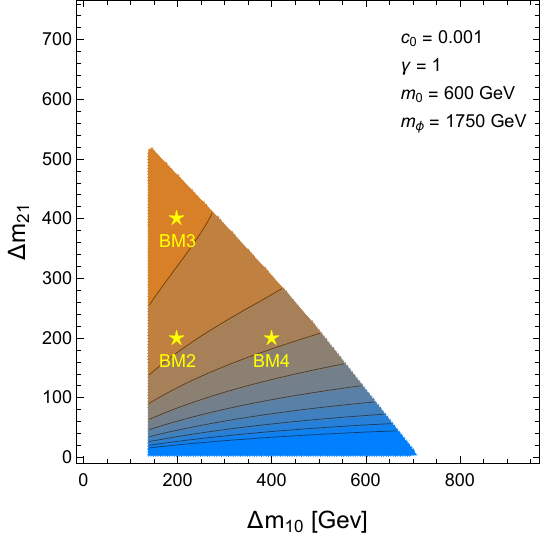}
  \includegraphics[clip, width=0.105\linewidth]{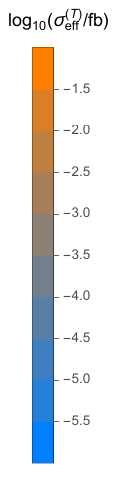}
\caption{Contours within the $(\Delta m_{10},\Delta_{21})$-plane of the effective cross-section 
  $\sigma_{\rm eff}^{(T)}$ defined in Eq.~(\protect\ref{eq:sigmaEffDef}) for processes 
  involving at least one tumbler at the $\sqrt{s} = 14$~TeV HL-LHC.~
  The results displayed in the left and right panels correspond to the 
  parameter assignments in corresponding panels of Fig.~\ref{fig:BRPlots}.
  As in Fig.~\protect\ref{fig:BRPlots}, results are shown only 
  within regions wherein all decay processes involved in the production of a tumbler are 
  kinematically allowed, where $c\tau_1$ and $c\tau_2$ both satisfy the criterion 
  $1~\mathrm{mm} < c\tau_n < 10$~m, and where $c\tau_\phi < 0.1\mathrm{~mm}$.  
  The four stars indicate the locations 
  of the parameter-space benchmarks defined in Table~\protect\ref{tab:benchmarks}. 
  \label{fig:sigmaEffs}}
\end{figure*}

For each of the four class of processes $\alpha$ itemized above, we define an effective 
cross-section $\sigma_{\rm eff}^{(\alpha)}$ which represents the sum of the individual 
contributions from all combinations of production and decay processes listed in
Table~\ref{tab:DecayChainList} that contribute to the overall event rate for processes 
in that class.  Each such individual contribution to $\sigma_{\rm eff}^{(\alpha)}$ is the 
product of the cross-section $\sigma_{a_1a_2}$ for the pair-production process 
$pp\rightarrow a_1 a_2$, where $a_i\in \{\phi,2,1,0\}$, and the two decay-chain 
probabilities $P_{a_1,c_1}$ and $P_{a_2,c_2}$ associated with the decay chains 
on each side of the event.  The index $c_i$ appearing in these probabilities represents the
sequence of particles produced from the decay of the corresponding initial particle $a_i$ and 
includes the null decay chain in the event that the initial particle is stable, in which case 
the corresponding decay-chain probability is unity.  In other words, our effective 
cross-section is    
\begin{equation}
  \sigma_{\rm eff}^{(\alpha)} ~\equiv~ \sum_{a_1} \sum_{a_2} 
    \sum_{c_1}\sum_{c_2} \big[\sigma_{a_1 a_2} P_{a_1,c_1}P_{a_2,c_2}
    \big]_\alpha~,
  \label{eq:sigmaEffDef}
\end{equation} 
where the subscript $\alpha$ on the brackets enclosing the summand indicates 
that only event topologies associated with the corresponding class of processes are       
included in the sum.  Indeed, it is the product of this effective cross-section and the 
integrated luminosity which yield the overall event count for the corresponding class of 
processes.

In Fig.~\ref{fig:sigmaEffs}, we show contours of the effective cross-section 
$\sigma_{\rm eff}^{(T)}$  for tumbler-class processes in $(\Delta m_{10},\Delta m_{21})$-space.
Cross-sections for all of the individual production processes were computed using the
\texttt{MG5\_aMC@NLO} code package~\cite{Alwall:2014hca} for a center-of-mass energy 
$\sqrt{s} = 14$~TeV.~  The results displayed in the left and right panels of the figure 
correspond to the parameter assignments in the corresponding panels of Fig.~\ref{fig:BRPlots}.
As in Fig.~\protect\ref{fig:BRPlots}, results are shown only within regions wherein all decay 
processes involved in the production of a tumbler are kinematically allowed, where the proper 
decay lengths $c\tau_1$ and $c\tau_2$ of the unstable LLPs both satisfy the criterion
$1~\mathrm{mm} < c\tau_n < 10$~m, and where the proper 
decay length of the mediator satisfies $c\tau_\phi < 0.1\mathrm{~mm}$.  However, no minimum
threshold for $P_{\phi 210}$ is imposed.  The four stars once again indicate the locations of 
the parameter-space benchmarks defined in Table~\protect\ref{tab:benchmarks}.

We observe that the contours of $\sigma_{\rm eff}^{({\rm T})}$ displayed in
Fig.~\ref{fig:sigmaEffs} have roughly the 
same shape as the contours of $P_{\phi 210}$ displayed in Fig.~\ref{fig:BRPlots}.   
This follows from the fact that $pp\ra \phi^\dagger\phi$ vastly dominates the event 
rate within our parameter-space region of interest.  As discussed in 
Sect.~\ref{sec:SurveyParamSpace}, the cross-section for this process depends essentially
on $m_\phi$ alone, and is therefore roughly uniform across the 
$(\Delta m_{10},\Delta m_{21})$-plane shown in each panel.  More importantly, however,
we also observe that an effective cross-section of order  
$\sigma_{\rm eff}^{({\rm T})}\sim \mathcal{O}(1$~--~$ 100~\mathrm{ab})$ for tumbler-class
processes can be achieved across a substantial region of our parameters space --- a region 
which includes the locations of all four of our parameter-space benchmarks.  
Given the integrated luminosity $\mathcal{L}_{\rm int} = 3000$~fb$^{-1}$ anticipated
for the full HL-LHC run, cross-sections of this order are in principle expected to give 
rise to a significant number of tumbler events at the HL-LHC.

In Table~\ref{tab:eventcounts}, we list the values of $\sigma_{\rm eff}^{({\rm T})}$ obtained
for each of our four benchmarks, along with the the respective effective cross-sections 
$\sigma_{\rm eff}^{({\rm DV})}$ and $\sigma_{\rm eff}^{(Nj)}$ for DV-class and multi-jet-class 
processes.  Also shown in the figure are the corresponding total numbers of tumbler events 
expected after Run~2 of the LHC ($\mathcal{L}_{\rm int} = 137$~fb$^{-1}$) and after the full
HL-LHC run ($\mathcal{L}_{\rm int} = 3000$~fb$^{-1}$).  We quote this number of events as   
$2\sigma_{\rm eff}^{({\rm T})}\mathcal{L}_{\rm int}$ in order to account for the contributions 
from both the CMS and ATLAS detectors.
While $\sigma_{\rm eff}^{({\rm T})}$ varies significantly across the 
$(\Delta m_{10},\Delta m_{21})$-plane shown in the panels of Fig.~\ref{fig:sigmaEffs}, we 
find that $\sigma_{\rm eff}^{({\rm DV})}$ and $\sigma_{\rm eff}^{(Nj)}$ are far less sensitive 
to the values of $\Delta m_{10}$ and $\Delta m_{21}$ within these same regions.  Indeed, we 
find that both of these effective cross-sections remain roughly within a single order of 
magnitude across this same region of $(\Delta m_{10},\Delta m_{21})$-space. 

\begin{table*}
\begin{tabular}{||c||c|c|c||c|c|||}
\hline\hline  \multirow{2}{*}{~Benchmark~} &  
  \multicolumn{3}{c||}{~$\sigma_{\mathrm{eff}}^{(\alpha)}$~(fb)~} & 
  \multicolumn{2}{c||}{Tumbler Events} \\
\cline{2-6} & ~~Tumblers~~ & ~~DV~~ & ~Multi-Jet~+~$\met$~ & 
  ~LHC Run 2 (137~fb$^{-1}$)~ & ~ HL-LHC (3000~fb$^{-1}$)~  \\
\hline 
BM1 & ~$1.5\times10^{-3}$~  & ~$5.3\times 10^{-2}$~ & $1.1\times 10^{-2}$  & 0.4 & 9.2  \\
BM2 & ~$4.3\times 10^{-3}$~ & $6.1\times 10^{-2}$ & $4.0\times 10^{-3}$  & 1.1 & 25.6  \\
BM3 & ~$1.3\times 10^{-2}$~ & $6.0\times 10^{-2}$ & $4.3\times 10^{-3}$  & 3.7 & 76.1  \\
BM4 & ~$1.4\times 10^{-3}$~ & $6.1\times 10^{-2}$ & $3.9\times 10^{-3}$  & 0.4 & 8.1 \\
\hline \hline
\end{tabular}
\caption{The effective cross-sections $\sigma_{\mathrm{eff}}^{(\alpha)}$ for tumbler-class, 
  DV-class, and multi-jet-class processes for our parameter-space benchmarks.  Also shown 
  are the total numbers of tumbler events expected after Run~2 of the LHC and after the full 
  HL-LHC run.  We quote this number of events as 
  $2\sigma_{\rm eff}^{({\rm T})}\mathcal{L}_{\rm int}$ in order to account for the 
  contributions from both the CMS and ATLAS detectors.
  \label{tab:eventcounts}}
\end{table*}

One of the primary messages of Table~\ref{tab:eventcounts} is that  
the effective cross-section $\sigma_{\rm eff}^{({\rm DV})}$ for each of our parameter-space 
benchmarks is $\sigma_{\rm eff}^{({\rm DV})} \lesssim 0.06$~fb$^{-1}$.  Such cross-sections
are consistent with the constraints from displaced-jet 
searches quoted above.  We have also confirmed, using the recasting tools 
associated with the {\tt MadAnalysis~5}~\cite{Conte:2012fm} package, that each of 
these benchmarks is consistent with the LLP-search results~\cite{Araz:2021akd} currently 
incorporated into the {\tt MadAnalysis} database.
We may therefore conclude that a significant number of both 
tumbler events and events involving DVs of any sort could potentially still be awaiting 
discovery at the LHC or at future colliders, even though no significant excess in such events 
has been observed to date.  Although the above cross-sections lie very close to the exclusion 
limits from displaced-jet searches, we also note that there are regions of our parameter
space wherein $\sigma_{\rm eff}^{({\rm DV})}$ lies even further below the bound from 
displaced-jet searches, tumblers still arise, and all additional constraints are satisfied. 

Looking ahead, in order to assess what the results in Table~\ref{tab:eventcounts} portend
in terms of the prospects for identifying a signal of new physics within the context of our 
model at the HL-LHC, we must take into account the relevant SM backgrounds.
Fortunately, one of the advantages of searching for signal processes which lead to DVs is that 
these backgrounds are typically extremely low.  
One such background arises from SM processes which involve genuine 
DVs --- for example, those associated with the decays of long-lived 
$B$- and $K$-mesons.  However, the visible particles produced by these 
decays tend to be highly collimated whenever they are highly energetic as a result of
the relatively small masses of the SM hadrons.  By contrast, the particles produced
by the decays of heavy LLPs into final states comprising significantly lighter particles
are typically far less collimated.  Indeed, this is the case for our example model 
when $\Delta m_{21}$ and $\Delta m_{10}$ are both 
$\mathcal{O}(100~{\rm GeV})$, Within this regime, cuts on variables which reflect 
the degree to which the visible particles produced at a DV are collimated --- such as 
the uncertainty in the distance $d_{\rm BV}$ between the 
primary vertex and the DV~\cite{Sirunyan:2018pwn,Sirunyan:2021kty} or, in the case
of our example tumbler model, the angle between the three-momenta of the two 
reconstructed jets --- can be quite effective in reducing this background 
without a significant loss in the number of signal events.

After the contribution involving genuine DVs is suppressed in this way,
the dominant contribution to the SM background in searches for displaced jets at 
the LHC is generally the one which arises as a consequence of multi-jet 
events in which poorly reconstructed tracks lead to the identification of spurious 
DVs~\cite{Sirunyan:2018pwn,Sirunyan:2021kty}.  
Most such events arise from purely strong-interaction processes.
Since both $\tau_1$ and $\tau_2$ satisfy $c \tau_n > \mathcal{O}(1\mathrm{~mm})$ 
for all of our benchmarks, the DVs that result from $\chi_1$ and $\chi_2$ decay 
are typically significantly farther that $0.1~\mu$m away from the primary 
vertex.  For displacements
of this size, events involving additional primary vertices from pile-up do not represent a 
significant background~\cite{Sirunyan:2021kty}.  Thus, a rough estimate of the background 
event rate at the end of the full HL-LHC run can be obtained simply by scaling 
the expected number of background events obtained from searches using Run~2 data after the 
application of all relevant cuts by the ratio of the corresponding integrated luminosities.  

The expected number of background events in any given displaced-jet search depends on 
the particular set of event-selection criteria employed, but the leading searches discussed 
above yield $0.1$~--~$0.7$ background events~\cite{Sirunyan:2020cao,Sirunyan:2021kty} 
at an integrated luminosity of around 137~fb$^{-1}$.  Thus, one would expect 
around $2.2$~--~$17.5$ background events at the end of the full HL-LHC run.
We also note that this background estimate is actually a conservative one, given that 
improvements in machine-learning approaches to LLP tagging have the potential to further 
reduce SM backgrounds without a significant loss in the signal-event rate~\cite{Sirunyan:2019nfw}. 
By contrast, the signal efficiency obtained for the same cuts is typically around 
$\epsilon_{\rm S} \sim 0.45$~--~$0.75$. Values within this range are obtained
in both Ref.~\cite{Sirunyan:2020cao} and Ref.~\cite{Sirunyan:2021kty} for 
event topologies analogous to the one we consider here.
Thus, given the results in Table~\ref{tab:eventcounts}, we see that a significant excess of
DV-class events would be observed at the HL-LHC for all of our benchmarks.  Moreover, for BM2
and BM3, this excess would include a substantial number of tumbler events.  The observation
of such an excess would clearly prompt significant additional investigation into how we might 
better probe the underlying physics responsible for this excess.  It is toward this 
question that we now turn.


\FloatBarrier
\section{Distinguishing Tumblers via Mass Reconstruction\label{sec:ResultsRecon}}


While we have shown that our model can give rise to a significant number of tumbler 
events at the LHC, we have also shown that it typically simultaneously gives rise to a 
far larger number of non-tumbler DV-class events --- a substantial fraction of which 
likewise involve more than one DV.~  Indeed, any of the processes listed in
Table~\ref{tab:DecayChainList} in which each decay chain involves only a single $\chi_1$ 
or $\chi_2$ particle gives rise to a pair of DVs.  At this stage of the analysis, such 
pairs of DVs are indistinguishable from tumblers.  Thus, in this sense, our model not 
only gives rise to tumblers but also simultaneously gives rise to a ``background'' of 
non-tumbler events, each involving a pair of DVs which arise from 
decays within different chains.  If a significant number of events involving 
multiple DVs is observed at the LHC either before or after the high-luminosity upgrade, 
it will therefore become imperative to develop methods of assessing whether or 
not a significant number of these events in fact involve tumblers.  

This concern is not unique to our model alone.
Indeed, there are also a variety of scenarios for physics beyond the SM in 
which events involving multiple DVs arise.  These
include SUSY models such as those constrained by the ATLAS and CMS searches in
Refs.~\cite{Sirunyan:2019ctn,Sirunyan:2019xwh,Aad:2020aze}, 
as well as hidden-valley models~\cite{Strassler:2006im} and other
scenarios which give rise to emerging jets~\cite{Schwaller:2015gea}.  While tumblers
can in fact arise within certain regimes in some of these models, many other models 
give rise to non-tumbler events exclusively.
This then provides further motivation for developing methods of distinguishing
between tumbler and non-tumbler events.  Without doing so, one can not truly 
claim to have detected a tumbler signature.

Fortunately the distinctive kinematics associated with tumbler decay chains 
provides a basis on which we may discriminate between tumbler and non-tumbler 
events at colliders.  In this section, we develop a set of event-selection criteria 
which are capable of efficiently discriminating between tumbler and non-tumbler events.
In the process, we shall also investigate the extent to which the masses and lifetimes
of the $\chi_n$ can be reconstructed from the kinematic and timing information provided
by a collider detector.

\subsection{Mass Reconstruction\label{sec:MassRec}}

In order to distinguish between tumbler events and other events which involve multiple DVs,
we employ an event-selection procedure which makes use of the distinctive kinematic 
structure associated with tumbler decay chains.  This procedure follows from the observation 
that if two DVs in a given event arise from successive decays along the same decay chain, 
it is in principle possible to reconstruct the masses
of the $\chi_n$ involved in that decay chain.  That such an 
event-by-event mass reconstruction is possible for tumblers is itself noteworthy.
Methods for reconstructing the masses of unstable particles in multi-step decay
chains which terminate in invisible particles typically rely on the identification of 
features such as cusps~\cite{Cho:2007qv,Han:2009ss,Agashe:2010gt,Cho:2014naa},
edges~\cite{Hinchliffe:1996iu,Lester:1999tx,Allanach:2000kt,Barr:2003rg,Miller:2005zp,Konar:2008ei,Burns:2009zi,Matchev:2009iw,Matchev:2009ad,Cho:2012er,Cho:2014naa,Kim:2015bnd,Debnath:2015wra,Debnath:2016gwz,Debnath:2018azt}, or peaks~\cite{Cho:2012er,Agashe:2012bn,Agashe:2013eba} in the distributions of kinematic variables --- features 
which emerge only in the aggregate, from a sizable population of events.  By contrast, 
when the vertices in the decay chain are macroscopically displaced from each other and 
from $V_P$, as they are for a tumbler, additional information can be brought to bear in 
reconstructing the masses of the unstable particles.   

The information we need in order to reconstruct the $m_n$ for a tumbler 
includes the three-momenta of the four displaced jets produced by the decays of $\chi_1$ 
and $\chi_2$, the three-momenta of the additional jets produced at the primary vertex, and
the timing information supplied by the ECAL or timing layer concerning the time at which 
these jets exit the tracker.  As discussed in Sect.~\ref{sec:TumblersTiming}, these 
three-momenta, in conjunction with timing information, are sufficient to reconstruct the 
times $t_P$, $t_S$, and $t_T$ and spatial locations $\vec{\mathbf{x}}_P$, $\vec{\mathbf{x}}_S$, 
and $\vec{\mathbf{x}}_T$ of the primary, secondary, and tertiary vertices.  Taken together, 
these measurements are then sufficient to determine the velocities 
$\vec{\boldsymbol{\beta}}_1 \equiv (\vec{\mathbf{x}}_T - \vec{\mathbf{x}}_S)/(t_T - t_S)$
and $\vec{\boldsymbol{\beta}}_2 \equiv (\vec{\mathbf{x}}_S - \vec{\mathbf{x}}_P)/(t_S - t_P)$ 
of $\chi_1$ and $\chi_2$, respectively.

Given these velocities, the $m_n$ can then be determined in a straightforward manner.  
Approximating the quarks as massless and noting that the energy $E_n$ and 
momentum $\vec {\mathbf{p}}_n$ of each $\chi_n$ are given by $E_n = \gamma_n m_n$ 
and $\vec{\mathbf{p}}_n = \gamma_n m_n\vec{\boldsymbol{\beta}}_n$, 
we find that the equations which represent four-momentum conservation
at $V_S$ may be written in the form 
\begin{eqnarray}
  \gamma_2 m_2 &~=~& \gamma_1 m_1 + |\vec{\mathbf{p}}_q| + |\vec{\mathbf{p}}_{\bar{q}}|
    \nonumber \\ 
  \gamma_2 m_2\vec{\boldsymbol{\beta}}_2 &~=~& \gamma_1 m_1\vec{\boldsymbol{\beta}}_1 
    + \vec{\mathbf{p}}_q + \vec{\mathbf{p}}_{\bar{q}}~.
  \label{eq:P4convS}
\end{eqnarray} 
Likewise, applying four-momentum conservation at $V_T$ yields 
\begin{eqnarray}
  \gamma_1 m_1 &~=~& \gamma_0 m_0 + |\vec{\mathbf{p}}_{q'}| + |\vec{\mathbf{p}}_{\bar{q}'}|
    \nonumber \\ 
  \gamma_1 m_1\vec{\boldsymbol{\beta}}_1 &~=~& \vec{\mathbf{p}}_0 
    + \vec{\mathbf{p}}_{q'} + \vec{\mathbf{p}}_{\bar{q}'}~.
  \label{eq:P4convT}
\end{eqnarray} 
Solving this system of equations for the three $m_n$, we obtain 
\begin{eqnarray}
  m_2 &~=~& \frac{\big|\vec{\mathbf{p}}_q + \vec{\mathbf{p}}_{\bar{q}} 
    - \vec{\boldsymbol{\beta}}_1\big(|\vec{\mathbf{p}}_q| +
    |\vec{\mathbf{p}}_{\bar{q}}|\big)\big|}
    {\gamma_2|\vec{\boldsymbol{\beta}}_1 - \vec{\boldsymbol{\beta}}_2|}\nonumber \\
  m_1 &~=~& 
    \frac{\big|\vec{\mathbf{p}}_q + \vec{\mathbf{p}}_{\bar{q}} 
    - \vec{\boldsymbol{\beta}}_2\big(|\vec{\mathbf{p}}_q| +
    |\vec{\mathbf{p}}_{\bar{q}}|\big)\big|}
    {\gamma_1|\vec{\boldsymbol{\beta}}_1 - \vec{\boldsymbol{\beta}}_2|}\nonumber \\
  m_0^2 &~=~&  m_1^2 -2\gamma_1 m_1
    \Big[|\vec{\mathbf{p}}_{q'}| + |\vec{\mathbf{p}}_{\bar{q}'}| 
      - \vec{\boldsymbol{\beta}}_1 \cdot (\vec{\mathbf{p}}_{q'} + \vec{\mathbf{p}}_{\bar{q}'} ) 
      \Big] \nonumber \\ 
      &&+ 2\big(|\vec{\mathbf{p}}_{q'}| |\vec{\mathbf{p}}_{\bar{q}'}| -
       \vec{\mathbf{p}}_{q'}\cdot\vec{\mathbf{p}}_{\bar{q}'}\big)~.
  \label{eq:solsallsummary}
\end{eqnarray}

Were it possible to measure with arbitrary precision both the magnitude of the momentum of 
each jet in a tumbler event and the time at which each jet exits the tracker, it would 
be possible to reconstruct the $m_n$ exactly from the relations in
Eq.~(\ref{eq:solsallsummary}).  In practice, of course, our ability to reconstruct these 
masses is limited by the precision with which the detector is capable of measuring these 
quantities.  Nevertheless, provided that these uncertainties are sufficiently small, it 
is highly likely that the $m_n$ values obtained when these reconstruction formulas 
are applied to the jets associated with a tumbler will satisfy certain basic self-consistency
criteria.  For example, these reconstructed $m_n$ values will be real, positive, and properly 
ordered in the sense that $m_2 > m_1 > m_0$. 

By contrast, when the mass-reconstruction formulas in Eq.~(\ref{eq:solsallsummary}) are 
applied to the jets associated with a pair of DVs in the same event which 
do {\it not}\/ arise from successive decays along the same decay chain, it is far less 
likely that they will yield a set of masses for the $\chi_n$ which satisfy these criteria.   
This consideration suggests that these mass-reconstruction formulas can be used in 
order to distinguish tumbler events from the far larger ``background'' of non-tumbler
events involving multiple DVs which also arises in our model --- and indeed arises
generically in scenarios wherein the LLPs involved in the tumbler decay chain have
identical quantum numbers.

In order to assess the extent to which we are able to distinguish tumbler events from
other events involving multiple DVs in this way, we perform a Monte-Carlo analysis.
Our specific procedure is as follows.  Using the \texttt{MG5\_aMC@NLO} code 
package~\cite{Alwall:2014hca}, and for each of our parameter-space benchmarks, we generate 
100,000 events for the initial pair-production process $pp\rightarrow \phi^\dagger\phi$ 
at a center-of-mass energy $\sqrt{s} = 14$~TeV.~  This process overwhelmingly dominates the 
event rate for both tumbler-class and all relevant DV-class processes.  The number of
events in this sample is of course far larger than the expected event count for this 
pair-production process at the HL-LHC.~  Indeed, our goal at this stage of the analysis is
simply to examine the detailed shapes of these distributions and thereby develop a nuanced
understanding of how different event-selection criteria impact these shapes.  It is therefore
advantageous for us to consider a large population of events and a relatively narrow 
bin width for each $m_n$ distribution.  Once we have such an understanding, we shall 
return to assess the extent to which the $m_n$ can be reconstructed with a population 
of events appropriate for near-future collider studies and a coarser set of bin widths.

After our events are generated, we then simulate 
the kinematics of the subsequent decay chains using our own Monte-Carlo code.  For each jet 
we record not only the magnitude and direction of its three-momentum vector, but also 
the time at which the jet exits the tracker.  We work at the parton level and do not consider
the effects of initial-state or final-state radiation, parton-showering, or hadronization.
We determine the locations $\vec{\mathbf{x}}_S$ and 
$\vec{\mathbf{x}}_T$ of the secondary and tertiary vertices in each event from the momenta of 
the jets produced at these vertices using the parton-level vertexing algorithm described in
Appendix~\ref{app:Vertexing}.~  We likewise determine the location $\vec{\mathbf{x}}_P$ of the 
primary vertex from the momenta of the two jets produced by the prompt decays of $\phi$ and
$\phi^\dagger$ at this vertex.  
Thus, while the beam spot at a collider like the HL-LHC has a characteristic 
spread of a few cm in the $z$-direction and a time spread of around 200~ps,  
our procedure for reconstructing the primary vertex will effectively remove these 
uncertainties.

Of course, this parton-level vertexing procedure does not incorporate any of the uncertainties 
involved in a full track-based reconstruction of the locations of the primary or displaced 
vertices in the event.  Moreover, it does not account for the measurement uncertainties in 
the momenta of the jets.  Thus, in order to account for these uncertainties --- which can be
significant --- when estimating the precision with which we might hope to measure the values 
of the $m_n$ from tumbler data, we proceed as follows.

We account for the timing uncertainty by smearing the time at which each jet exits the tracker 
using a Gaussian smearing function with standard deviation $\sigma_t$.
We likewise account for the uncertainty in the {\it magnitude}\/ of the jet momenta by smearing 
the magnitude of each momentum vector according to a Gaussian smearing function whose standard 
deviation $\sigma_E(E_j)$ varies with the energy $E_j$ of the jet.  Since the jet-energy 
resolution of a collider detector also depends on the pseudorapidity $\eta_j$ 
of the jet, we adopt a conservative approach and model our $\sigma_E(E_j)$
after the jet-energy resolution obtained in Ref.~\cite{Bayatian:2006nff} for jets with 
$1.4 < \eta_j < 3.0$ in the endcap region rather than the barrel region of the CMS detector.

The uncertainties $\sigma_\eta$ and $\sigma_\phi$ in the pseudorapidity and 
azimuthal angle that characterize the {\it direction}\/ of each jet within a given event 
affect the reconstructed values of the $m_n$ in two ways.  The first is directly through 
$\vec{\mathbf{p}}_q$, $\vec{\mathbf{p}}_{\bar{q}}$, $\vec{\mathbf{p}}_{q'}$, and
$\vec{\mathbf{p}}_{\bar{q}'}$ themselves in Eq.~(\ref{eq:solsallsummary}).  The second is
indirectly through their effect on the reconstructed vertex positions $\vec{\mathbf{x}}_P$,
$\vec{\mathbf{x}}_S$, and $\vec{\mathbf{x}}_T$, which in turn affects the 
reconstructed LLP velocities $\vec{\boldsymbol{\beta}}_1$ and $\vec{\boldsymbol{\beta}}_2$.
Since the CMS detector is capable of measuring the directions of the momentum vectors of 
hadronic jets with excellent precision~\cite{Bayatian:2006nff}, the first effect turns 
out to be subleading in terms of its effect on the $m_n$ in comparison with the effect 
of jet-energy smearing.  By contrast, the second effect can have a more significant impact
on the $m_n$.  Indeed, $\sigma_\eta$ and $\sigma_\phi$ can dominate the uncertainty in 
$\vec{\boldsymbol{\beta}}_1$ and $\vec{\boldsymbol{\beta}}_2$ when $\sigma_t$ is small. 

Our method for simulating the effect of these uncertainties shall be the following.
Since $\sigma_E$ dominates the uncertainty in the $m_n$ that arises directly from the jet 
momenta, we shall simply take $\sigma_\eta = \sigma_\phi = 0$ in what follows.
However, in order to account for the effect of these angular uncertainties and other
uncertainties which enter into the track-based reconstruction of DVs at a 
real collider detector, we also shift each of the three 
vertex positions $\vec{\mathbf{x}}_P$, $\vec{\mathbf{x}}_S$, and $\vec{\mathbf{x}}_T$ that 
we obtain from our fitting procedure by an independent random offset vector.  The
magnitude of this offset vector is distributed according to a single-sided Gaussian function
with standard deviation $\sigma_r$, while its direction is distributed spherically uniformly. 
Since the estimated uncertainty in the vertex displacements for the CMS detector after the 
HL-LHC upgrade is roughly $\mathcal{O}(10$~--~$30~\mu\mathrm{m})$~\cite{Butler:2019rpu}, 
we take $\sigma_r = 30$~$\mu$m in what follows.

\begin{figure*}
  \includegraphics[clip, width=0.32\textwidth]{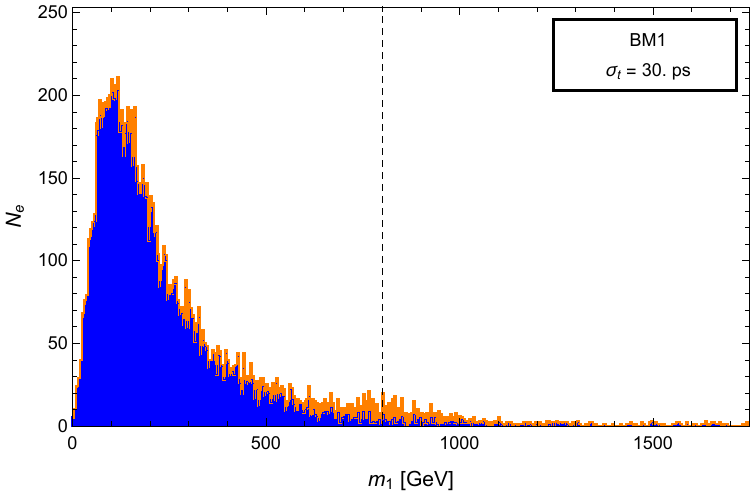}
  \includegraphics[clip, width=0.32\textwidth]{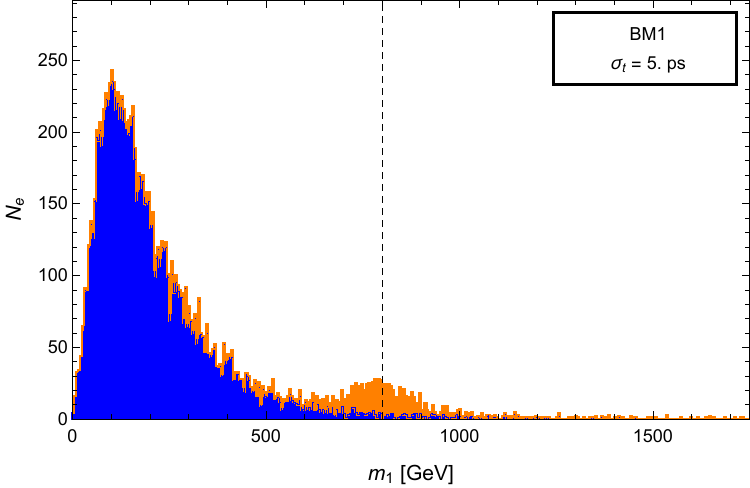}
  \includegraphics[clip, width=0.32\textwidth]{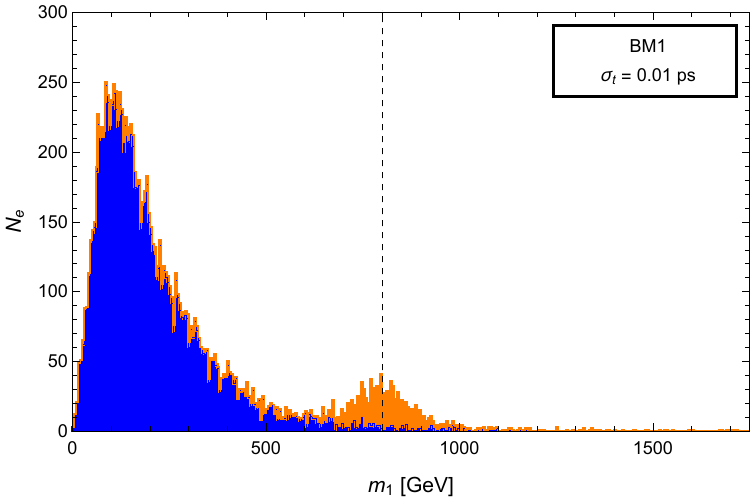}\\
  \includegraphics[clip, width=0.32\textwidth]{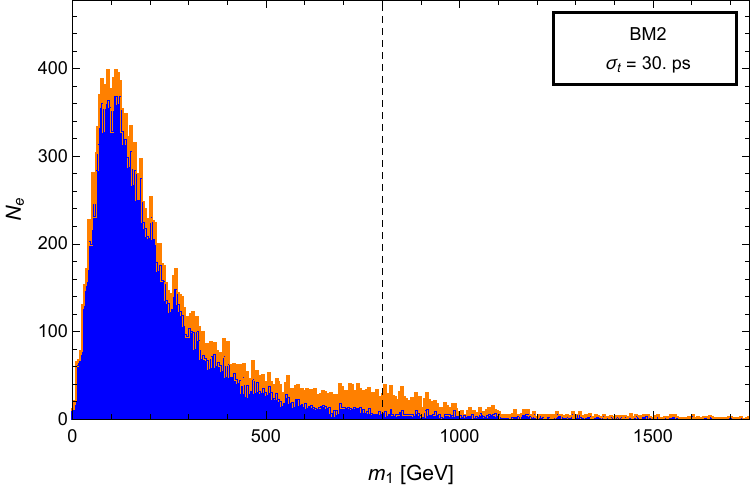}
  \includegraphics[clip, width=0.32\textwidth]{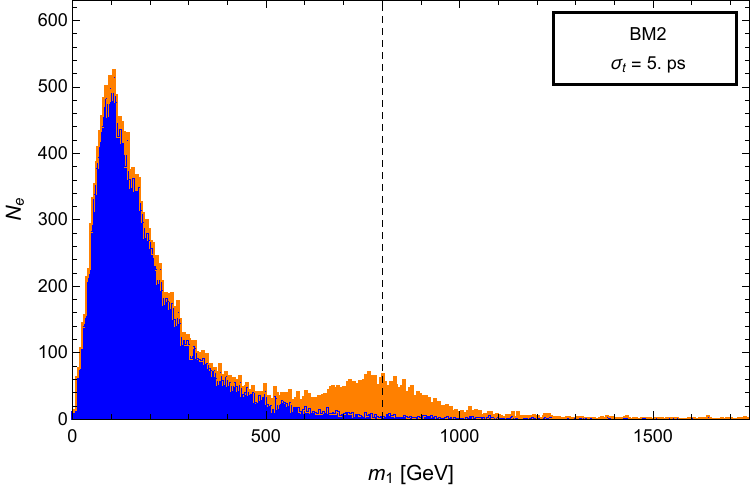}
  \includegraphics[clip, width=0.32\textwidth]{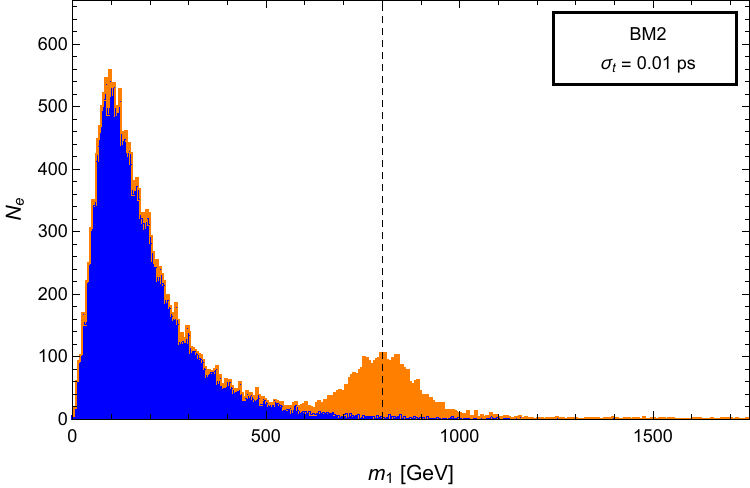}\\
  \includegraphics[clip, width=0.32\textwidth]{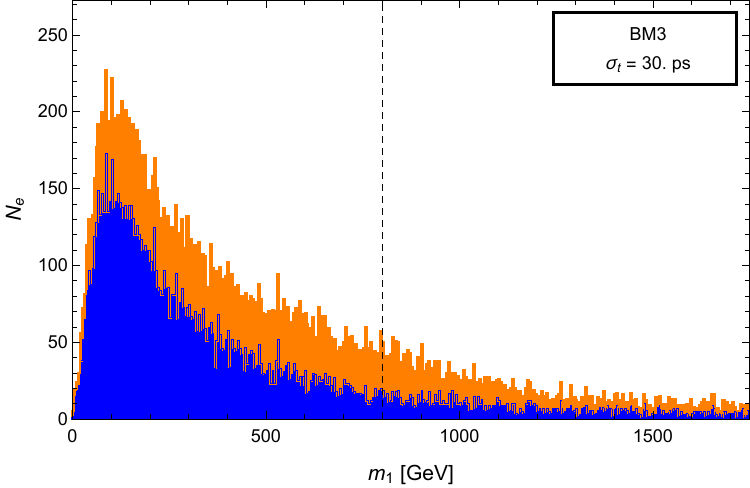}
  \includegraphics[clip, width=0.32\textwidth]{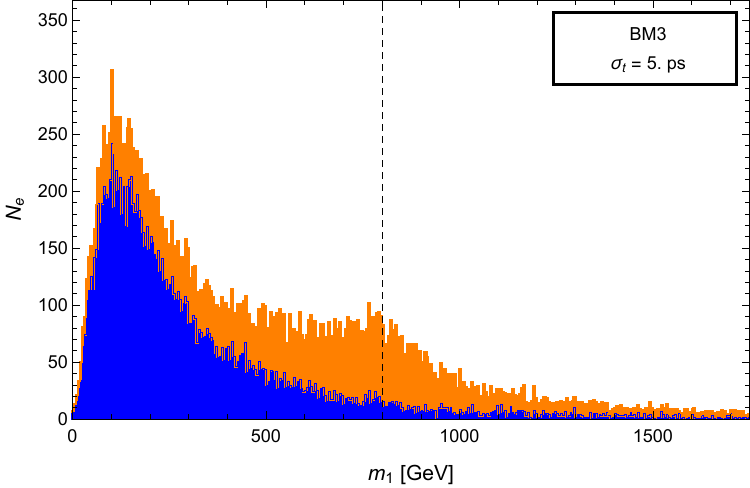}
  \includegraphics[clip, width=0.32\textwidth]{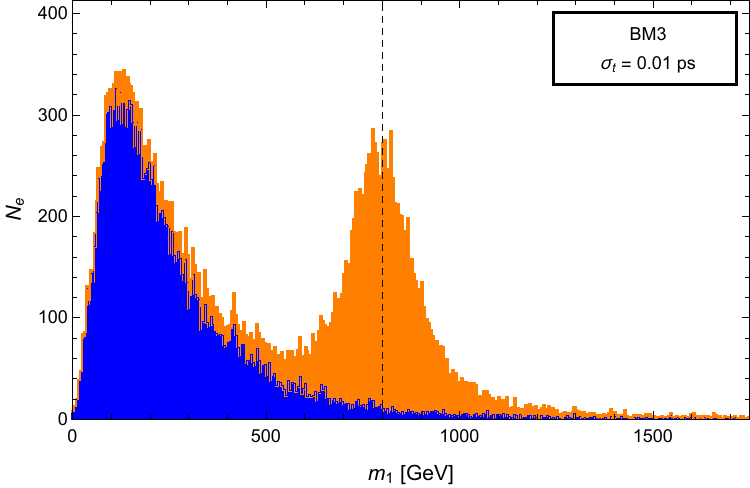}\\
  \includegraphics[clip, width=0.32\textwidth]{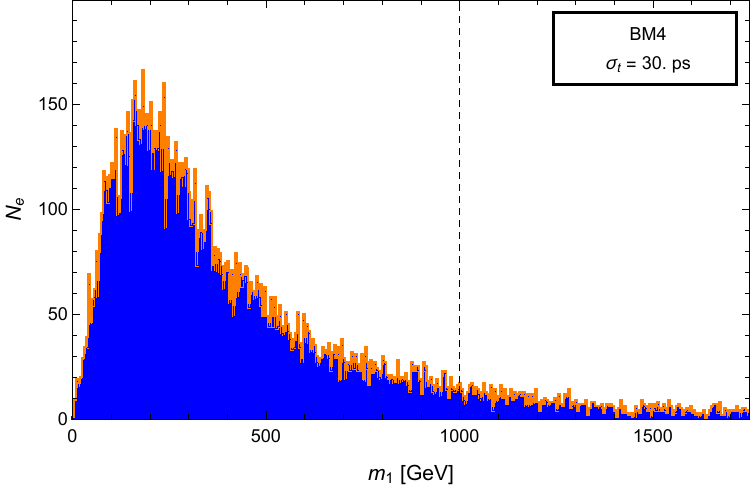}
  \includegraphics[clip, width=0.32\textwidth]{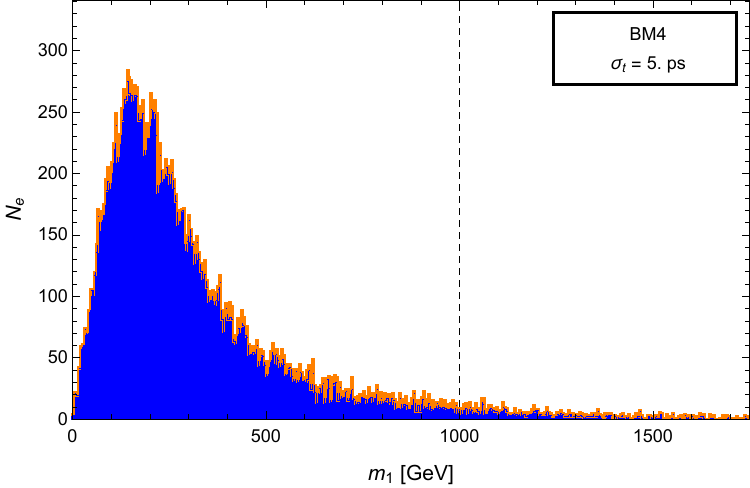}
  \includegraphics[clip, width=0.32\textwidth]{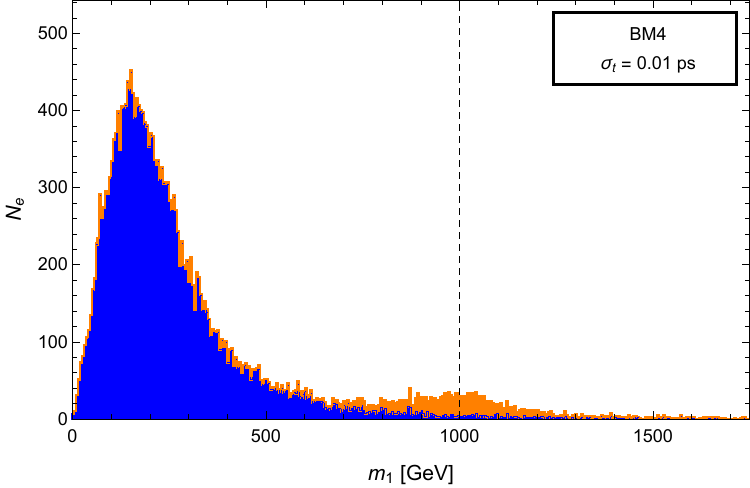}
\caption{The distribution of values of the mass $m_1$
  for the sample of Monte-Carlo events described in the text, as reconstructed from 
  tumbler kinematics.  The orange portion of each histogram bar represents the contribution 
  from tumbler events, while the blue portion represents the contribution from
  events with multiple DVs which do not involve a tumbler.  From top to bottom, the rows 
  in the figure correspond to the parameter-space benchmarks BM1 -- BM4 defined in
  Table~\protect\ref{tab:benchmarks}.~  The dashed black vertical line in each panel 
  indicates the actual value of $m_1$ for the corresponding benchmark.  The results shown 
  in the left, center, and right columns correspond respectively to the values 
  $\sigma_t = 30$~ps, $\sigma_t = 5$~ps, and $\sigma_t = 0.01$~ps for the timing uncertainty 
  of the detector.  Since the efficiency of the cuts depends on the benchmark and varies 
  with $\sigma_t$, the scale of the vertical axis has been varied from panel to panel in 
  order to facilitate comparison between the distributions.
  \label{fig:DoubleHistm1Nom0Cut}}
\end{figure*}

In order to extract a set of values for the $m_n$ from a given sample of of events, as well 
as an estimate of the uncertainties in these values, we proceed as follows.
We begin by requiring that the decays of all unstable dark-sector particles in the event
occur within the tracker region of our hypothetical detector.  Modeling this detector after
the CMS detector, we take this region to be a cylinder of radius $r=1.161$~m, centered at 
the interaction point $z=0$ and extending longitudinally within the range 
$-2.5\mathrm{~m} < z < 2.5\mathrm{~m}$, whose axis of symmetry runs along the beam.
We note that events which satisfy this requirement necessarily involve a significant number
of energetic jets --- including two highly-energetic prompt jets from the decays of 
$\phi$ and $\phi^\dagger$ --- and typically also significant $\met$.  The overwhelming 
majority of such events therefore satisfy one or more of several Level-1 triggers 
appropriate for a detector in high-luminosity collider environment~\cite{Contardo:2015bmq}.

In order to assess the impact of this requirement on our results,
we focus on the events which have the event topology given in the fourth line of
Table~\ref{tab:DecayChainList} --- \ie, events wherein the two decay chains are 
$\phi\rightarrow \chi_2\rightarrow \chi_1\rightarrow \chi_0$ and 
$\phi \rightarrow \chi_0$.  Events of this sort, which involve a single tumbler
but no additional unstable particles on the other side of the event, provide a
the clearest picture of where these decays tend to occur.  
In Table~\ref{tab:decaylocs}, for each of our parameter-space benchmarks,
we provide the fraction of events in our Monte-Carlo sample with this event 
topology in which the $\chi_1$ decays within each layer of the 
detector.  We observe that while a non-negligible fraction of these particles 
decay outside the tracker for all of these benchmarks except BM4, which has 
a far smaller value of $\tau_1$ than the other three benchmarks, the $\chi_1$ 
particle decays within the tracker the majority of the time.  
By contrast, $\tau_2$ is sufficiently short for all of our benchmarks 
that the probability for $\chi_2$ to decay outside the tracker is negligible.
The results shown in Table~\ref{tab:decaylocs} indicate that the
requirement that all unstable LLPs in the event decay within the tracker,
rather than elsewhere within the detector, will not have a significant impact
on our results.  Moreover, they also indicate that the fraction of events in 
which $\chi_1$ escapes the detector entirely before decaying is quite small
for all of our benchmarks.

\begin{table}
\begin{tabularx}{\linewidth}{||C||c|c|c|c||}
\hline\hline  \multirow{2}{*}{~Benchmark} &  
  \multirow{2}{*}{~Tracker~} & ~ECAL~+~ & ~Muon~ & ~Outside~ \\
  & & ~HCAL~ & ~Chamber~ & ~Detector~ \\
  \hline
BM1 & 0.56 & 0.26 & 0.15 & 0.03  \\
BM2 & 0.74 & 0.19 & 0.06 & 0.01 \\
BM3 & 0.77 & 0.17 & 0.06 & 0.01  \\
BM4 & 1.00 & 0.00 & 0.00 & 0.00  \\
\hline \hline
\end{tabularx}
\caption{The fraction of the events with the topology given in the 
  fourth line of Table~\ref{tab:DecayChainList} (\ie, with decay chains
  $\phi\rightarrow \chi_2\rightarrow \chi_1\rightarrow \chi_0$ and 
  $\phi \rightarrow \chi_0$) in which the last unstable particle decays 
  within each layer of the detector for each of our benchmarks BM1 -- BM4.
  \label{tab:decaylocs}}
\end{table}

We also require that the event contain at least two DVs.  We compute the time $t_i$ at 
which each such vertex $V_i$ occurred from the momentum and timing 
information obtained for the pair of displaced jets produced at that vertex.
For each combination of DVs $V_i$ and $V_j$ in the event which are appropriately time-ordered,
in the sense that $t_i < t_j$, we reconstruct a set of $m_n$ values using 
Eq.~(\ref{eq:solsallsummary}).  We then check whether this set of $m_n$ values, taken together
with the corresponding values of $|\vec{\boldsymbol{\beta}}_1|$, 
$|\vec{\boldsymbol{\beta}}_2|$, and the magnitude of the three-momentum vector $\vec{\mathbf{p}}_0$
obtained from Eq.~(\ref{eq:P4convT}), satisfy the following criteria, to which we shall 
henceforth refer as our {\it reconstruction criteria}\/:
\begin{itemize}
  \item $m_1$ and $m_2$ are real and positive
  \item $m_0^2$ is real 
  \item $|\vec{\mathbf{p}}_0|$ is real and positive
  \item $0 < |\vec{\boldsymbol{\beta}}_n| < 1$ for $n=1,2$
  \item $m_2^2 > m_1^2 > m_0^2$.
\end{itemize}
For reasons to be discussed shortly, we shall not require that $m_0^2 > 0$ at this stage of the 
analysis.  If any appropriately time-ordered combination of DVs in the event yields a set 
of masses which satisfy these criteria, we retain the event; if not, we reject it.
If multiple combinations of DVs within the same event satisfy all of these criteria, we 
take the set of $m_n$ for the combination which yields the largest value of $m_2$ to be the 
set of $m_n$ for the event.  

In order to illustrate the effect of these cuts, we shall begin by focusing on the 
reconstruction of $m_1$.  In Fig.~\ref{fig:DoubleHistm1Nom0Cut}, we show the distribution 
of reconstructed $m_1$ values for the set of events which survive these 
cuts for each of our parameter-space benchmarks.  The histogram in each panel of the 
figure is obtained by binning these $m_1$ values into bins of width $\Delta m_n = 5$~GeV.~
The blue portion of each histogram bar represents the contribution to that bin from
non-tumbler processes, whereas the orange portion represents the contribution
from processes which involve tumblers.  From top to bottom, the rows in the 
figure correspond to our parameter-space benchmarks BM1 -- BM4.  The dashed black vertical 
line in each panel indicates the actual value of $m_1$ for the corresponding benchmark.
The results shown in the left, center, and right columns correspond respectively to the 
values $\sigma_t = 30$~ps, $\sigma_t = 5$~ps, and $\sigma_t = 0.01$~ps for the timing 
uncertainty of the detector.  The first of these $\sigma_t$ values represents the timing 
uncertainty associated with the barrel timing layer to be installed within the CMS detector 
as part of the HL-LHC upgrade.  The second is a value chose to reflect a moderate improvement 
in this timing uncertainty, while the third is an extremely small value representative of 
the regime in which jet-energy and vertex-position smearing dominates the uncertainty in 
the mass reconstruction.  Since the efficiency of the cuts depends on the benchmark and varies 
with $\sigma_t$, the scale of the vertical axis has been varied from panel to panel in 
order to facilitate comparison between the distributions.

First, we observe from Fig.~\ref{fig:DoubleHistm1Nom0Cut} that the number of 
residual non-tumbler events is still quite significant even after the imposition of 
these preliminary cuts.  Moreover, we observe that this distribution has a well-defined 
shape that peaks at low values of $m_1$ and 
falls off rapidly as $m_1$ increases.  By contrast, the $m_1$ distribution for the tumbler
events exhibits a well-defined peak centered around the actual value of $m_1$, as well as
an additional population of events with $m_1$ values well below this peak.  
This additional population of events arises in part due to smearing effects and in part due 
to the combinatorial background which arises from incorrect identifications of the vertices 
$V_S$ and $V_T$ in events which contain more than two DVs.  The relative size of the 
peak in the $m_1$ distribution for the tumbler events at low $\sigma_t$ is primarily 
controlled by $P_{\phi 210}$.  Indeed, we observe that this peak is more pronounced for BM3,
which has by far the largest value of $P_{\phi 210}$, than for our other three benchmarks.

The presence of this peak in the $m_1$ distribution is a unique and distinctive feature of tumbler 
events.  As we shall see, similar peaks appear in the distributions of $m_0$ and $m_2$ 
for tumbler events as well.  An observation of these peaks, taken together, would constitute 
compelling evidence for tumblers.  It is in this way, then, that our mass-reconstruction
procedure furnishes a method through which tumblers can unambiguously be detected.

We also observe from Fig.~\ref{fig:DoubleHistm1Nom0Cut} that as $\sigma_t$ decreases,
the peak in the tumbler distribution becomes both narrower and more pronounced for all of our 
benchmarks.  Indeed, this is to be expected, since increasing $\sigma_t$  
renders the reconstructed values of $t_P$, $t_S$, and $t_T$ less reliable.  However, 
a greater reduction in timing uncertainty is required to resolve this peak  
for some of our benchmarks than for others.  For example, the peak obtained for BM4 remains
effectively washed out even for $\sigma_t = 5$~ps.  We can
make sense of these differences in sensitivity to $\sigma_t$ by comparing the lifetimes 
$\tau_1$ and $\tau_2$ quoted for each of our benchmarks in Table~\ref{tab:benchquants} 
to the value of $\sigma_t$ itself.  For BM1 and BM2, 
$\tau_1 \gg \tau_2 \sim \mathcal{O}(100\mathrm{~ps})$, and thus the effect of 
the timing uncertainty on the times $t_S$ and $t_T$ reconstructed for the DVs in a 
tumbler event will be negligible for either of these benchmarks when 
$\sigma_t \ll 100$~ps.  By contrast, for BM3 and BM4, 
$\tau_2 \sim \mathcal{O}(10\mathrm{~ps})$,
which implies that the effect of the timing uncertainty on $t_S$ will only be negligible when
$\sigma_t \ll 10$~ps.  Furthermore, for BM4, $\tau_1 \sim \mathcal{O}(100\mathrm{~ps})$ is 
also quite small, and thus the timing uncertainty has a non-negligible impact on  
$t_T$ as well unless $\sigma_t \ll 100$~ps.  As a result, the reconstructed value of 
$m_1$ is more sensitive to the value of $\sigma_t$ for BM4 than they are for BM3, and
are more sensitive to this value for BM3 than they are for BM1 and BM2.

In order to further suppress the contribution from non-tumbler events, we 
shall impose one additional cut on the data.  In particular, in addition to the
criteria described above, we shall also impose one additional reconstruction criterion: 
\begin{itemize}
  \item $m_0^2 > 0$.
\end{itemize}
We have separated out this particular criterion from the others because it merits special 
attention.  In particular, as we shall demonstrate,
not only does requiring that $m_0^2 > 0$ induce a dramatic enhancement in the ratio 
of tumbler to non-tumbler events, but it also gives rise to an additional feature in the 
distribution of reconstructed $m_1$ values --- a feature which reveals additional information 
about the mass spectrum of the $\chi_n$, and in particular about the mass 
splitting $\Delta m_{10}$.

\begin{figure}[t]
  \includegraphics[clip, width=0.45\textwidth]{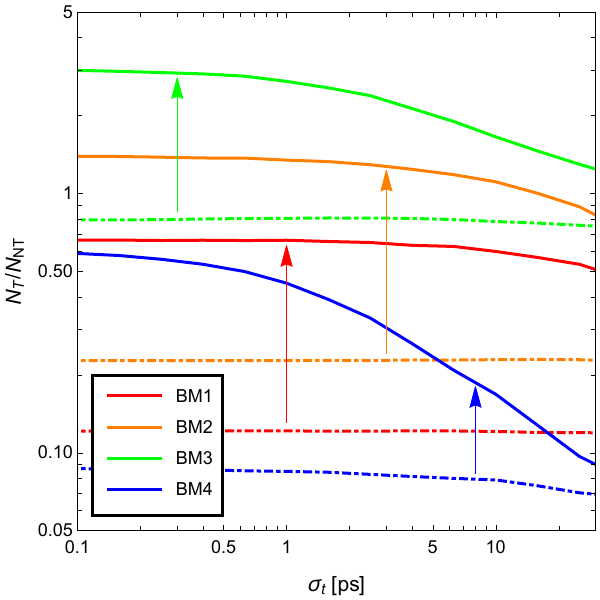}
\caption{The ratio of the number $N_{\rm T}$ of tumbler events to the number $N_{\rm NT}$
  of non-tumbler events for each of our parameter-space benchmarks, shown as a function
  of the timing uncertainty $\sigma_t$.  The dash-dotted curves in each panel represent 
  the corresponding efficiencies obtained {\it without}\/ imposing the $m_0^2 > 0$ 
  criterion, whereas the solid curves represent the corresponding efficiencies obtained 
  with the $m_0^2 > 0$ criterion included.  The vertical arrows in each case therefore 
  indicate the improvements induced by imposing the $m_0^2>0$ cut.
  \label{fig:EfficiencyPlot}}
\end{figure}

In order to quantify the effect of the $m_0^2 > 0$ criterion on the ratio of 
of tumbler to non-tumbler events for each of our four parameter-space benchmarks, in 
Fig.~\ref{fig:EfficiencyPlot} we plot the ratio of the number $N_{\rm T}$ of tumbler events
to the number $N_{\rm NT}$ of non-tumbler events obtained for each of our parameter-space 
benchmarks after cuts as a function of $\sigma_t$.  The dash-dotted curves represent the 
$N_{\rm T}/N_{\rm NT}$ ratios obtained after the imposition of all of our event-selection 
criteria {\it except}\/ the $m_0^2 > 0$ criterion.  By contrast, the solid curves represent the 
$N_{\rm T}/N_{\rm NT}$ ratios obtained after the $m_0^2 > 0$ criterion is also imposed.

It is evident from Fig.~\ref{fig:EfficiencyPlot} that the imposition of the $m_0^2 > 0$
criterion has a significant impact on $N_{\rm T}/N_{\rm NT}$.  When $\sigma_t$ is relatively 
large, as on the right side of this figure, this enhancement factor is already significant 
for our first three benchmarks, even up to $\sigma_t=30$~pb.  By contrast, as $\sigma_t$ 
decreases (towards the left side of this figure), this ratio is enhanced even further, 
ultimately reaching a factor of $\sim 10$ for all of the benchmarks.  The only exception 
to this behavior arises for BM4.   For BM4, the value of $\sigma_t$ has a proportionally 
greater effect on the times reconstructed for the DVs and the difference between 
$\tau_1$ and $\tau_2$ is far smaller than for our other benchmarks.  As a result of these 
differences, the effect of smearing $\sigma_t$ is more likely to result in a set of 
reconstructed masses which fail our reconstruction criteria for BM4 than it is for our 
other three benchmarks. 

We now turn to discuss the impact of the $m_0^2 > 0$ criterion on the {\it shapes}\/ of the 
$m_n$ distributions obtained from our mass-reconstruction procedure, and in particular on
the shape of the $m_1$ distribution --- the distribution on which this criterion has the 
greatest impact.  Indeed, since the presence of identifiable peaks in each of the three 
reconstructed $m_n$ distributions is the characteristic feature that distinguishes a 
population of tumbler events from a population of non-tumbler events, the shapes of these
distributions are of crucial importance.

The shapes of these distributions also allow us to determine the masses of the LLPs 
involved in the tumbler.  In order to assess the precision with which this can be done, 
we need a method of estimating the width of the peak in the corresponding mass distribution.
We shall do this in the following way.   We begin by constructing a template
for the non-tumbler contribution to each $m_n$
distribution after the application of our event-selection 
criteria.  We construct each such template by performing a smoothing procedure on
the non-tumbler contribution to the $m_n$ distribution obtained from an additional
sample of Monte-Carlo events --- a smoothing 
procedure wherein we replace the number of events in each histogram bin with the mean value 
of the event counts in all bins whose central $m_n$ values are within 25~GeV of the central
$m_n$ value for that bin.  We then subtract this template from the 
corresponding $m_n$ distribution in order to obtain the contribution 
to the $m_n$ distribution from the tumbler events alone.  We then perform a fit 
of this ``background-subtracted'' $m_n$ distribution to the rescaled Gaussian 
function
\begin{equation}
  f(m_n) ~=~ \frac{N_{m_n}}{\sqrt{2\pi\sigma_{m_n}^2}}
    \exp\left[-\frac{(m_n-\langle m_n\rangle)^2}{2\sigma_{m_n}^2}\right]~.
  \label{eq:RescaledGaussian}
\end{equation}
We take the values of $\langle m_n\rangle$ and $\sigma_{m_n}$ as our best estimates for $m_n$ 
and its uncertainty.  While more sophisticated 
modeling of the shape of the mass peak would of course improve upon these results,
this procedure provides a reasonably reliable indicator of the extent to which 
one might hope to extract a meaningful measurement of each $m_n$ for a given set
of model parameters at the LHC or at future colliders.

\begin{figure*}
  \includegraphics[clip, width=0.32\textwidth]{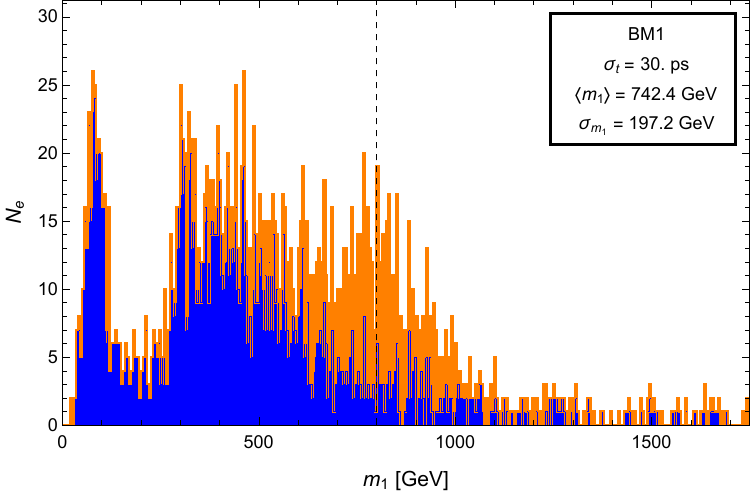}
  \includegraphics[clip, width=0.32\textwidth]{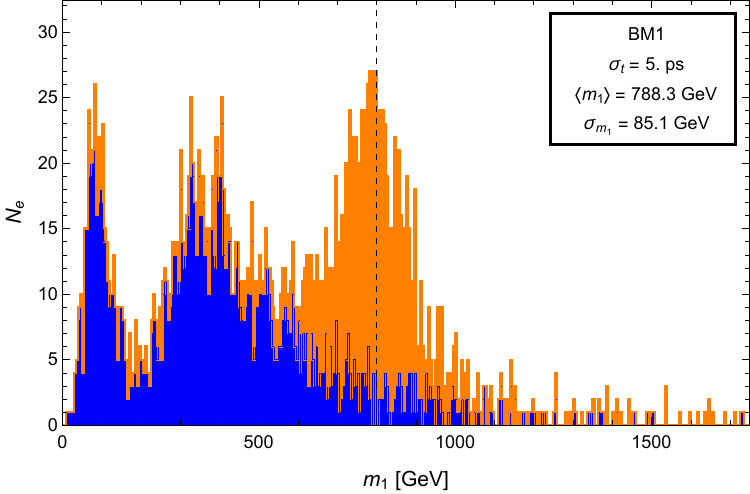}
  \includegraphics[clip, width=0.32\textwidth]{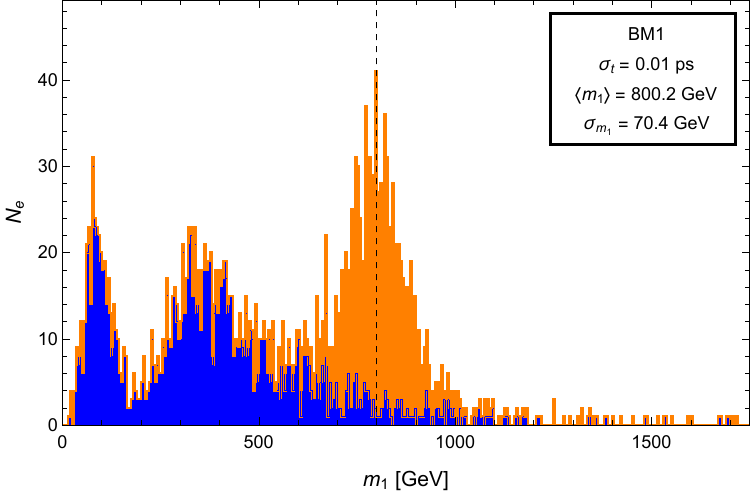}\\
  \includegraphics[clip, width=0.32\textwidth]{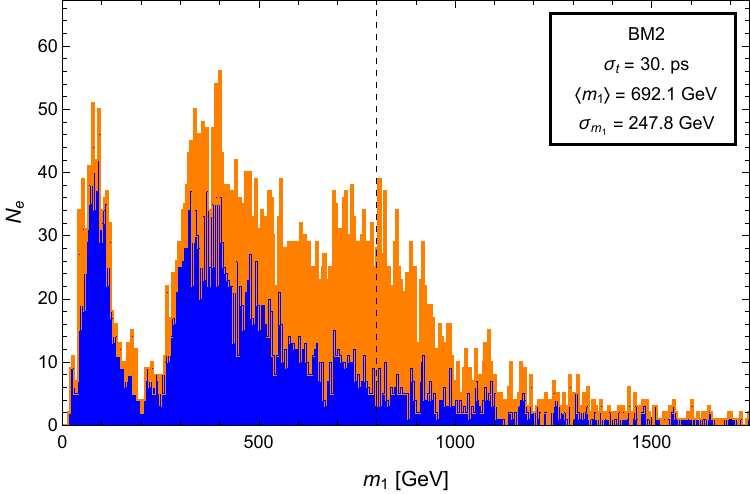}
  \includegraphics[clip, width=0.32\textwidth]{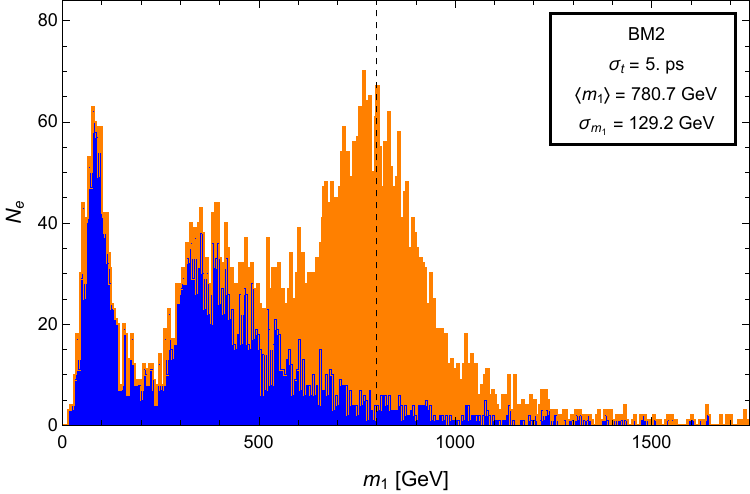}
  \includegraphics[clip, width=0.32\textwidth]{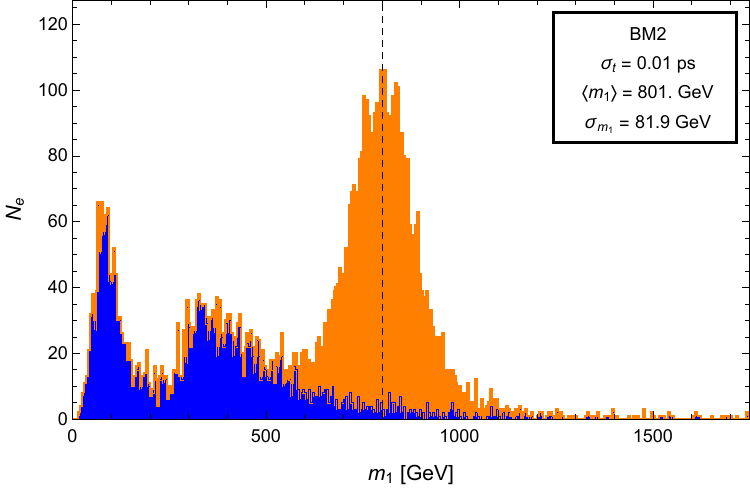}\\
  \includegraphics[clip, width=0.32\textwidth]{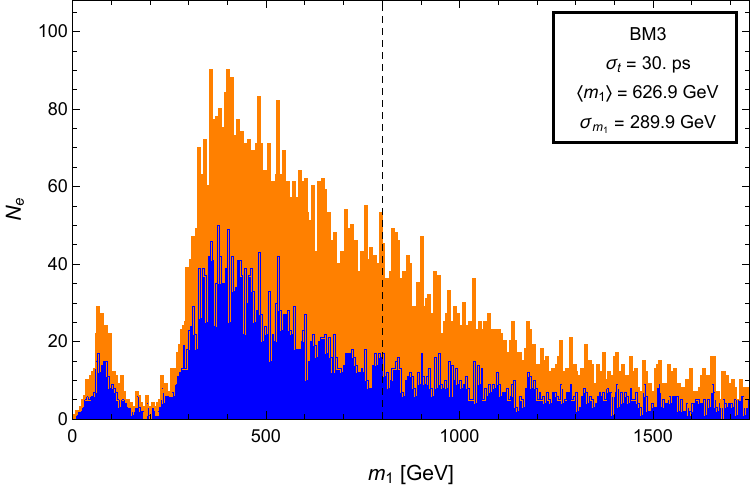}
  \includegraphics[clip, width=0.32\textwidth]{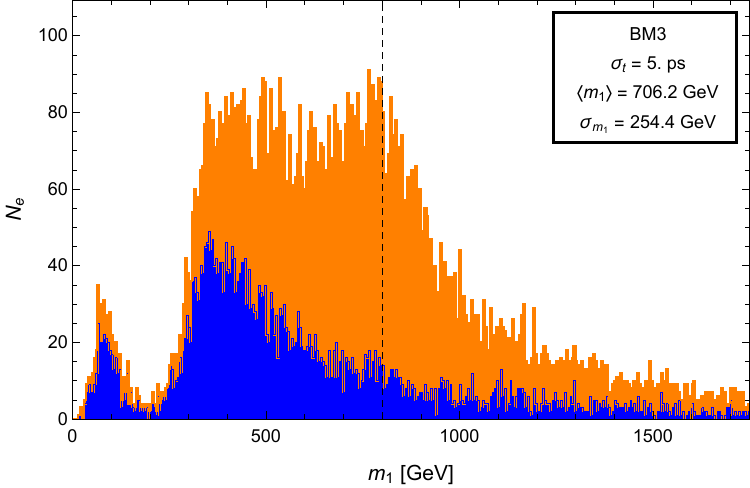}
  \includegraphics[clip, width=0.32\textwidth]{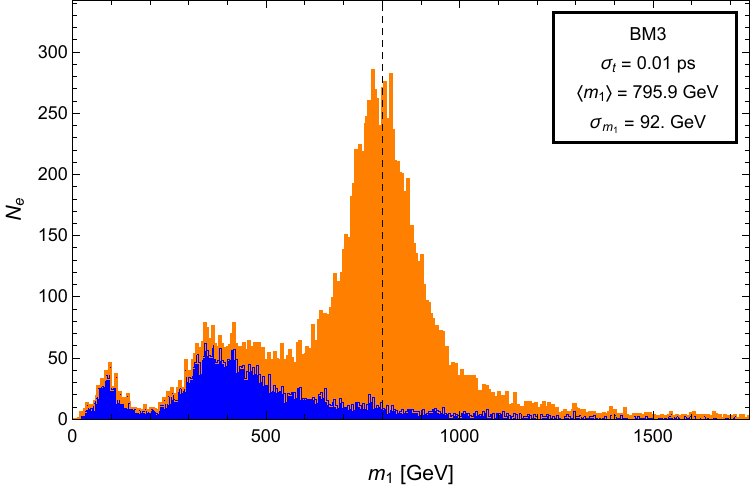}\\
  \includegraphics[clip, width=0.32\textwidth]{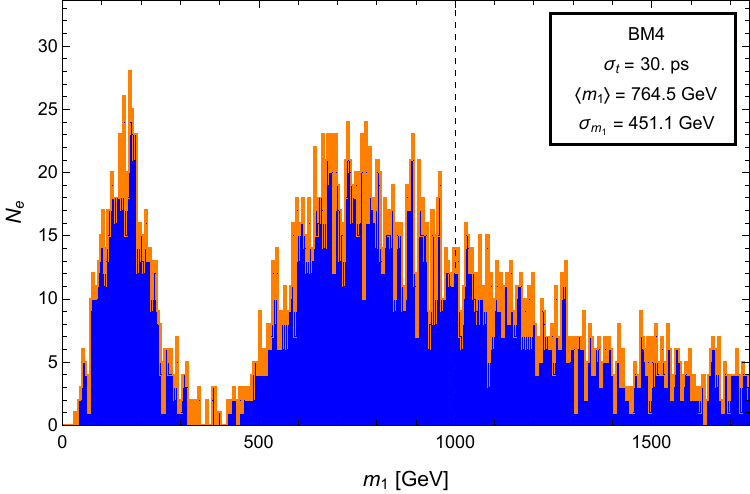}
  \includegraphics[clip, width=0.32\textwidth]{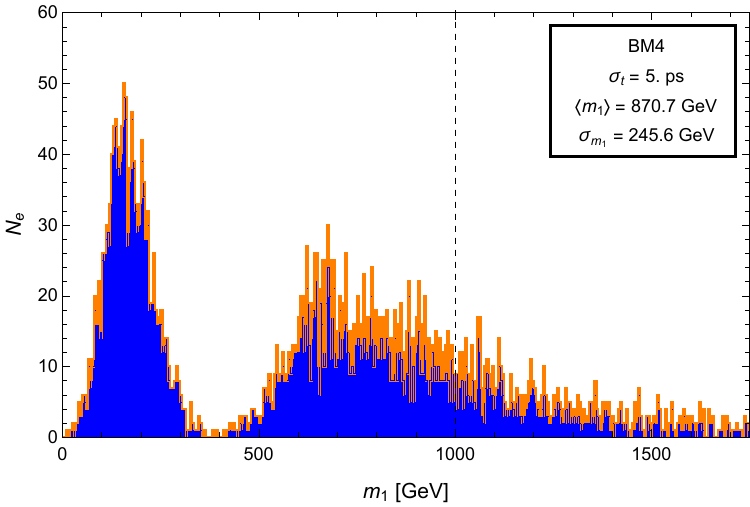}
  \includegraphics[clip, width=0.32\textwidth]{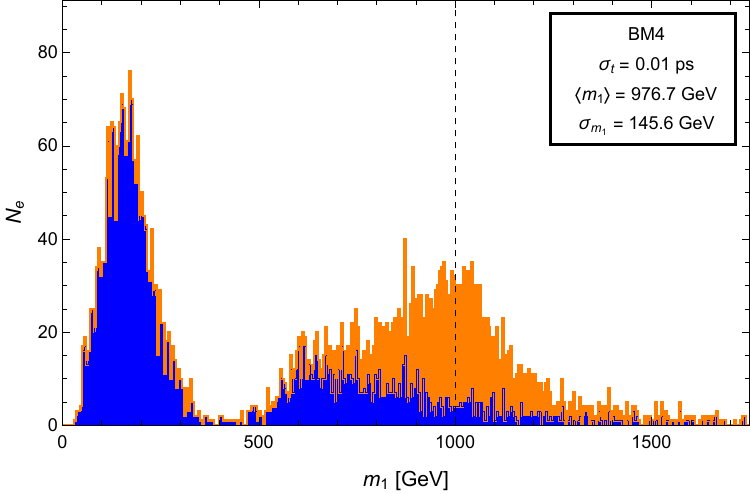}
\caption{Same as Fig.~\protect\ref{fig:DoubleHistm1Nom0Cut}, but after the imposition
  of the additional $m_0^2 > 0$ reconstruction criterion.
  \label{fig:DoubleHistm1}}
\end{figure*}

In Fig.~\ref{fig:DoubleHistm1}, we display the $m_1$ distributions obtained
for our benchmarks after the application of all of our event-selection criteria,
{\it including}\/ the $m_0^2 > 0$ criterion.  Thus, all differences between the $m_1$ 
distribution shown in each panel of this figure and the distribution shown in the corresponding
panel of Fig.~\ref{fig:DoubleHistm1Nom0Cut} are solely due to the effect of this 
criterion.  The $\langle m_1\rangle$ and $\sigma_{m_1}$ values we obtain from our
fitting procedure for the distribution shown in each panel are also indicated.

We observe that the non-tumbler contribution to each of the $m_1$ distributions
shown in Fig.~\ref{fig:DoubleHistm1} is significantly reduced relative to the 
corresponding distribution in Fig.~\ref{fig:DoubleHistm1Nom0Cut}.  However, somewhat 
surprisingly, we also see that each of these distributions now manifests a visible dip or 
trough at a particular reconstructed value 
of $m_1$ well below this peak.  The origin of this dip can be understood as follows.
First, we see from Eq.~(\ref{eq:solsallsummary}) that events which fail to satisfy the 
$m_0^2 > 0$ criterion are events for which 
\begin{equation}
    m_1^2 - 2m_1 E_{jj}^\ast + m_{jj}^2 ~\leq~ 0~,
\end{equation}
where we have used the fact that the center-of-mass energy reconstructed for the 
$q'\bar{q}'$ system is given by $E_{jj}^\ast \equiv \gamma_1 [|\vec{\mathbf{p}}_{q'}| +
|\vec{\mathbf{p}}_{\bar{q}'}| -\vec{\boldsymbol{\beta}}_1\cdot (\vec{\mathbf{p}}_{q'} +
\vec{\mathbf{p}}_{\bar{q}'})]$ and the fact that the invariant mass of this system is
given by $m_{jj}^2 = 2(|\vec{\mathbf{p}}_{q'}| |\vec{\mathbf{p}}_{\bar{q}'}| -
\vec{\mathbf{p}}_{q'}\cdot\vec{\mathbf{p}}_{\bar{q}'})$ in order to write this 
condition more compactly.  Thus, for any particular values of $E_{jj}^\ast$ 
and $m_{jj}$, the range of reconstructed $m_1$ values excluded by the $m_0^2 > 0$
criterion is
\begin{equation}
    E_{jj}^\ast - \sqrt{(E_{jj}^\ast)^2 -m_{jj}^2 } ~\leq~ m_1 ~\leq~
    E_{jj}^\ast + \sqrt{(E_{jj}^\ast)^2 -m_{jj}^2 }~.
  \label{eq:DipExclusionRange}
\end{equation}
We note that this range of excluded $m_1$ values always contains the point 
$m_1 = E_{jj}^\ast$.

We also observe that constraints which follow from standard three-body-decay kinematics
restrict the true values of $E_{jj}^\ast$ and $m_{jj}$ to lie within the respective ranges   
$0 \leq m_{jj} \leq m_1 - m_0$ and $(m_1^2 - m_0^2)/(2m_1) \leq E \leq m_1 - m_0$.
Of course, the reconstructed values of $E_{jj}^\ast$ and $m_{jj}$ will in general differ 
from these true values due to timing, jet-energy, and vertex-position smearing, and
can in principle lie outside these ranges.  However, in the regime in which 
$\sigma_t$ is negligible compared to $\tau_1$ and $\tau_2$, we find that the vast
majority of reconstructed values for $E_{jj}^\ast$ and $m_{jj}$ lie within or only 
slightly outside these ranges.  For all of our parameter-space benchmarks, we note that 
the range of kinematically-allowed $E_{jj}^\ast$ values is fairly narrow.  For BM1 -- BM3, 
this range is $175\mathrm{~GeV} \leq E_{jj}^\ast \leq 200\mathrm{~GeV}$; for BM4, this range
is $320\mathrm{~GeV} \leq E_{jj}^\ast \leq 400\mathrm{~GeV}$.  As a result, when the 
reconstructed value of $m_1$ for an event lies within this narrow range of $E_{jj}^\ast$ 
values, Eq.~(\ref{eq:DipExclusionRange}) implies that the event will typically be excluded.  
Indeed, we observe a dramatic suppression in each of the distributions shown in
Fig.~\ref{fig:DoubleHistm1} across the corresponding range of $m_1$ values.

It is worth remarking that this dip in the $m_1$ distribution arises solely as a consequence 
of the decay kinematics at the final vertex $V_T$ along the tumbler decay chain.  Thus, the
kinematic considerations which lead to the dip are insensitive to the full structure 
of that decay chain.  We would therefore expect the contribution to the $m_1$ distribution 
from non-tumbler events in which a $\chi_1$ particle appears in either one or both of the 
decay chains to exhibit a similar dip.  Indeed, we observe that a dip appears in both tumbler 
and non-tumbler contributions to the $m_1$ distributions in Fig.~\ref{fig:DoubleHistm1}.  
It is also worth remarking that the location and width of the dip provide additional 
information about the mass spectrum of the $\chi_n$.  Indeed, we have seen that both 
$E_{jj}^\ast$ and $m_{jj}$ are bounded from above by $\Delta m_{10}$.  Thus, in principle,
correlations between the properties of the dip and the locations of the tumbler peaks in the 
$m_n$ distributions can be exploited to improve the precision with which the $m_n$ can be
measured.

\begin{figure*}
  \includegraphics[clip, width=0.32\textwidth]{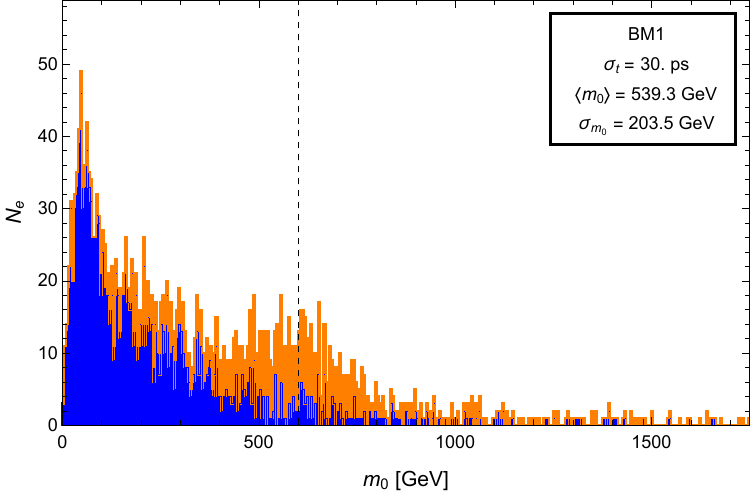}
  \includegraphics[clip, width=0.32\textwidth]{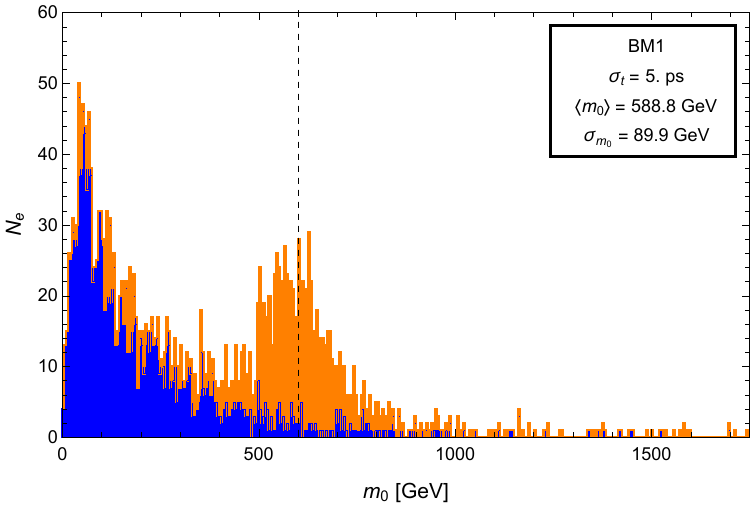}
  \includegraphics[clip, width=0.32\textwidth]{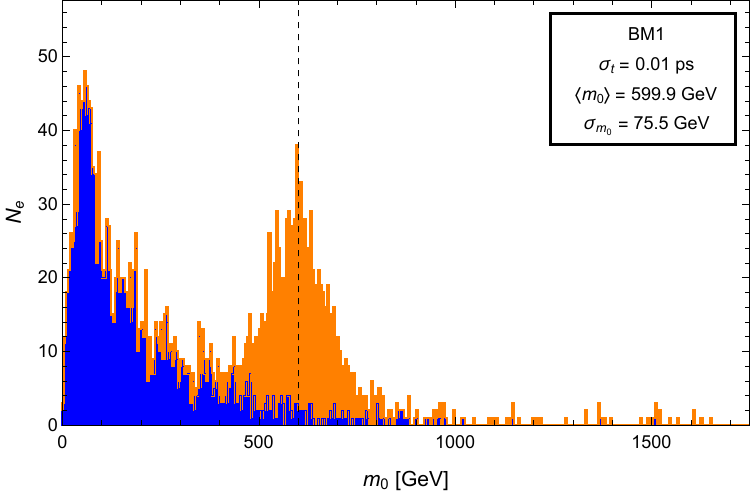}\\
  \includegraphics[clip, width=0.32\textwidth]{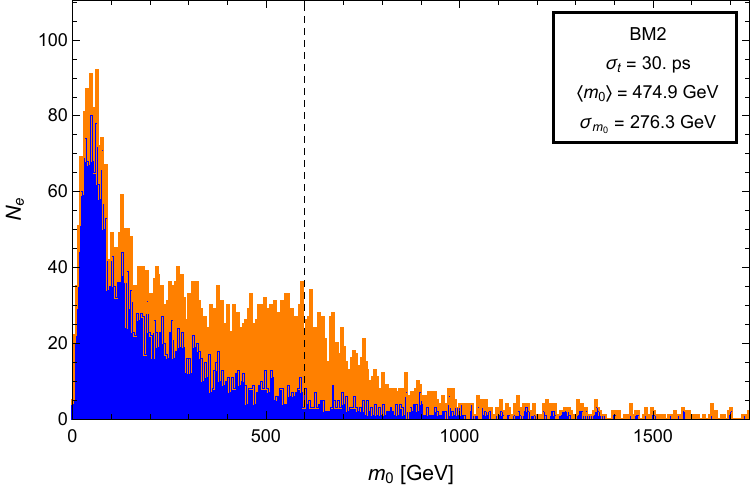}
  \includegraphics[clip, width=0.32\textwidth]{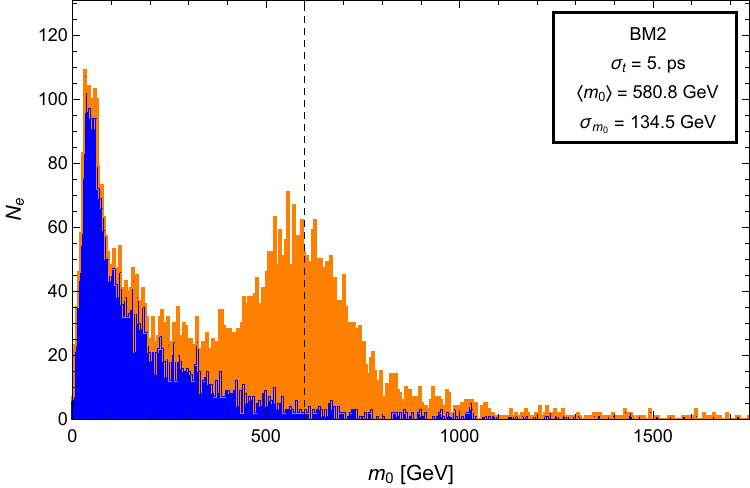}
  \includegraphics[clip, width=0.32\textwidth]{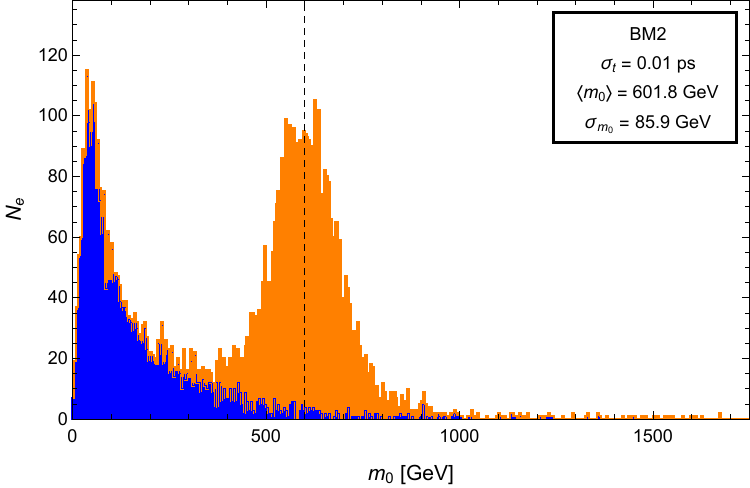}\\
  \includegraphics[clip, width=0.32\textwidth]{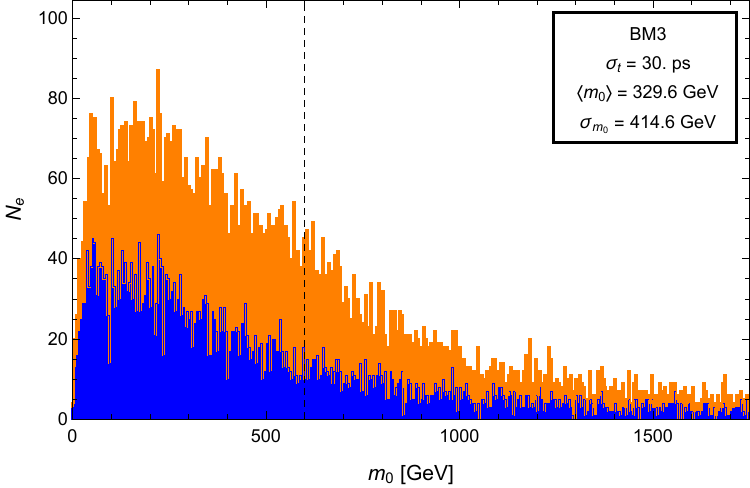}
  \includegraphics[clip, width=0.32\textwidth]{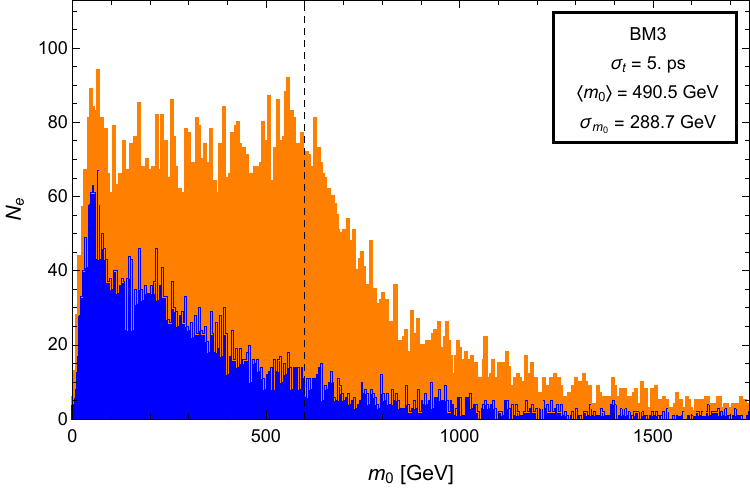}
  \includegraphics[clip, width=0.32\textwidth]{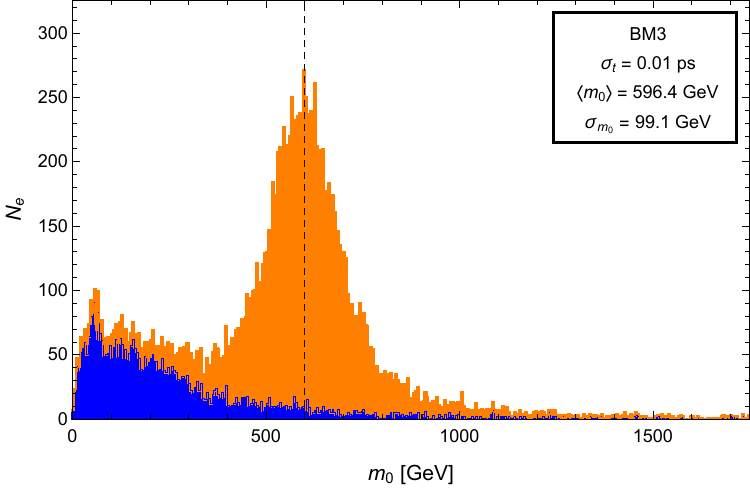}\\
  \includegraphics[clip, width=0.32\textwidth]{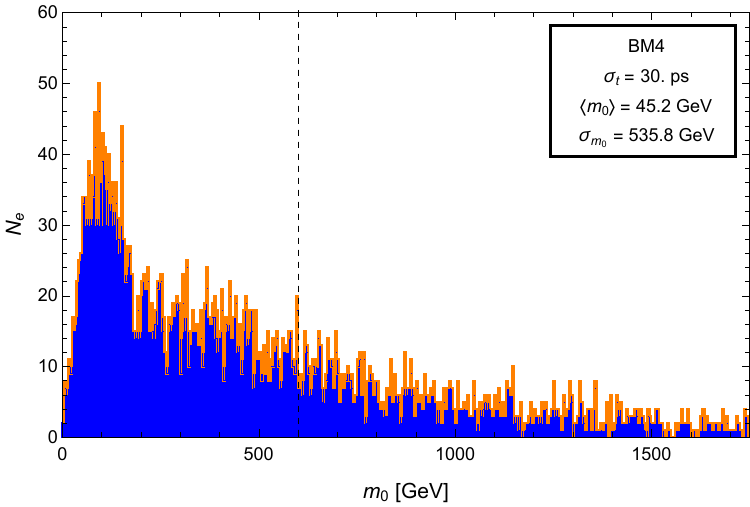}
  \includegraphics[clip, width=0.32\textwidth]{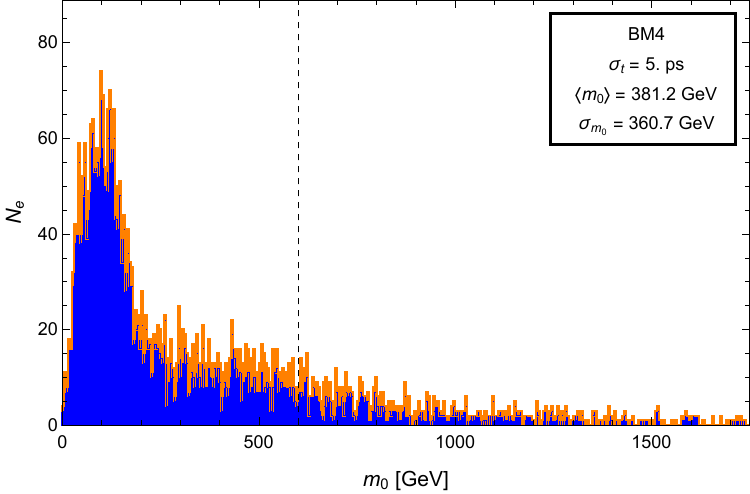}
  \includegraphics[clip, width=0.32\textwidth]{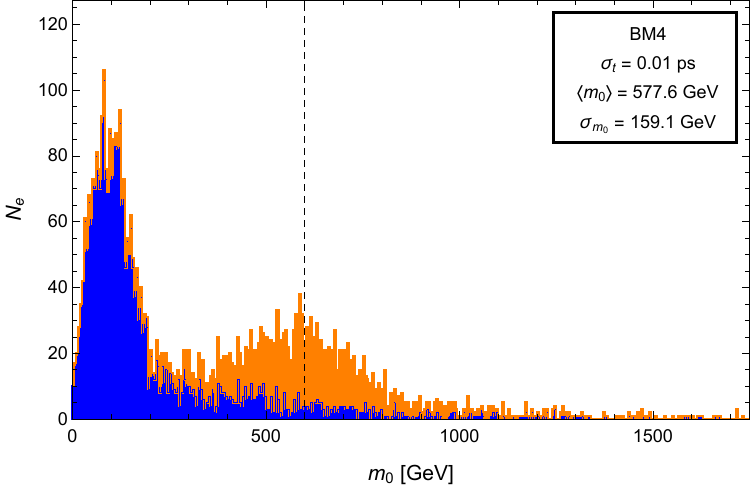}
\caption{Same as Fig.~\protect\ref{fig:DoubleHistm1}, except that the distributions
  shown are for $m_0$ rather than $m_1$.
  \label{fig:DoubleHistm0}}
\end{figure*}

\begin{figure*}
  \includegraphics[clip, width=0.32\textwidth]{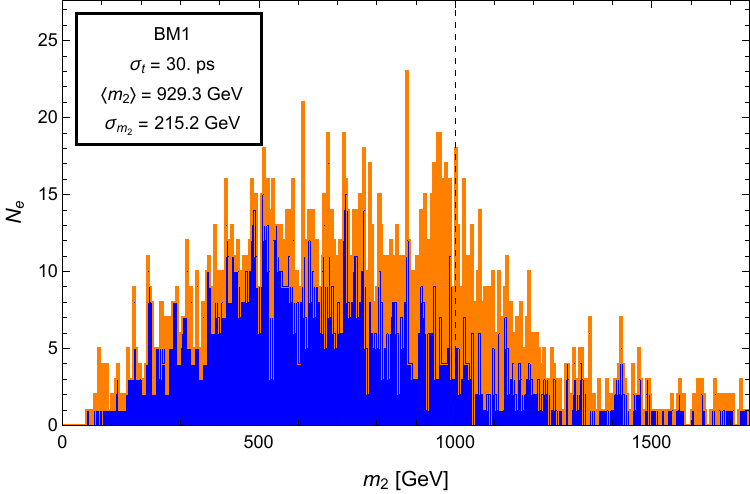}
  \includegraphics[clip, width=0.32\textwidth]{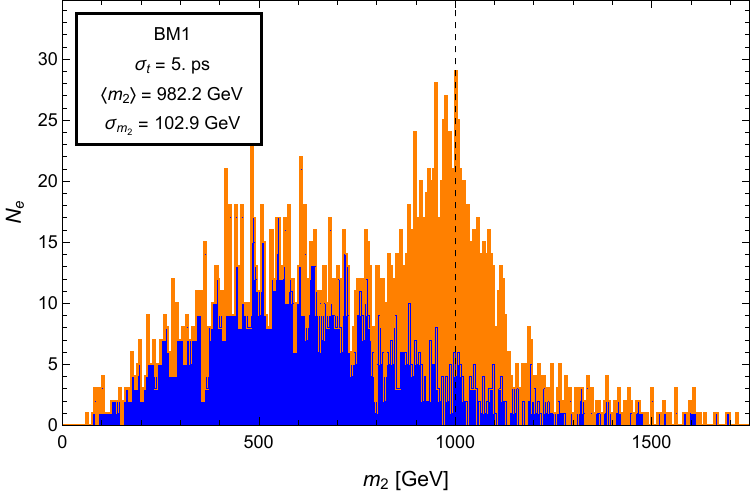}
  \includegraphics[clip, width=0.32\textwidth]{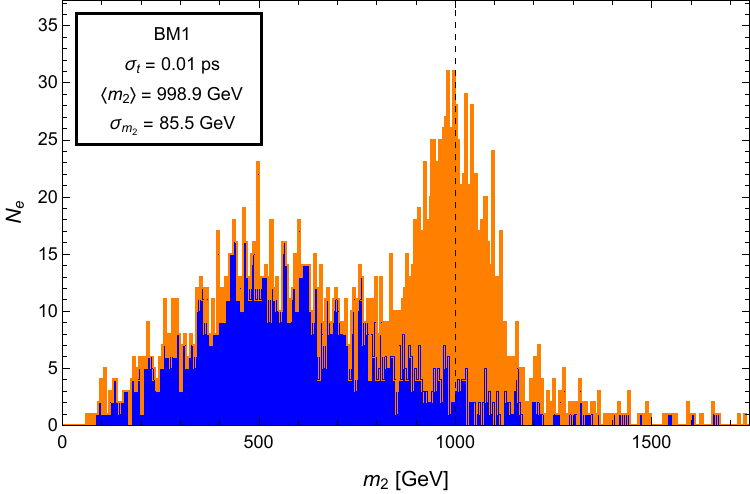}\\
  \includegraphics[clip, width=0.32\textwidth]{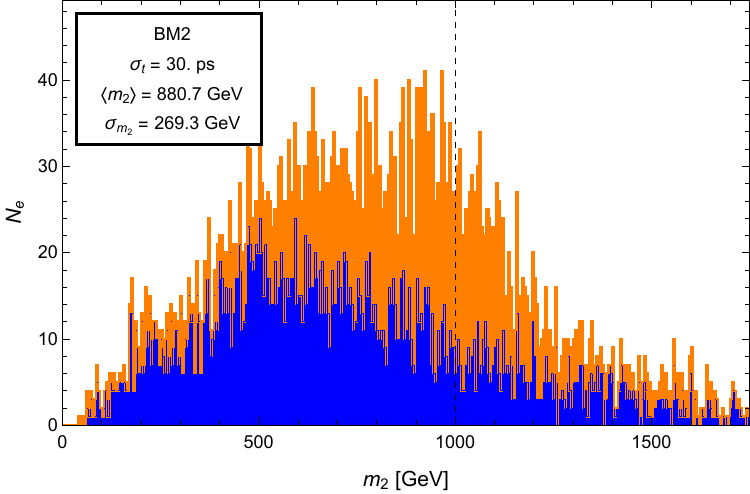}
  \includegraphics[clip, width=0.32\textwidth]{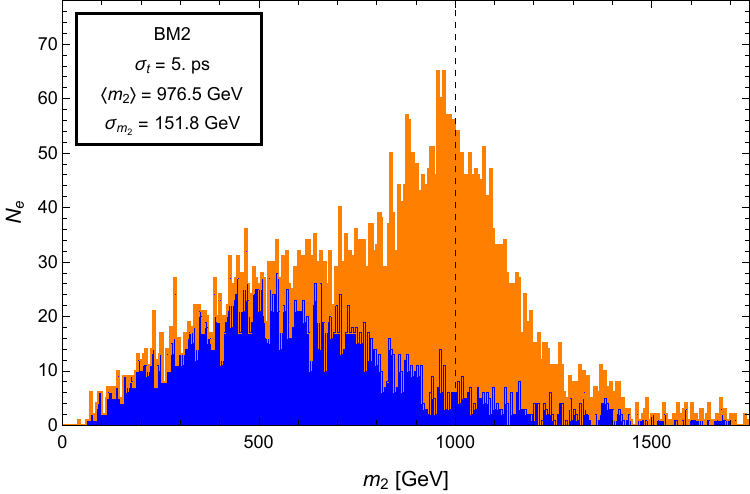}
  \includegraphics[clip, width=0.32\textwidth]{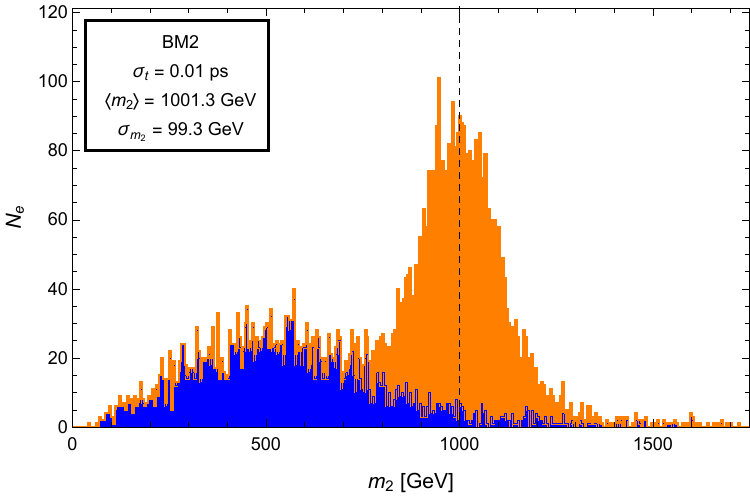}\\
  \includegraphics[clip, width=0.32\textwidth]{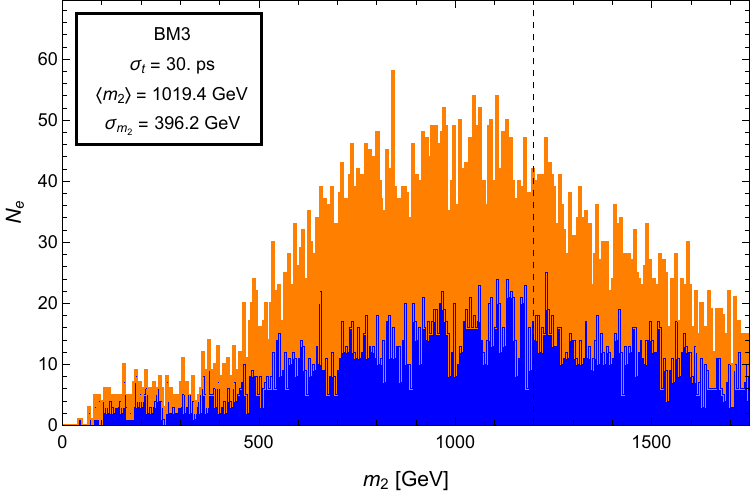}
  \includegraphics[clip, width=0.32\textwidth]{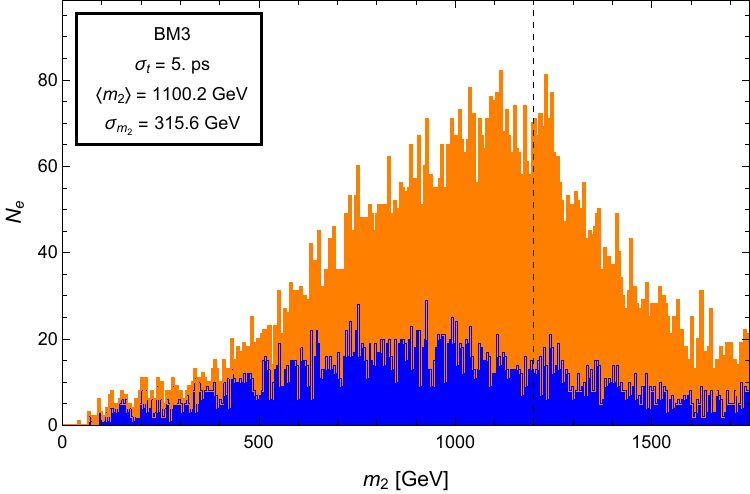}
  \includegraphics[clip, width=0.32\textwidth]{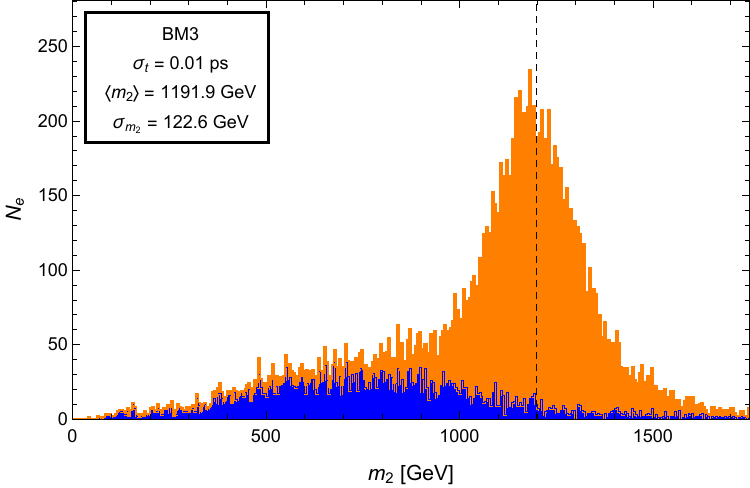}\\
  \includegraphics[clip, width=0.32\textwidth]{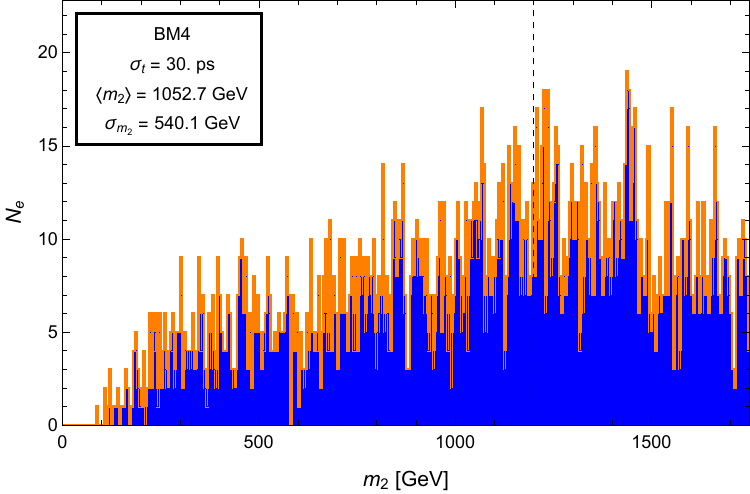}
  \includegraphics[clip, width=0.32\textwidth]{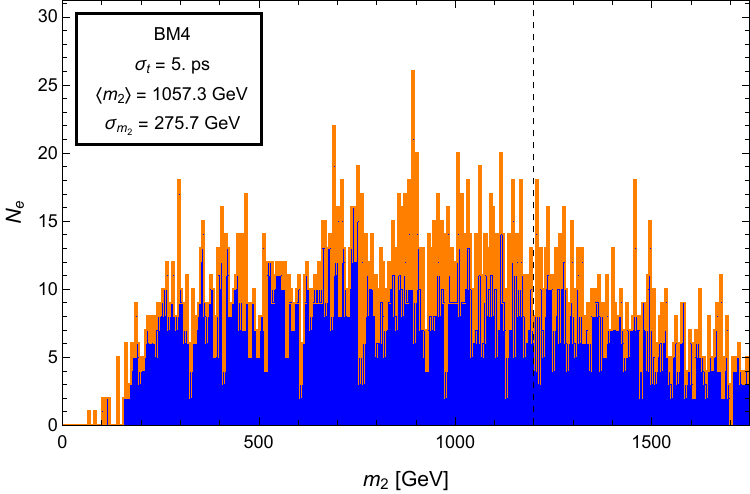}
  \includegraphics[clip, width=0.32\textwidth]{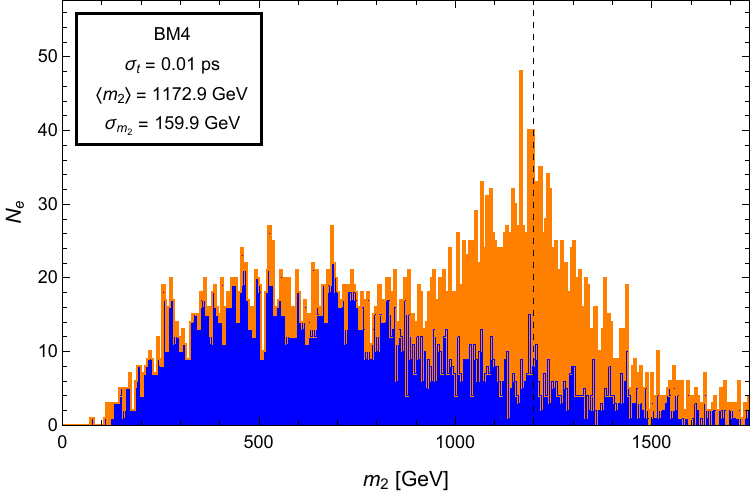}
\caption{Same as Fig.~\protect\ref{fig:DoubleHistm1}, except that the distributions
  shown are for $m_2$ rather than $m_1$.
  \label{fig:DoubleHistm2}}
\end{figure*}

We also observe from Fig.~\ref{fig:DoubleHistm1} that the width $\sigma_{m_1}$ of the 
tumbler peak obtained for each benchmark depends quite sensitively on $\sigma_t$.  
When $\sigma_t$ is fairly large, as shown in the left and center columns of this figure, 
timing uncertainty tends to dominate the widths of the peaks in the tumbler distributions.  
By contrast, when $\sigma_t$ is sufficiently small, as in the right column of this figure, 
the widths of these peaks are instead dominated by $\sigma_E$ and $\sigma_r$.  The value of 
$\sigma_t$ at which this transition occurs for each of our benchmarks depends once again on
$\tau_1$ and $\tau_2$.  Nevertheless, it is clear that the identification of a 
tumbler peak in the $m_1$ distribution will be extremely challenging with a 
timing resolution on the order of the $\sigma_t = 30$~ps that the CMS barrel timing 
layer will be able to provide at the beginning of the upcoming HL-LHC run. 
However, it is also clear that a reduction in timing uncertainty by even a factor of
a few relative to this value would significantly enhance the capabilities of the HL-LHC or 
future colliders --- both in terms of distinguishing tumblers from other signatures of new 
physics involving multiple DVs and in terms of extracting information about the mass spectrum 
of the particles involved.  Indeed, the results shown in Fig.~\ref{fig:DoubleHistm1} are an
indication that we are on the doorstep of being able to probe the underlying physics which 
gives rise to DVs at a much deeper level.  

Thus far, we have focused on the reconstruction of the mass $m_1$.
In Figs.~\ref{fig:DoubleHistm0} and~\ref{fig:DoubleHistm2}, we show the corresponding
distributions of reconstructed $m_0$ and $m_2$ values for our four 
benchmarks, respectively, after the application of our event-selection criteria,
including the $m_0^2 > 0$ criterion.  As with the $m_1$ distributions, there are no 
significant discernible peaks when $\sigma_t$ is larger than 
$\mathcal{O}(1$~--~$5\mathrm{~ps})$.  This is true for all benchmarks.  However, as 
$\sigma_t$ decreases, a discernible peak begins to appear in both the $m_0$ and $m_2$ 
distributions, ultimately becoming higher and narrower as $\sigma_t$ drops.  Moreover, 
for each benchmark, these peaks are centered around the true values of the corresponding 
masses.  However, unlike the distributions shown in Fig.~\ref{fig:DoubleHistm1},
the distributions in Figs.~\ref{fig:DoubleHistm0} and~\ref{fig:DoubleHistm2} do not 
exhibit a discernible dip at any particular value of the corresponding reconstructed $m_n$.

Taken together, the results shown in Figs.~\ref{fig:DoubleHistm1}~--~\ref{fig:DoubleHistm2}
attest that our mass-reconstruction procedure is quite effective in discriminating between 
tumbler and non-tumbler events, provided that the timing uncertainty is sufficiently small
that the peaks in the $m_n$ distributions can be resolved.  On the one hand, it is clear from 
these figures that conclusively identifying tumblers at the HL-LHC with the 
$\sigma_t \approx 30$~ps timing resolution the CMS timing layer is anticipated to provide 
would prove challenging indeed.  On the other hand, it is also clear that a moderate reduction 
in timing uncertainty from $\sigma_t \approx 30$~ps to $\sigma_t \approx 5$~ps would have a 
dramatic effect on our ability
to probe the underlying structure of the decay chains that give rise to events involving
multiple DVs.  As we demonstrated in Sect.~\ref{sec:Discovery}, a robust excess in the 
relevant detection channels could yet be observed at the LHC.~  If such an excess is 
in fact observed, improvements in timing precision, in conjunction with event-selection 
procedures like the one we have developed here, will play a pivotal role in determining
whether or not this excess arises as a consequence of successive decays within the same 
decay chain.

\subsection{Lifetime Reconstruction\label{sec:LifetimeRec}}

\begin{figure*}
  \includegraphics[clip, width=0.32\textwidth]{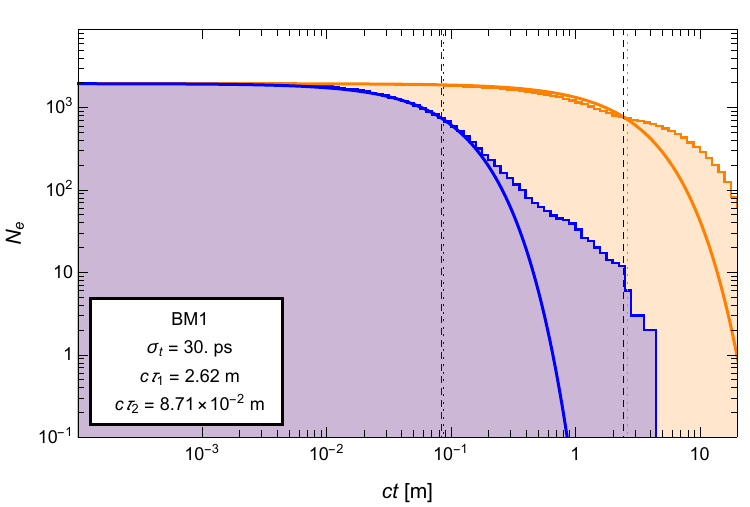}
  \includegraphics[clip, width=0.32\textwidth]{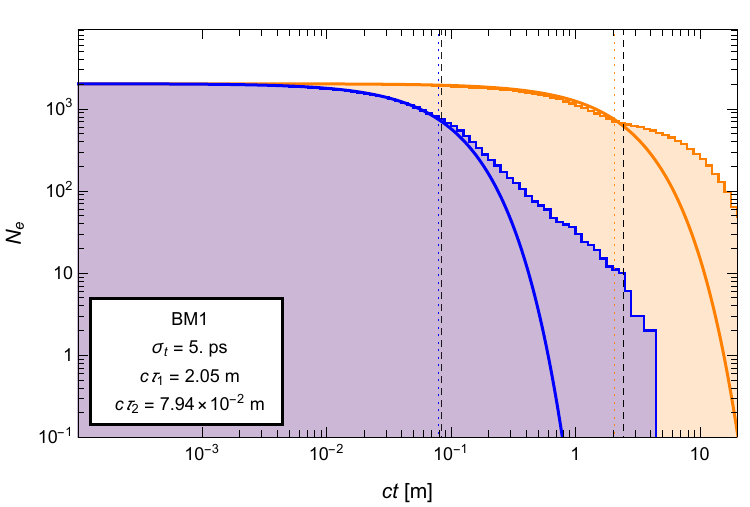}
  \includegraphics[clip, width=0.32\textwidth]{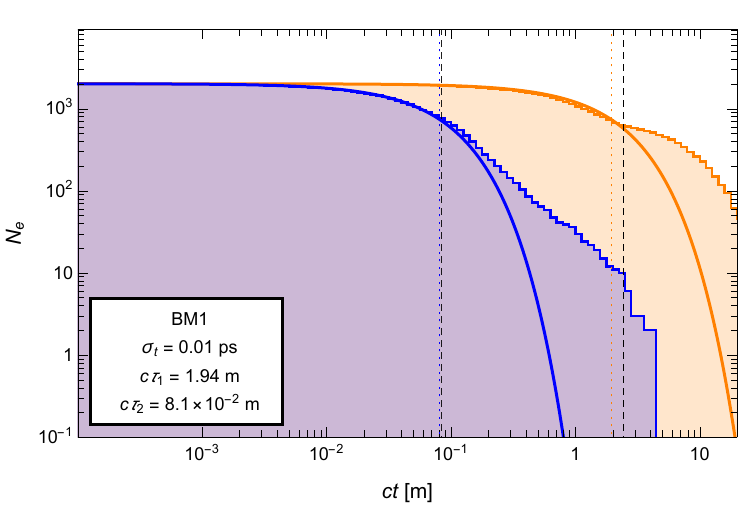}\\
  \includegraphics[clip, width=0.32\textwidth]{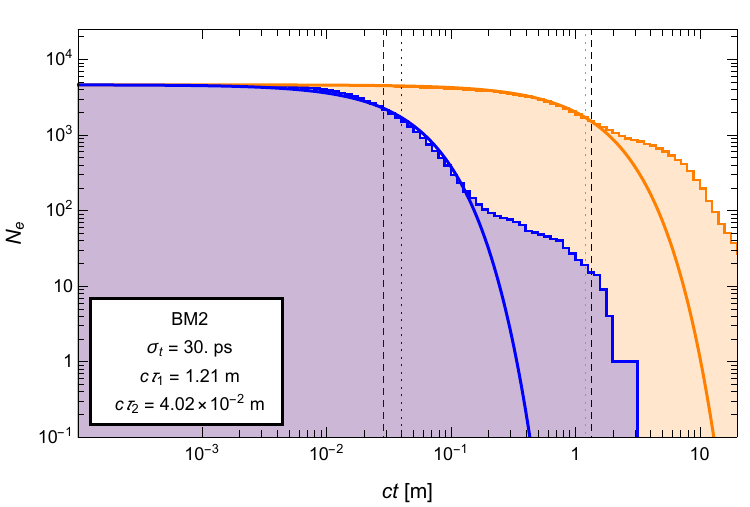}
  \includegraphics[clip, width=0.32\textwidth]{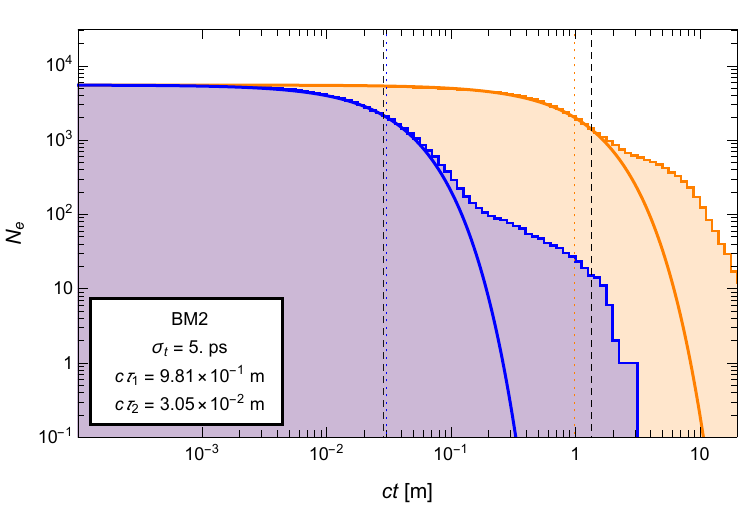}
  \includegraphics[clip, width=0.32\textwidth]{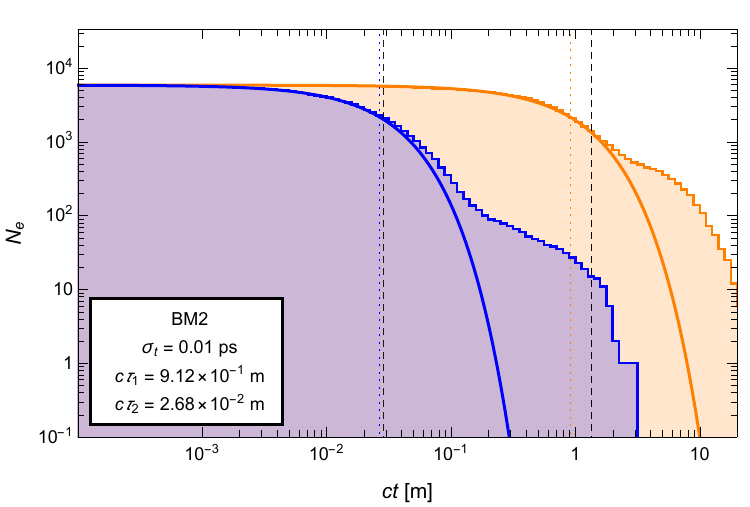}\\
  \includegraphics[clip, width=0.32\textwidth]{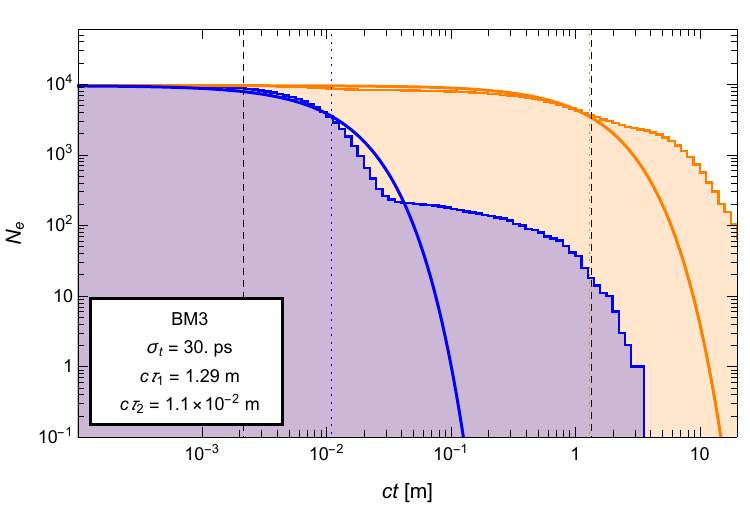}
  \includegraphics[clip, width=0.32\textwidth]{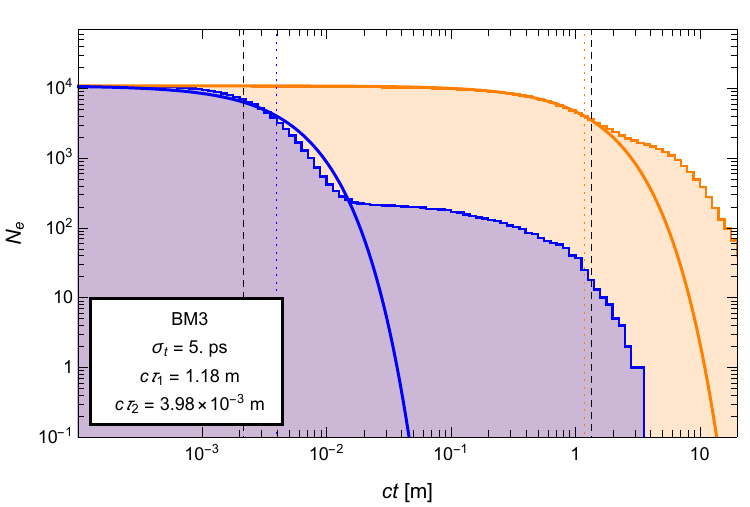}
  \includegraphics[clip, width=0.32\textwidth]{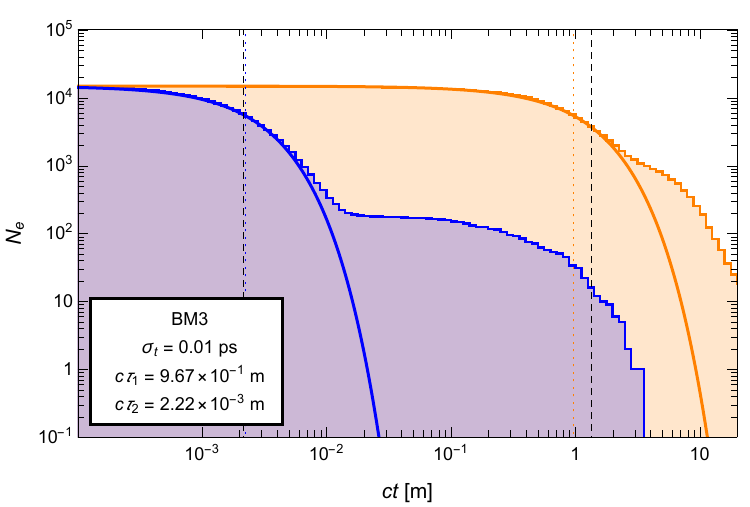}\\
  \includegraphics[clip, width=0.32\textwidth]{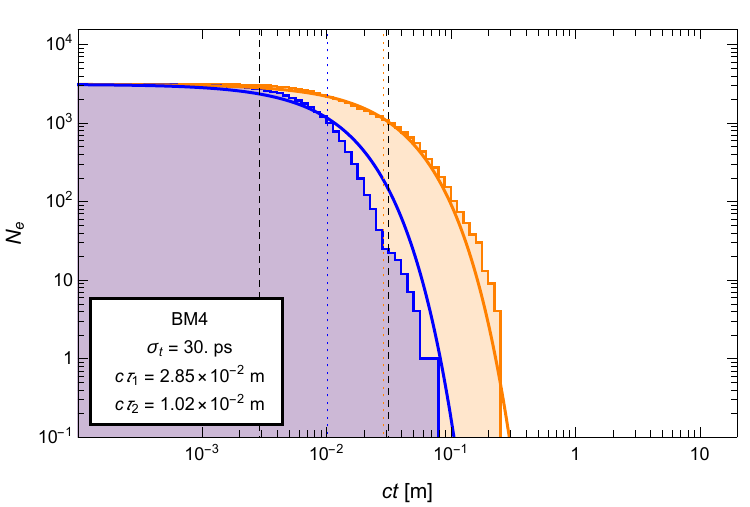}
  \includegraphics[clip, width=0.32\textwidth]{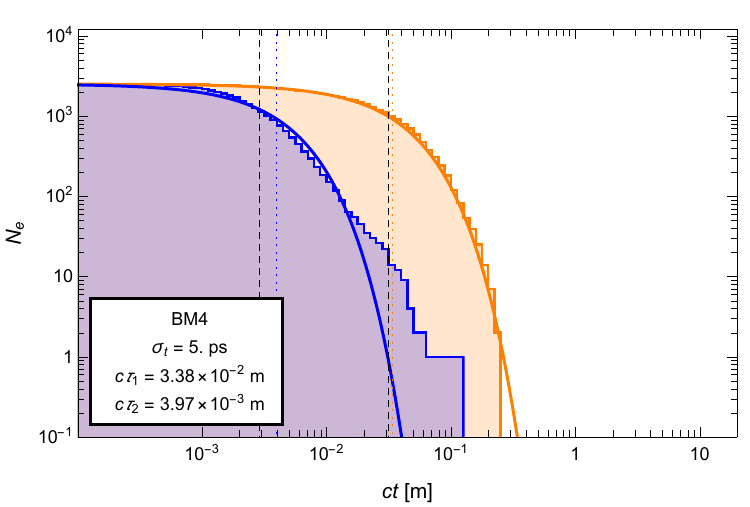}
  \includegraphics[clip, width=0.32\textwidth]{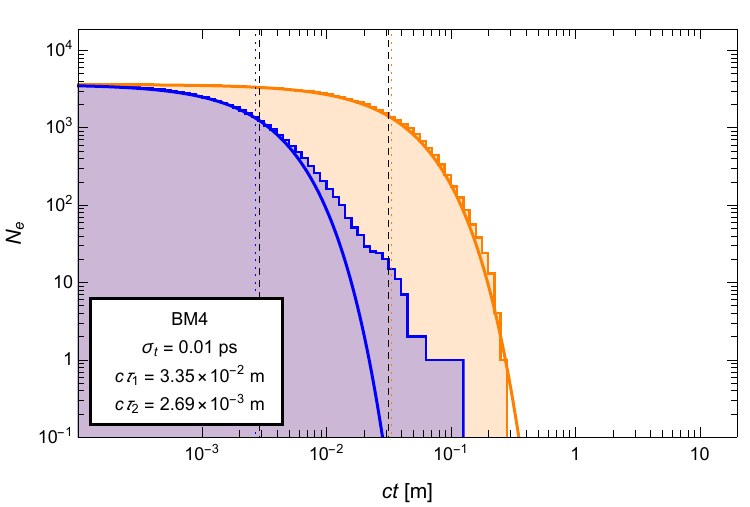}
\caption{Distributions of the number of events $N_1(t)$ (orange histogram) and $N_2(t)$ 
  (blue histogram) for which the corresponding LLP $\chi_1$ or $\chi_2$ 
  has not yet decayed a proper time $t$ after it was initially produced, displayed as a 
  function of the corresponding proper decay distance $ct$.  
  From top to bottom, the rows in the figure correspond to the parameter-space benchmarks 
  BM1 -- BM4 defined in Table~\protect\ref{tab:benchmarks}.~ 
  The results shown in the left, center, and right columns correspond respectively to the 
  values $\sigma_t = 30$~ps, $\sigma_t = 5$~ps, and $\sigma_t = 0.01$~ps for the timing 
  uncertainty of the detector.  Exponential-decay curves constructed 
  using the best-fit values of $c\tau_1$ (thick orange curve) and $c\tau_2$ (thick blue curve) 
  are also shown in each panel.  The dotted orange and blue
  vertical lines correspond to the best-fit values of $c \tau_1$ and $c\tau_2$, respectively,
  while the dashed black vertical lines indicate the actual values of these proper decay 
  lengths.  The best-fit values of $c \tau_1$ and $c\tau_2$ are also quoted in the box in the 
  lower left corner of each panel. 
  \label{fig:LifetimeFit}}
\end{figure*}

We now assess the degree to which we can likewise measure the respective lifetimes 
$\tau_1$ and $\tau_2$ of the unstable LLPs involved in the tumbler
decay chain.  For any given tumbler, the proper-time intervals $t_1$ and $t_2$ between 
the production and decay of each of these particles are given by
$t_1 = (t_T - t_S)/\gamma_1$ and $t_2 = (t_S - t_P)/\gamma_2$, where 
$\gamma_n \equiv (1-|\vec{\boldsymbol{\beta}}_n|)^{-1/2}$ is the usual relativistic factor.
In order to estimate the proper lifetime $\tau_n$ of each particle from a given 
sample of events, we first select events which satisfy the same criteria
we imposed in our mass-reconstruction analysis.  We then define $N_n(t)$ to 
represent the number 
of events in the sample for which $t_n > t$.  We then perform a least-squares fit of the 
function $f(t) = N_n(0)\exp(-t/\tau_n)$ to the events in the sample 
and interpret the value of $\tau_n$ as our estimate for the proper lifetime of $\chi_n$.
Since the goodness-of-fit statistic for this non-linear fit is more sensitive to deviations 
in which $t$ is small and $N_n(t)$ is large, the resulting value of $\tau_n$ is 
typically insensitive to the small, residual contribution to $N_n(t)$ at large $t$ from 
non-tumbler events which nevertheless survive our mass-reconstruction cuts.
 
In Fig.~\ref{fig:LifetimeFit}, we show the results of such a fit for the parameter-space
benchmarks defined in Table~\ref{tab:benchmarks}.~  The orange and blue 
histograms in each panel respectively represent the $N_1(t)$ and $N_2(t)$ distributions 
obtained for a Monte-Carlo data sample that once again initially consists 
of 100,000 events prior the imposition of our 
event-selection criteria.  However, only the $N_1(t)$ and $N_2(t)$ for events which 
pass all of these cuts are included in the histograms.  
The thick orange and blue curves represent the exponential-decay
functions obtained for our best-fit values of $c\tau_1$ and $c\tau_2$, respectively.  From 
top to bottom, the rows in the figure correspond to our parameter-space benchmarks BM1 -- BM4.  
The results shown in the left, center, and right columns of Fig.~\ref{fig:LifetimeFit} once
again correspond respectively to the timing uncertainties $\sigma_t = 30$~ps, 
$\sigma_t = 5$~ps, and $\sigma_t = 0.01$~ps.

We begin by noting that each $N_n(t)$ distribution shown in 
Fig.~\ref{fig:LifetimeFit} clearly includes contributions from two distinct
populations of events.  The first of these populations, which is far larger than
the second and dominates $N_n(t)$ when $ct$ is small, comprises genuine tumbler
events.  The second population, which includes 
events with much higher $ct$ values, consists primarily of residual non-tumbler 
events.  Since this second population is quite small, it does not have a dramatic 
impact on the best-fit value of the corresponding $c\tau_n$.

The results shown in Fig.~\ref{fig:LifetimeFit} demonstrate that for relatively large 
$\sigma_t$ values, the accuracy with which these lifetimes can be measured differs 
among the different benchmarks.  For example, the extent to which our fitting 
procedure overestimates the value of $\tau_2$ for BM3 and BM4 is significant, 
whereas this effect is less severe for BM1 and BM2.  This is once again primarily 
a reflection of the fact that $\tau_2$ is far shorter 
for BM3 and BM4 that it is for these other benchmarks, and hence the effect of timing 
uncertainty on the results for BM3 and BM4 becomes significant at a far lower value of 
$\sigma_t$.

Somewhat counterintuitively, however, we also observe that our fit 
systematically underestimates the value of $\tau_1$ for BM1 -- BM3 by as much as
a factor of two when $\sigma_t$ is small.  This is a consequence of $\tau_1$ being 
sufficiently large for these benchmarks that a small but non-negligible fraction of 
the $\chi_1$ particles produced by $\chi_2$ decays themselves decay outside the 
timing layer.  Since events in which these $\chi_1$ particles decay outside the 
timing layer are 
of course not included in any of our event samples, the $N_1(t)$ distribution is 
slightly skewed toward lower lifetimes.  Thus, as $\sigma_t$ decreases, the best-fit 
value of $c\tau_1$ approaches a value slightly below the actual proper decay length.
This effect is not particularly significant for BM4, however, since $\tau_1$ is far 
shorter and the fraction of events in which $\chi_1$ escapes the detector before 
it decays is therefore far smaller.  That said, we emphasize that reasonably 
reliable measurements of both $\tau_1$ and $\tau_2$ can nevertheless be made for 
all four of our benchmarks when $\sigma_t = 0.01$~ps, even for the simple, 
physically-motivated functional fit we 
have performed here.  An alternative functional fit which accounts for this 
finite-volume effect could yield even better estimates of the LLP lifetimes.   


\begin{figure*}[t]
  \includegraphics[clip, width=0.32\textwidth]{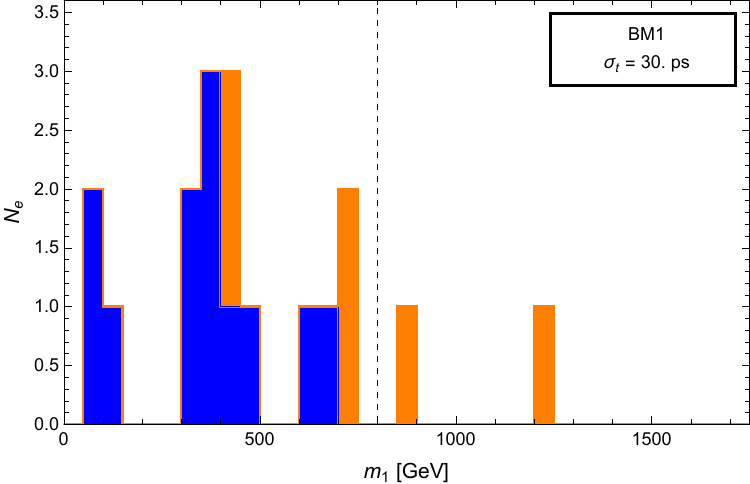}
  \includegraphics[clip, width=0.32\textwidth]{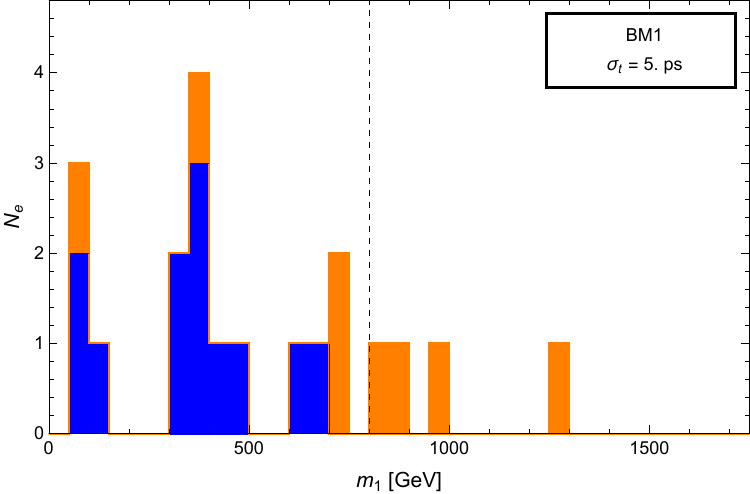}
  \includegraphics[clip, width=0.32\textwidth]{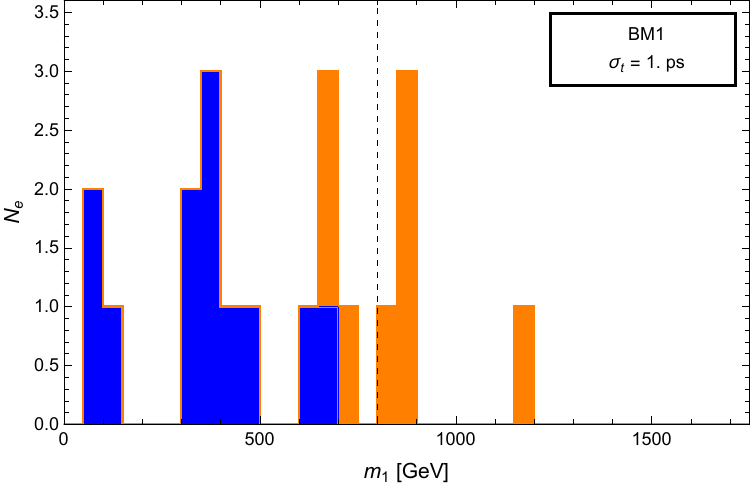}\\
  \includegraphics[clip, width=0.32\textwidth]{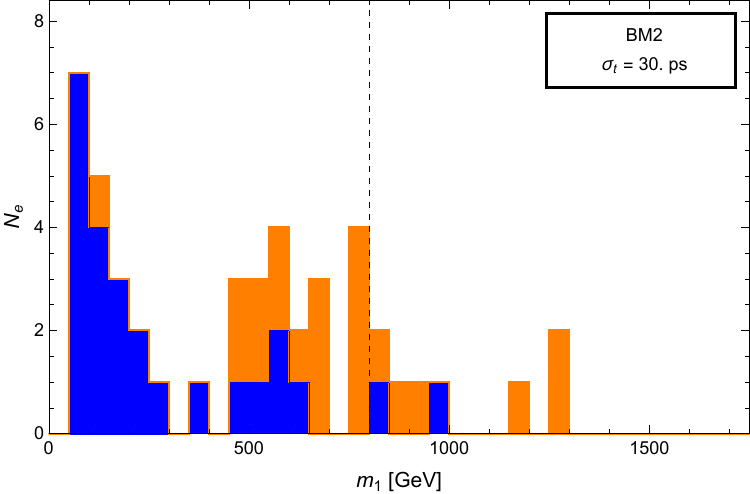}
  \includegraphics[clip, width=0.32\textwidth]{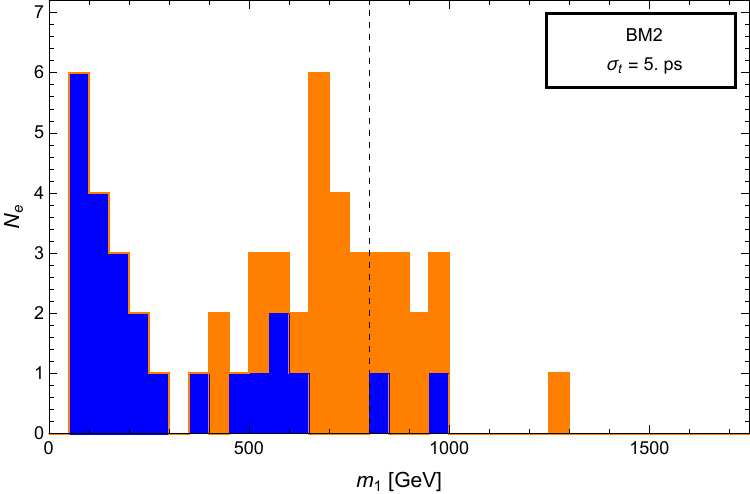}  
  \includegraphics[clip, width=0.32\textwidth]{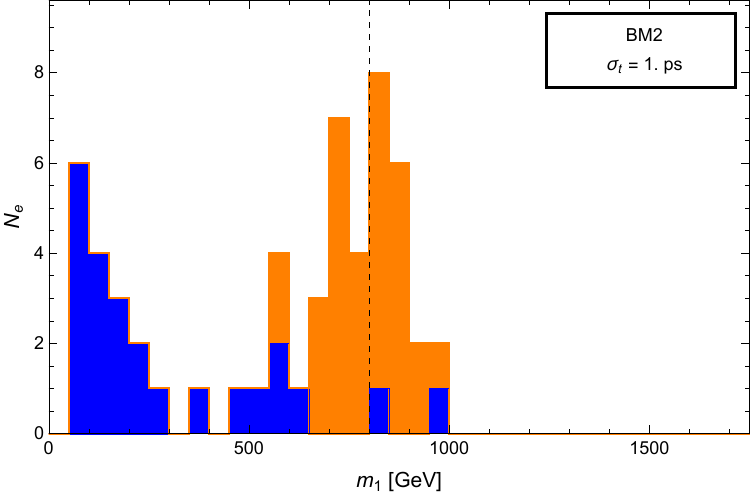}\\
  \includegraphics[clip, width=0.32\textwidth]{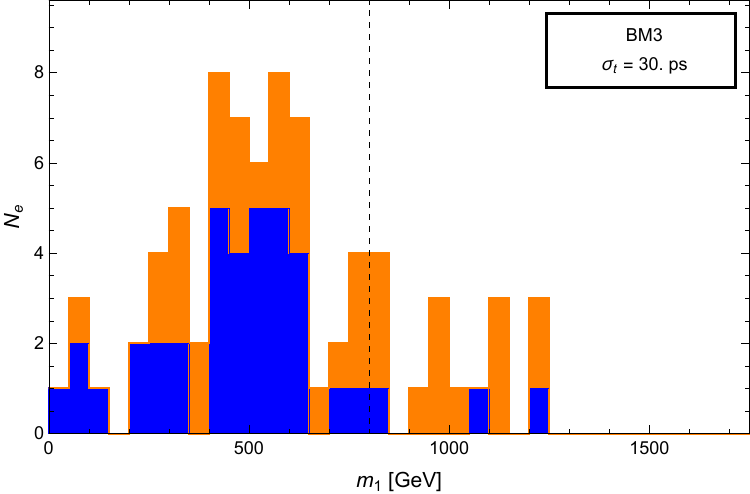}
  \includegraphics[clip, width=0.32\textwidth]{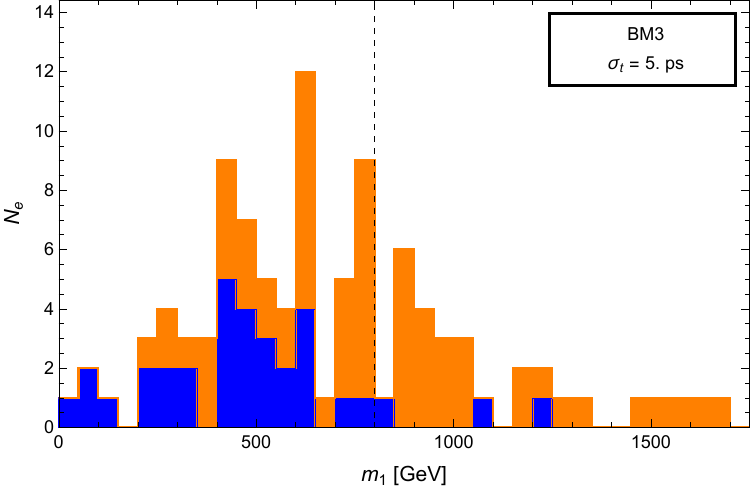}
  \includegraphics[clip, width=0.32\textwidth]{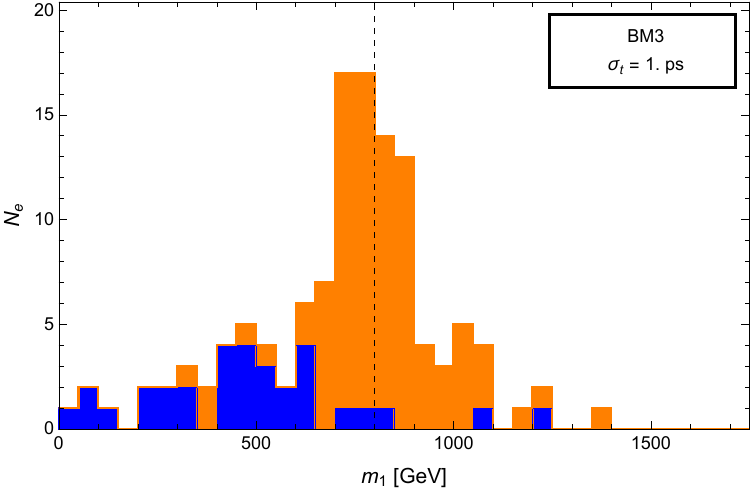}\\
  \includegraphics[clip, width=0.32\textwidth]{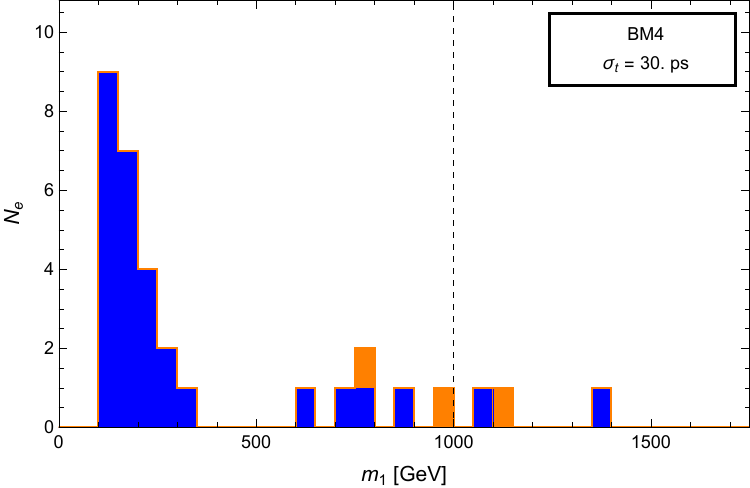}
  \includegraphics[clip, width=0.32\textwidth]{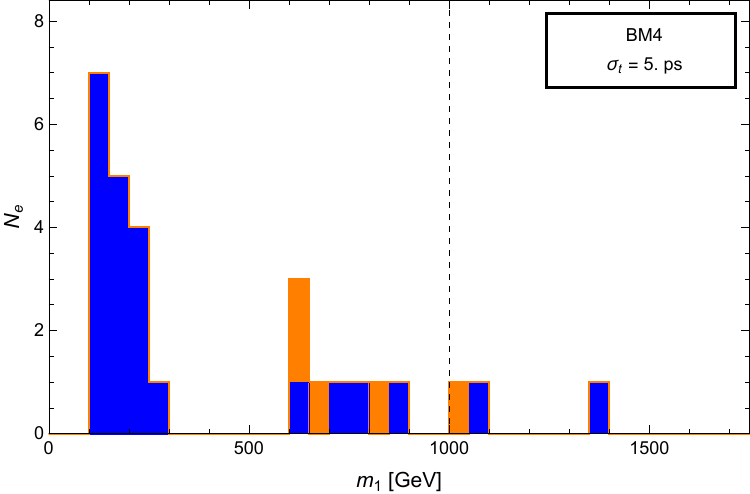}
  \includegraphics[clip, width=0.32\textwidth]{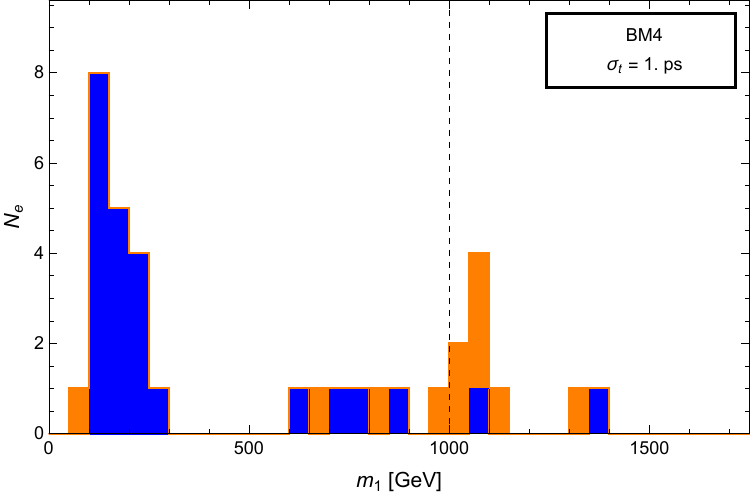}
\caption{Same as Fig.~\ref{fig:DoubleHistm1}, but for a smaller sample of 
  Monte-Carlo events.  In particular, the numbers of both tumbler and non-tumbler DV
  events included in each data sample before any cuts are applied are equal to the 
  expected numbers of events at a collider essentially identical to the HL-LHC, but
  with an integrated luminosity twice that anticipated for the full HL-LHC run.
  The results shown in the left, center, and right columns correspond respectively to the 
  values $\sigma_t = 30$~ps, $\sigma_t = 5$~ps, and $\sigma_t = 1$~ps for the timing 
  uncertainties of the two general-purpose detectors at this collider, and
  we have adopted a bin width of 50~GeV when constructing each histogram.
  \label{fig:DoubleHistFCm1}}
\end{figure*}
 
\subsection{Tumbler Searches with Limited Event Counts\label{sec:LimEvCts}}
 
We now consider the extent to which we are able to resolve the 
characteristic tumbler peaks in the distributions of the reconstructed $m_n$
values given a far smaller number of events --- a number which might 
realistically be obtained at the HL-LHC or at other, near-future colliders.
For this purpose, we shall consider a hypothetical collider essentially identical 
to the HL-LHC.  The two general-purpose detectors at this collider are each assumed 
to be equipped with a barrel timing layer with timing uncertainty
$\sigma_t$, but to be otherwise similar in design and performance to the CMS detector.  
We assume an integrated luminosity $\mathcal{L}_{\rm int} = 6000$~fb$^{-1}$ in each
detector --- an integrated luminosity equal to twice that anticipated 
for the HL-LHC over its full run.  Thus, the total event count for tumbler events
before cuts is taken to be $2\mathcal{L}_{\rm int}\sigma^{({\rm T})}$, and the 
total number of non-tumbler events including at least one DV is calculated in
an analogous manner.

In Fig.~\ref{fig:DoubleHistFCm1}, we show the distribution of reconstructed 
$m_1$ values for a Monte-Carlo data set consisting of the expected number 
of events for each of our parameter-space benchmarks at such a pair of 
collider detectors.  Only events which survive all of our cuts --- including 
the $m_0 > 0$ criterion --- are included in each distribution shown.
From top to bottom, the rows in the figure correspond to the parameter-space 
benchmarks defined in Table~\protect\ref{tab:benchmarks}.~ 
The results shown in the left, center, and right columns correspond respectively 
to the timing uncertainties $\sigma_t = 30$~ps, $\sigma_t = 5$~ps, and 
$\sigma_t = 1$~ps.  As in Fig.~\protect\ref{fig:DoubleHistm1}, the orange and blue
portions of each histogram represent the contributions from tumbler and non-tumbler 
events, respectively, while the dashed black vertical line in each panel 
indicates the actual value of $m_1$ for the corresponding benchmark.  However,
we have adopted a coarser bin width of 50~GeV than we did when constructing 
each histogram in Fig.~\protect\ref{fig:DoubleHistm1}.

Perhaps the most important message of Fig.~\ref{fig:DoubleHistFCm1} is that 
the characteristic tumbler peak in the $m_1$ distribution around the true value 
of $m_1$ is evident for many of our benchmarks for $\sigma_t \lesssim 5$~ps.  
Indeed for BM2 and BM3, this peak is particularly striking.  This once again 
demonstrates that an order-of-magnitude enhancement in timing resolution could 
yield compelling evidence of tumblers --- even with an integrated luminosity 
that could easily be achieved at future colliders.


\FloatBarrier
\section{Conclusions\label{sec:Conclusion}}


In this paper, we have described a novel potential signature of new physics at 
colliders.  This signature involves processes which we call tumblers --- processes
in which multiple successive decays of LLPs within the same decay
chain give rise to multiple DVs within the same event.  We have investigated the
prospects for observing tumblers at the LHC 
both before and after the high-luminosity upgrade. 
Despite the stringent constraints that current LHC data impose on processes involving 
DVs, we have shown in the context of a concrete model that a significant number of 
tumbler events could yet be observed at the LHC.~  However, scenarios which give rise to 
a significant number of tumbler events also often give rise to a significant 
number of non-tumbler events which also involve multiple DVs.
In order to address this issue, we have developed an event-selection procedure which 
permits us to discriminate efficiently between tumbler and non-tumbler events on the basis 
of the distinctive kinematics associated with tumbler decay chains.  This procedure 
incorporates the timing 
information provided by the collider detector regarding the SM particles produced by these 
decay chains.  As a result, the degree to which this procedure is capable of distinguishing
tumbler from non-tumbler events depends crucially on the timing resolution of the detector.
Interestingly, we have shown that a modest enhancement in timing precision beyond the 
$\sigma \approx 30$~ps timing resolution that will be provided by the CMS timing layer at the 
outset of the forthcoming HL-LHC upgrade could have a crucial impact on the prospects for
discerning tumblers amongst possible signals of new physics involving multiple DVs.
Moreover, via this same procedure, we have shown that it is also possible 
to reconstruct the masses and lifetimes of these LLPs.  Once again, the precision 
to which these masses and lifetimes can be measured depends crucially on the timing 
uncertainty of the detector.

Several comments are in order.  First, we have made a number of simplifications
concerning the manner in which DVs are identified and reconstructed in our analysis.  
In so doing, we have accounted for the relevant uncertainties in a manner sufficient 
to provide a reasonable estimate of the detector capabilities necessary in order to detect 
a robust signature of tumblers.  That said, a precise, quantitative estimate of the discovery 
reach for tumblers at a particular detector would require a more detailed, track-based 
analysis which incorporates information about the tracker geometry.
Moreover, advances in detector technology may enhance the performance of 
particular regions of a collider detector with regard to DV reconstruction.
For example, during the forthcoming high-luminosity upgrade, a High-Granularity Calorimeter 
(HGCal) with a timing resolution of $\sim 40$~ps will be installed within the endcap region of 
the CMS detector.  This HGCal will make it possible to reconstruct DVs produced by particles
whose decay products are emitted anywhere within the endcap region of the detector 
with excellent precision, even at trigger level~\cite{Liu:2020vur}.  Such detector 
capabilities would improve the geometric acceptance for events involving DVs and 
therefore enhance the discovery reach for tumblers.

Second, in this paper, we have employed the mass-reconstruction procedure introduced in 
Sect.~\ref{sec:ResultsRecon} as our primary mechanism for distinguishing between tumbler and 
non-tumbler events.  However, there may be more efficient methods of distinguishing between 
these two types of events.  Various possibilities along these lines are under
investigation~\cite{TumblersVsAntlers}.

Third, we have focused in this paper on the case in which the tumbler decay chains involve 
only three particles: $\chi_0$, $\chi_1$, and $\chi_2$.  Indeed, this is the minimum number 
of $\chi_n$ needed in order to give rise to a tumbler.  However, tumblers 
can also arise in more 
complicated scenarios in which the number $N$ of $\chi_n$ particles is larger --- perhaps 
substantially so.  It is therefore interesting to consider how the tumbler phenomenology of  
the $N=3$ model analyzed in this paper generalizes for larger values of $N$.  In keeping
with our established notation, we shall assume that these additional $\chi_n$, where 
$n = 3,\ldots, N-1$, are all heavier than $\chi_2$.  We shall nevertheless continue
to assume that $m_\phi > m_{N-1}$.  Several observations can then immediately be made.

One possibility is that the lifetimes $\tau_n$ of the additional 
$\chi_n$ are sufficiently short that these particles decay promptly.  In this regime,
tumbler events which arise as a consequence of $pp\ra\phi^\dagger \phi$ production 
will often include additional prompt jets which can be traced back to the primary 
vertex.  When the number of such jets is large, both triggering and the reconstruction
of DVs from kinematic information becomes more challenging.
Furthermore, when $N$ becomes large, the contribution to the total event 
rate from processes of the form $pp\ra \phi\chi_n$ and $pp\ra \chi_m\chi_n$ increases 
simply as a result of the multiplicity of the LLPs.~  For sufficiently large
$N$, the contribution from these processes to the effective cross-section
$\sigma_{\mathrm{eff}}^{(\mathrm{T})}$ for tumbler events --- and to the 
effective cross-sections for other classes of processes as well --- can overwhelm
the contribution from $pp\ra \phi^\dagger\phi$.  

In cases in which the $\tau_n$ for one or more of the additional states are
within the DV regime, further complications arise.  The reconstruction of the 
$m_n$ and $\tau_n$ in this case becomes more challenging, since the tumblers themselves 
can involve different sequences of $\chi_n$, even for decay chains involving only two DVs.   
Moreover, tumblers involving more than two DV can also arise.  Nevertheless, the 
methods we have developed in Sect.~\ref{sec:ResultsRecon} can be generalized in a 
straightforward manner.  It is still the case, for example, that the momentum and timing 
information for the jets produced by a tumbler involving more than two individual decay steps 
is sufficient to permit the reconstruction of the $m_n$ and $\tau_n$ of the LLPs involved 
in the corresponding decay chain.  In particular, the reconstructed $m_n$ distributions 
corresponding to the maximal tumbler decay chain --- \ie, the chain involving the largest 
possible number of individual displaced decay steps --- will each exhibit a peak around 
the true value of $m_n$.  However, the reconstructed mass distributions of {\it non}\/-maximal 
such decay chains will manifest a more complicated peak structure as a result of different decay 
sequences involving the same number of steps.  For example, a two-step decay sequence
from $\chi_3$ to $\chi_0$ could proceed via $\chi_3 \ra \chi_2 \ra \chi_0$ or 
$\chi_3 \ra \chi_1 \ra \chi_0$.  Following the procedures we have outlined
in this paper for such two-step decays, a reconstruction of the mass of the intermediate state 
would then result in {\it two}\/ peaks: one centered around $m_1$ and one centered around 
$m_2$.  Of course, given the limited spatial extent of the tracker and the fact that lighter 
$\chi_n$ are typically longer-lived than the heavier $\chi_n$, decay chains involving 
large numbers of steps may be difficult to resolve in this manner. 

Fourth, one could also consider more complicated event topologies involving LLPs which 
are themselves produced at DVs.  Indeed, tumblers are merely the simplest example of such 
an event topology.  More complicated event topologies in which multiple LLPs are 
produced at the same DV are also possible.
Such possibilities would result in a proliferation of decay chains, ultimately leading 
to ``showers'' of LLPs within the collider environment.  Of course, whether or not 
these showers are detectable as such depends on the lifetimes of the particles 
involved.

Fifth, in addition to considering changes in the topology of the decay chains, one might also 
consider changes in the properties of the individual decays themselves, such as their decay 
products.  In this paper we have focused on models in which each decay within the tumbler 
produces two quarks, ultimately leading to two jets.  However, it is also possible to consider 
models in which only a single quark is produced at each DV.~  In such cases, the
techniques we have employed in this paper for reconstructing DVs would not be appropriate.
However, as discussed above, DVs can still be reconstructed via a track-based analysis, even
in such cases.  Likewise, it is possible to consider models in which the SM particles produced 
by LLP decays include charged leptons as well as quarks and/or gluons.  Methods for 
reconstructing DVs likewise exist for such cases. 

Sixth, our primary aim in this paper has been to demonstrate that the observation 
of a tumbler signature is a viable possibility at the HL-LHC or other near-future 
colliders.  Thus, while we have shown that there do exist regions of the 
parameter space of our example model which are consistent with current 
constraints, we have not undertaken a detailed analysis of 
exactly where the exclusion contours lie within that parameter space.  
Recasting tools such as 
{\tt MadAnalysis~5}~\cite{Conte:2012fm,Araz:2021akd},
{\tt SModelS}~\cite{Ambrogi:2018ujg}, 
{\tt CheckMATE}~\cite{Desai:2021jsa}, 
and the computational resources associated 
with the {\tt RECAST} framework~\cite{Cranmer:2010hk,ATLAS:2020viz} 
can assist in establishing the locations of these exclusion 
contours.  That said, a dedicated study along these lines would be valuable, 
in light of the numerous subtleties involved in recasting the results of 
searches involving DV signatures in order to constrain more general classes 
of new-physics scenarios.
We leave such a study for future work.

Finally, in this paper, we have focused on the case in which both LLPs involved in our
(two-step) tumbler decay chain decay within the collider tracker.  One could also 
consider the case in which the decay of one or both of these LLPs occurs 
within the calorimeters or the muon chamber.  Indeed, searches have been 
performed by the ATLAS Collaboration~\cite{Aad:2019xav} 
for events involving multiple displaced decays in which one such decay 
occurs within the tracker and the other occurs within these outer layers 
of the detector.  Moreover, one could also consider the case in which the 
lighter LLP escapes the main detector entirely and decays 
within an external detector designed specifically for the purpose of 
observing LLP decays, such as MATHUSLA~\cite{Chou:2016lxi} or 
FASER~\cite{Feng:2017uoz}.  By incorporating information from such dedicated 
LLP detectors, one would potentially be able to extend an 
analysis of the sort we have performed in this paper across a broader range 
of LLP lifetimes.  In fact, MATHUSLA may even be capable of detecting 
evidence of a tower of LLPs, as discussed in 
Refs.~\cite{Curtin:2018mvb,Curtin:2018ees}.

\begin{acknowledgments}

We would like to thank Gabriel Facini and Zhen Liu for discussions. 
TL wishes to thank the EXCEL Scholars Program for Undergraduate Research at Lafayette 
College, which helped to facilitate this research.   
The research activities of KRD are supported in part by the Department of Energy 
under Grant DE-FG02-13ER41976 (DE-SC0009913) and by the National Science Foundation 
through its employee IR/D program.  The research activities of DK are supported 
in part by the Department of Energy under Grant DE-SC0010813.  
The research activities of TL and BT are supported in 
part by the National Science Foundation under Grant PHY-1720430.
The research activities of BT are also supported in part by the National Science 
Foundation under Grant No. PHY-2014104.  The opinions and conclusions expressed 
herein are those of the authors, and do not represent any funding agencies.

\end{acknowledgments}

\appendix

\FloatBarrier
\section{~Vertexing Procedure\label{app:Vertexing}}


We consider a pair of SM particles $A$ and $B$ which we assume to have been produced 
at the same vertex within a collider detector.  We refer to the lab-frame three-momenta of 
these particles as $\vec{\mathbf{p}}_A$ and $\vec{\mathbf{p}}_B$, and we refer to
the lab-frame coordinates at which they exit the tracker as $\vec{\mathbf{r}}_A$ and 
$\vec{\mathbf{r}}_B$.  The trajectories of these particles lie along two lines which are 
described parametrically by  
\begin{eqnarray}
  \vec{\mathbf{R}}_A(a) &~\equiv~& \vec{\mathbf{r}}_A + a\vec{\mathbf{p}}_A
    \nonumber \\
  \vec{\mathbf{R}}_B(b) &~\equiv~& \vec{\mathbf{r}}_B
    + b\vec{\mathbf{p}}_B~.
  \label{eq:TracebackLines}
\end{eqnarray} 
A value of $a$ or $b$ identifies a particular location along the corresponding line.  

Given that $A$ and $B$ are produced at the same vertex, the 
lines in Eq.~(\ref{eq:TracebackLines}) will intersect at the vertex location, assuming
$\vec{\mathbf{r}}_A$, $\vec{\mathbf{r}}_B$, $\vec{\mathbf{p}}_A$, and 
$\vec{\mathbf{p}}_B$ are all measured with infinite precision.
However, in an actual experiment, measurement uncertainties in these quantities will 
typically result in the lines passing very close to each other, but not actually intersecting.  
We can obtain a best estimate 
for the intersection point by identifying the values of $a$ and $b$ for which the vector 
$\vec{\mathbf{D}}(a,b) \equiv \vec{\mathbf{R}}_A(a) - \vec{\mathbf{R}}_B(b)$
is perpendicular to both lines --- \ie, for which
\begin{eqnarray}
  \vec{\mathbf{D}}(a,b) \cdot \vec{\mathbf{p}}_A &~=~& 0 \nonumber \\
  \vec{\mathbf{D}}(a,b) \cdot \vec{\mathbf{p}}_B &~=~& 0~.
  \label{eq:DabEqs}
\end{eqnarray} 
 
Solving the system of equations in Eq.~(\ref{eq:DabEqs}) for $a$ and $b$, we find that
\begin{eqnarray}
    a &~=~& \frac{-\vec{\mathbf{p}}_A \cdot (\vec{\mathbf{r}}_A 
      - \vec{\mathbf{r}}_B)|\vec{\mathbf{p}}_B|^2
      + \vec{\mathbf{p}}_B \cdot (\vec{\mathbf{r}}_A - \vec{\mathbf{r}}_B) 
      (\vec{\mathbf{p}}_A\cdot \vec{\mathbf{p}}_B) }
      {|\vec{\mathbf{p}}_A|^2|\vec{\mathbf{p}}_B|^2 
      - (\vec{\mathbf{p}}_A\cdot \vec{\mathbf{p}}_B)^2} 
      \nonumber \\
    b &~=~& \frac{\vec{\mathbf{p}}_B \cdot (\vec{\mathbf{r}}_A 
      - \vec{\mathbf{r}}_B)|\vec{\mathbf{p}}_A|^2
      - \vec{\mathbf{p}}_A \cdot (\vec{\mathbf{r}}_A 
      - \vec{\mathbf{r}}_B) (\vec{\mathbf{p}}_A\cdot \vec{\mathbf{p}}_B) }
      {|\vec{\mathbf{p}}_A|^2|\vec{\mathbf{p}}_B|^2 
      - (\vec{\mathbf{p}}_A\cdot \vec{\mathbf{p}}_B)^2}~. \nonumber \\
  \label{eq:CloseToIntersectingab}
\end{eqnarray}
Evaluating $\vec{\mathbf{R}}_A(a)$ and $\vec{\mathbf{R}}_B(b)$ at these values of $a$ and
$b$ and taking the midpoint between them, we obtain an estimate for the location of the 
corresponding vertex.

We emphasize that this vertexing procedure not only provides a way of pinpointing the
location of a vertex from the measured momenta of a pair of particles produced at that vertex, 
but can also be used in order to assess whether or not two particles in the event were in 
fact produced at the same vertex.  In cases in which the two particles were in fact 
produced at the same vertex, the magnitude of the vector $\vec{\mathbf{D}}(a,b)$, 
when evaluated at the values of $a$ and $b$ in Eq.~(\ref{eq:CloseToIntersectingab}), will be 
extremely small.  By contrast, if the particles were not in fact produced at the same vertex,
$|\vec{\mathbf{D}}(a,b)|$, when evaluated at the corresponding values of $a$ and $b$, typically 
will be far larger.  

The processes that we have considered in this paper yield up to ten jets emanating from up 
to five displaced vertices when both decay chains are included.  For the reasons discussed above, 
it is very unlikely that identifying unrelated pairs of jets as coming from the same vertex will 
result in small minimum values of $|D(a,b)|$.  Thus by considering different pairwise combinations 
of jets and evaluating their minimum values of $|D(a,b)|$, it should be relatively straightforward 
to correctly identify those that emanate from the same vertex.   We therefore expect 
the combinatorial background from misidentifications of jet pairs to be negligible.  

\bibliographystyle{apsrev4-2} 
\bibliography{references}

\begin{thebibliography}{77}%
\makeatletter
\providecommand \@ifxundefined [1]{%
 \@ifx{#1\undefined}
}%
\providecommand \@ifnum [1]{%
 \ifnum #1\expandafter \@firstoftwo
 \else \expandafter \@secondoftwo
 \fi
}%
\providecommand \@ifx [1]{%
 \ifx #1\expandafter \@firstoftwo
 \else \expandafter \@secondoftwo
 \fi
}%
\providecommand \natexlab [1]{#1}%
\providecommand \enquote  [1]{``#1''}%
\providecommand \bibnamefont  [1]{#1}%
\providecommand \bibfnamefont [1]{#1}%
\providecommand \citenamefont [1]{#1}%
\providecommand \href@noop [0]{\@secondoftwo}%
\providecommand \href [0]{\begingroup \@sanitize@url \@href}%
\providecommand \@href[1]{\@@startlink{#1}\@@href}%
\providecommand \@@href[1]{\endgroup#1\@@endlink}%
\providecommand \@sanitize@url [0]{\catcode `\\12\catcode `\$12\catcode
  `\&12\catcode `\#12\catcode `\^12\catcode `\_12\catcode `\%12\relax}%
\providecommand \@@startlink[1]{}%
\providecommand \@@endlink[0]{}%
\providecommand \url  [0]{\begingroup\@sanitize@url \@url }%
\providecommand \@url [1]{\endgroup\@href {#1}{\urlprefix }}%
\providecommand \urlprefix  [0]{URL }%
\providecommand \Eprint [0]{\href }%
\providecommand \doibase [0]{https://doi.org/}%
\providecommand \selectlanguage [0]{\@gobble}%
\providecommand \bibinfo  [0]{\@secondoftwo}%
\providecommand \bibfield  [0]{\@secondoftwo}%
\providecommand \translation [1]{[#1]}%
\providecommand \BibitemOpen [0]{}%
\providecommand \bibitemStop [0]{}%
\providecommand \bibitemNoStop [0]{.\EOS\space}%
\providecommand \EOS [0]{\spacefactor3000\relax}%
\providecommand \BibitemShut  [1]{\csname bibitem#1\endcsname}%
\let\auto@bib@innerbib\@empty
\bibitem [{\citenamefont {Alimena}\ \emph {et~al.}(2020)\citenamefont {Alimena}
  \emph {et~al.}}]{Alimena:2019zri}%
  \BibitemOpen
  \bibfield  {author} {\bibinfo {author} {\bibfnamefont {J.}~\bibnamefont
  {Alimena}} \emph {et~al.},\ }\href {https://doi.org/10.1088/1361-6471/ab4574}
  {\bibfield  {journal} {\bibinfo  {journal} {J. Phys. G}\ }\textbf {\bibinfo
  {volume} {47}},\ \bibinfo {pages} {090501} (\bibinfo {year} {2020})},\
  \Eprint {https://arxiv.org/abs/1903.04497} {arXiv:1903.04497 [hep-ex]}
  \BibitemShut {NoStop}%
\bibitem [{\citenamefont {Fischer}\ \emph {et~al.}(2021)\citenamefont {Fischer}
  \emph {et~al.}}]{Fischer:2021sqw}%
  \BibitemOpen
  \bibfield  {author} {\bibinfo {author} {\bibfnamefont {O.}~\bibnamefont
  {Fischer}} \emph {et~al.},\ }\href@noop {} {\  (\bibinfo {year} {2021})},\
  \Eprint {https://arxiv.org/abs/2109.06065} {arXiv:2109.06065 [hep-ph]}
  \BibitemShut {NoStop}%
\bibitem [{\citenamefont {Acosta}\ \emph {et~al.}(2021)\citenamefont {Acosta}
  \emph {et~al.}}]{Alimena:2021mdu}%
  \BibitemOpen
  \bibfield  {author} {\bibinfo {author} {\bibfnamefont {D.}~\bibnamefont
  {Acosta}} \emph {et~al.},\ }\href@noop {} {\  (\bibinfo {year} {2021})},\
  \Eprint {https://arxiv.org/abs/2110.14675} {arXiv:2110.14675 [hep-ex]}
  \BibitemShut {NoStop}%
\bibitem [{\citenamefont {Curtin}\ \emph {et~al.}(2019)\citenamefont {Curtin}
  \emph {et~al.}}]{Curtin:2018mvb}%
  \BibitemOpen
  \bibfield  {author} {\bibinfo {author} {\bibfnamefont {D.}~\bibnamefont
  {Curtin}} \emph {et~al.},\ }\href {https://doi.org/10.1088/1361-6633/ab28d6}
  {\bibfield  {journal} {\bibinfo  {journal} {Rept. Prog. Phys.}\ }\textbf
  {\bibinfo {volume} {82}},\ \bibinfo {pages} {116201} (\bibinfo {year}
  {2019})},\ \Eprint {https://arxiv.org/abs/1806.07396} {arXiv:1806.07396
  [hep-ph]} \BibitemShut {NoStop}%
\bibitem [{\citenamefont {Schwaller}\ \emph {et~al.}(2015)\citenamefont
  {Schwaller}, \citenamefont {Stolarski},\ and\ \citenamefont
  {Weiler}}]{Schwaller:2015gea}%
  \BibitemOpen
  \bibfield  {author} {\bibinfo {author} {\bibfnamefont {P.}~\bibnamefont
  {Schwaller}}, \bibinfo {author} {\bibfnamefont {D.}~\bibnamefont
  {Stolarski}},\ and\ \bibinfo {author} {\bibfnamefont {A.}~\bibnamefont
  {Weiler}},\ }\href {https://doi.org/10.1007/JHEP05(2015)059} {\bibfield
  {journal} {\bibinfo  {journal} {JHEP}\ }\textbf {\bibinfo {volume} {05}},\
  \bibinfo {pages} {059}},\ \Eprint {https://arxiv.org/abs/1502.05409}
  {arXiv:1502.05409 [hep-ph]} \BibitemShut {NoStop}%
\bibitem [{\citenamefont {Gray}\ and\ \citenamefont {Tabarelli~de
  Fatis}()}]{Gray:2017}%
  \BibitemOpen
  \bibfield  {author} {\bibinfo {author} {\bibfnamefont {L.}~\bibnamefont
  {Gray}}\ and\ \bibinfo {author} {\bibfnamefont {T.}~\bibnamefont
  {Tabarelli~de Fatis}},\ }\href@noop {} {\ }\bibinfo {note}
  {CERN-LHCC-2017-027, LHCC-P-009}\BibitemShut {NoStop}%
\bibitem [{\citenamefont {Butler}\ and\ \citenamefont {Tabarelli~de
  Fatis}()}]{Butler:2019rpu}%
  \BibitemOpen
  \bibfield  {author} {\bibinfo {author} {\bibfnamefont {J.~N.}\ \bibnamefont
  {Butler}}\ and\ \bibinfo {author} {\bibfnamefont {T.}~\bibnamefont
  {Tabarelli~de Fatis}} (\bibinfo {collaboration} {CMS}),\ }\href@noop {} {\
  }\bibinfo {note} {CERN-LHCC-2019-003, CMS-TDR-020}\BibitemShut {NoStop}%
\bibitem [{\citenamefont {Martin}(2007)}]{Martin:2007gf}%
  \BibitemOpen
  \bibfield  {author} {\bibinfo {author} {\bibfnamefont {S.~P.}\ \bibnamefont
  {Martin}},\ }\href {https://doi.org/10.1103/PhysRevD.75.115005} {\bibfield
  {journal} {\bibinfo  {journal} {Phys. Rev. D}\ }\textbf {\bibinfo {volume}
  {75}},\ \bibinfo {pages} {115005} (\bibinfo {year} {2007})},\ \Eprint
  {https://arxiv.org/abs/hep-ph/0703097} {arXiv:hep-ph/0703097} \BibitemShut
  {NoStop}%
\bibitem [{\citenamefont {Strassler}\ and\ \citenamefont
  {Zurek}(2007)}]{Strassler:2006im}%
  \BibitemOpen
  \bibfield  {author} {\bibinfo {author} {\bibfnamefont {M.~J.}\ \bibnamefont
  {Strassler}}\ and\ \bibinfo {author} {\bibfnamefont {K.~M.}\ \bibnamefont
  {Zurek}},\ }\href {https://doi.org/10.1016/j.physletb.2007.06.055} {\bibfield
   {journal} {\bibinfo  {journal} {Phys. Lett. B}\ }\textbf {\bibinfo {volume}
  {651}},\ \bibinfo {pages} {374} (\bibinfo {year} {2007})},\ \Eprint
  {https://arxiv.org/abs/hep-ph/0604261} {arXiv:hep-ph/0604261} \BibitemShut
  {NoStop}%
\bibitem [{\citenamefont {Cohen}\ \emph {et~al.}(2015)\citenamefont {Cohen},
  \citenamefont {Lisanti},\ and\ \citenamefont {Lou}}]{Cohen:2015toa}%
  \BibitemOpen
  \bibfield  {author} {\bibinfo {author} {\bibfnamefont {T.}~\bibnamefont
  {Cohen}}, \bibinfo {author} {\bibfnamefont {M.}~\bibnamefont {Lisanti}},\
  and\ \bibinfo {author} {\bibfnamefont {H.~K.}\ \bibnamefont {Lou}},\ }\href
  {https://doi.org/10.1103/PhysRevLett.115.171804} {\bibfield  {journal}
  {\bibinfo  {journal} {Phys. Rev. Lett.}\ }\textbf {\bibinfo {volume} {115}},\
  \bibinfo {pages} {171804} (\bibinfo {year} {2015})},\ \Eprint
  {https://arxiv.org/abs/1503.00009} {arXiv:1503.00009 [hep-ph]} \BibitemShut
  {NoStop}%
\bibitem [{\citenamefont {Park}\ and\ \citenamefont
  {Zhang}(2019)}]{Park:2017rfb}%
  \BibitemOpen
  \bibfield  {author} {\bibinfo {author} {\bibfnamefont {M.}~\bibnamefont
  {Park}}\ and\ \bibinfo {author} {\bibfnamefont {M.}~\bibnamefont {Zhang}},\
  }\href {https://doi.org/10.1103/PhysRevD.100.115009} {\bibfield  {journal}
  {\bibinfo  {journal} {Phys. Rev. D}\ }\textbf {\bibinfo {volume} {100}},\
  \bibinfo {pages} {115009} (\bibinfo {year} {2019})},\ \Eprint
  {https://arxiv.org/abs/1712.09279} {arXiv:1712.09279 [hep-ph]} \BibitemShut
  {NoStop}%
\bibitem [{\citenamefont {Knapen}\ \emph {et~al.}(2017)\citenamefont {Knapen},
  \citenamefont {Pagan~Griso}, \citenamefont {Papucci},\ and\ \citenamefont
  {Robinson}}]{Knapen:2016hky}%
  \BibitemOpen
  \bibfield  {author} {\bibinfo {author} {\bibfnamefont {S.}~\bibnamefont
  {Knapen}}, \bibinfo {author} {\bibfnamefont {S.}~\bibnamefont {Pagan~Griso}},
  \bibinfo {author} {\bibfnamefont {M.}~\bibnamefont {Papucci}},\ and\ \bibinfo
  {author} {\bibfnamefont {D.~J.}\ \bibnamefont {Robinson}},\ }\href
  {https://doi.org/10.1007/JHEP08(2017)076} {\bibfield  {journal} {\bibinfo
  {journal} {JHEP}\ }\textbf {\bibinfo {volume} {08}},\ \bibinfo {pages}
  {076}},\ \Eprint {https://arxiv.org/abs/1612.00850} {arXiv:1612.00850
  [hep-ph]} \BibitemShut {NoStop}%
\bibitem [{\citenamefont {D'Agnolo}\ and\ \citenamefont
  {Low}(2019)}]{DAgnolo:2019cio}%
  \BibitemOpen
  \bibfield  {author} {\bibinfo {author} {\bibfnamefont {R.~T.}\ \bibnamefont
  {D'Agnolo}}\ and\ \bibinfo {author} {\bibfnamefont {M.}~\bibnamefont {Low}},\
  }\href {https://doi.org/10.1007/JHEP08(2019)163} {\bibfield  {journal}
  {\bibinfo  {journal} {JHEP}\ }\textbf {\bibinfo {volume} {08}},\ \bibinfo
  {pages} {163}},\ \Eprint {https://arxiv.org/abs/1902.05535} {arXiv:1902.05535
  [hep-ph]} \BibitemShut {NoStop}%
\bibitem [{\citenamefont {Dienes}\ \emph {et~al.}(2020)\citenamefont {Dienes},
  \citenamefont {Kim}, \citenamefont {Song}, \citenamefont {Su}, \citenamefont
  {Thomas},\ and\ \citenamefont {Yaylali}}]{Dienes:2019krh}%
  \BibitemOpen
  \bibfield  {author} {\bibinfo {author} {\bibfnamefont {K.~R.}\ \bibnamefont
  {Dienes}}, \bibinfo {author} {\bibfnamefont {D.}~\bibnamefont {Kim}},
  \bibinfo {author} {\bibfnamefont {H.}~\bibnamefont {Song}}, \bibinfo {author}
  {\bibfnamefont {S.}~\bibnamefont {Su}}, \bibinfo {author} {\bibfnamefont
  {B.}~\bibnamefont {Thomas}},\ and\ \bibinfo {author} {\bibfnamefont
  {D.}~\bibnamefont {Yaylali}},\ }\href
  {https://doi.org/10.1103/PhysRevD.101.075024} {\bibfield  {journal} {\bibinfo
   {journal} {Phys. Rev. D}\ }\textbf {\bibinfo {volume} {101}},\ \bibinfo
  {pages} {075024} (\bibinfo {year} {2020})},\ \Eprint
  {https://arxiv.org/abs/1910.01129} {arXiv:1910.01129 [hep-ph]} \BibitemShut
  {NoStop}%
\bibitem [{\citenamefont {Giromini}\ \emph {et~al.}(2008)\citenamefont
  {Giromini}, \citenamefont {Happacher}, \citenamefont {Kim}, \citenamefont
  {Kruse}, \citenamefont {Pitts}, \citenamefont {Ptohos},\ and\ \citenamefont
  {Torre}}]{Giromini:2008xh}%
  \BibitemOpen
  \bibfield  {author} {\bibinfo {author} {\bibfnamefont {P.}~\bibnamefont
  {Giromini}}, \bibinfo {author} {\bibfnamefont {F.}~\bibnamefont {Happacher}},
  \bibinfo {author} {\bibfnamefont {M.~J.}\ \bibnamefont {Kim}}, \bibinfo
  {author} {\bibfnamefont {M.}~\bibnamefont {Kruse}}, \bibinfo {author}
  {\bibfnamefont {K.}~\bibnamefont {Pitts}}, \bibinfo {author} {\bibfnamefont
  {F.}~\bibnamefont {Ptohos}},\ and\ \bibinfo {author} {\bibfnamefont
  {S.}~\bibnamefont {Torre}},\ }\href@noop {} {\  (\bibinfo {year} {2008})},\
  \Eprint {https://arxiv.org/abs/0810.5730} {arXiv:0810.5730 [hep-ph]}
  \BibitemShut {NoStop}%
\bibitem [{\citenamefont {Strassler}(2008)}]{Strassler:2008jq}%
  \BibitemOpen
  \bibfield  {author} {\bibinfo {author} {\bibfnamefont {M.~J.}\ \bibnamefont
  {Strassler}},\ }\href@noop {} {\  (\bibinfo {year} {2008})},\ \Eprint
  {https://arxiv.org/abs/0811.1560} {arXiv:0811.1560 [hep-ph]} \BibitemShut
  {NoStop}%
\bibitem [{\citenamefont {Strassler}(2006)}]{Strassler:2006qa}%
  \BibitemOpen
  \bibfield  {author} {\bibinfo {author} {\bibfnamefont {M.~J.}\ \bibnamefont
  {Strassler}},\ }\href@noop {} {\  (\bibinfo {year} {2006})},\ \Eprint
  {https://arxiv.org/abs/hep-ph/0607160} {arXiv:hep-ph/0607160} \BibitemShut
  {NoStop}%
\bibitem [{\citenamefont {Juknevich}(2010)}]{Juknevich:2009gg}%
  \BibitemOpen
  \bibfield  {author} {\bibinfo {author} {\bibfnamefont {J.~E.}\ \bibnamefont
  {Juknevich}},\ }\href {https://doi.org/10.1007/JHEP08(2010)121} {\bibfield
  {journal} {\bibinfo  {journal} {JHEP}\ }\textbf {\bibinfo {volume} {08}},\
  \bibinfo {pages} {121}},\ \Eprint {https://arxiv.org/abs/0911.5616}
  {arXiv:0911.5616 [hep-ph]} \BibitemShut {NoStop}%
\bibitem [{\citenamefont {Juknevich}\ \emph {et~al.}(2009)\citenamefont
  {Juknevich}, \citenamefont {Melnikov},\ and\ \citenamefont
  {Strassler}}]{Juknevich:2009ji}%
  \BibitemOpen
  \bibfield  {author} {\bibinfo {author} {\bibfnamefont {J.~E.}\ \bibnamefont
  {Juknevich}}, \bibinfo {author} {\bibfnamefont {D.}~\bibnamefont
  {Melnikov}},\ and\ \bibinfo {author} {\bibfnamefont {M.~J.}\ \bibnamefont
  {Strassler}},\ }\href {https://doi.org/10.1088/1126-6708/2009/07/055}
  {\bibfield  {journal} {\bibinfo  {journal} {JHEP}\ }\textbf {\bibinfo
  {volume} {07}},\ \bibinfo {pages} {055}},\ \Eprint
  {https://arxiv.org/abs/0903.0883} {arXiv:0903.0883 [hep-ph]} \BibitemShut
  {NoStop}%
\bibitem [{\citenamefont {Craig}\ \emph {et~al.}(2015)\citenamefont {Craig},
  \citenamefont {Katz}, \citenamefont {Strassler},\ and\ \citenamefont
  {Sundrum}}]{Craig:2015pha}%
  \BibitemOpen
  \bibfield  {author} {\bibinfo {author} {\bibfnamefont {N.}~\bibnamefont
  {Craig}}, \bibinfo {author} {\bibfnamefont {A.}~\bibnamefont {Katz}},
  \bibinfo {author} {\bibfnamefont {M.}~\bibnamefont {Strassler}},\ and\
  \bibinfo {author} {\bibfnamefont {R.}~\bibnamefont {Sundrum}},\ }\href
  {https://doi.org/10.1007/JHEP07(2015)105} {\bibfield  {journal} {\bibinfo
  {journal} {JHEP}\ }\textbf {\bibinfo {volume} {07}},\ \bibinfo {pages}
  {105}},\ \Eprint {https://arxiv.org/abs/1501.05310} {arXiv:1501.05310
  [hep-ph]} \BibitemShut {NoStop}%
\bibitem [{\citenamefont {Lanni}\ \emph {et~al.}()\citenamefont {Lanni},
  \citenamefont {Pontecorvo} \emph {et~al.}}]{Lanni:2020}%
  \BibitemOpen
  \bibfield  {author} {\bibinfo {author} {\bibfnamefont {F.}~\bibnamefont
  {Lanni}}, \bibinfo {author} {\bibfnamefont {L.}~\bibnamefont {Pontecorvo}},
  \emph {et~al.} (\bibinfo {collaboration} {ATLAS}),\ }\href@noop {} {\
  }\bibinfo {note} {CERN-LHCC-2020-007, ATLAS-TDR-031}\BibitemShut {NoStop}%
\bibitem [{\citenamefont {del Re}(2015)}]{delRe:2015hla}%
  \BibitemOpen
  \bibfield  {author} {\bibinfo {author} {\bibfnamefont {D.}~\bibnamefont {del
  Re}},\ }\href {https://doi.org/10.1088/1742-6596/587/1/012003} {\bibfield
  {journal} {\bibinfo  {journal} {J. Phys. Conf. Ser.}\ }\textbf {\bibinfo
  {volume} {587}},\ \bibinfo {pages} {012003} (\bibinfo {year}
  {2015})}\BibitemShut {NoStop}%
\bibitem [{\citenamefont {Liu}\ \emph {et~al.}(2019)\citenamefont {Liu},
  \citenamefont {Liu},\ and\ \citenamefont {Wang}}]{Liu:2018wte}%
  \BibitemOpen
  \bibfield  {author} {\bibinfo {author} {\bibfnamefont {J.}~\bibnamefont
  {Liu}}, \bibinfo {author} {\bibfnamefont {Z.}~\bibnamefont {Liu}},\ and\
  \bibinfo {author} {\bibfnamefont {L.-T.}\ \bibnamefont {Wang}},\ }\href
  {https://doi.org/10.1103/PhysRevLett.122.131801} {\bibfield  {journal}
  {\bibinfo  {journal} {Phys. Rev. Lett.}\ }\textbf {\bibinfo {volume} {122}},\
  \bibinfo {pages} {131801} (\bibinfo {year} {2019})},\ \Eprint
  {https://arxiv.org/abs/1805.05957} {arXiv:1805.05957 [hep-ph]} \BibitemShut
  {NoStop}%
\bibitem [{\citenamefont {Liu}\ \emph {et~al.}(2020)\citenamefont {Liu},
  \citenamefont {Liu}, \citenamefont {Wang},\ and\ \citenamefont
  {Wang}}]{Liu:2020vur}%
  \BibitemOpen
  \bibfield  {author} {\bibinfo {author} {\bibfnamefont {J.}~\bibnamefont
  {Liu}}, \bibinfo {author} {\bibfnamefont {Z.}~\bibnamefont {Liu}}, \bibinfo
  {author} {\bibfnamefont {L.-T.}\ \bibnamefont {Wang}},\ and\ \bibinfo
  {author} {\bibfnamefont {X.-P.}\ \bibnamefont {Wang}},\ }\href
  {https://doi.org/10.1007/JHEP11(2020)066} {\bibfield  {journal} {\bibinfo
  {journal} {JHEP}\ }\textbf {\bibinfo {volume} {11}},\ \bibinfo {pages}
  {066}},\ \Eprint {https://arxiv.org/abs/2005.10836} {arXiv:2005.10836
  [hep-ph]} \BibitemShut {NoStop}%
\bibitem [{\citenamefont {Flowers}\ \emph {et~al.}(2020)\citenamefont
  {Flowers}, \citenamefont {Meier}, \citenamefont {Rogan}, \citenamefont
  {Kang},\ and\ \citenamefont {Park}}]{Kang:2019ukr}%
  \BibitemOpen
  \bibfield  {author} {\bibinfo {author} {\bibfnamefont {Z.}~\bibnamefont
  {Flowers}}, \bibinfo {author} {\bibfnamefont {Q.}~\bibnamefont {Meier}},
  \bibinfo {author} {\bibfnamefont {C.}~\bibnamefont {Rogan}}, \bibinfo
  {author} {\bibfnamefont {D.~W.}\ \bibnamefont {Kang}},\ and\ \bibinfo
  {author} {\bibfnamefont {S.~C.}\ \bibnamefont {Park}},\ }\href
  {https://doi.org/10.1007/JHEP03(2020)132} {\bibfield  {journal} {\bibinfo
  {journal} {JHEP}\ }\textbf {\bibinfo {volume} {03}},\ \bibinfo {pages}
  {132}},\ \Eprint {https://arxiv.org/abs/1903.05825} {arXiv:1903.05825
  [hep-ph]} \BibitemShut {NoStop}%
\bibitem [{\citenamefont {Cottin}(2018)}]{Cottin:2018hyf}%
  \BibitemOpen
  \bibfield  {author} {\bibinfo {author} {\bibfnamefont {G.}~\bibnamefont
  {Cottin}},\ }\href {https://doi.org/10.1007/JHEP03(2018)137} {\bibfield
  {journal} {\bibinfo  {journal} {JHEP}\ }\textbf {\bibinfo {volume} {03}},\
  \bibinfo {pages} {137}},\ \Eprint {https://arxiv.org/abs/1801.09671}
  {arXiv:1801.09671 [hep-ph]} \BibitemShut {NoStop}%
\bibitem [{\citenamefont {Bae}\ \emph {et~al.}(2020)\citenamefont {Bae},
  \citenamefont {Park},\ and\ \citenamefont {Zhang}}]{Bae:2020dwf}%
  \BibitemOpen
  \bibfield  {author} {\bibinfo {author} {\bibfnamefont {K.~J.}\ \bibnamefont
  {Bae}}, \bibinfo {author} {\bibfnamefont {M.}~\bibnamefont {Park}},\ and\
  \bibinfo {author} {\bibfnamefont {M.}~\bibnamefont {Zhang}},\ }\href
  {https://doi.org/10.1103/PhysRevD.101.115036} {\bibfield  {journal} {\bibinfo
   {journal} {Phys. Rev. D}\ }\textbf {\bibinfo {volume} {101}},\ \bibinfo
  {pages} {115036} (\bibinfo {year} {2020})},\ \Eprint
  {https://arxiv.org/abs/2001.02142} {arXiv:2001.02142 [hep-ph]} \BibitemShut
  {NoStop}%
\bibitem [{\citenamefont {Sirunyan}\ \emph
  {et~al.}(2019{\natexlab{a}})\citenamefont {Sirunyan} \emph
  {et~al.}}]{Sirunyan:2019gut}%
  \BibitemOpen
  \bibfield  {author} {\bibinfo {author} {\bibfnamefont {A.~M.}\ \bibnamefont
  {Sirunyan}} \emph {et~al.} (\bibinfo {collaboration} {CMS}),\ }\href
  {https://doi.org/10.1016/j.physletb.2019.134876} {\bibfield  {journal}
  {\bibinfo  {journal} {Phys. Lett. B}\ }\textbf {\bibinfo {volume} {797}},\
  \bibinfo {pages} {134876} (\bibinfo {year} {2019}{\natexlab{a}})},\ \Eprint
  {https://arxiv.org/abs/1906.06441} {arXiv:1906.06441 [hep-ex]} \BibitemShut
  {NoStop}%
\bibitem [{\citenamefont {Sirunyan}\ \emph
  {et~al.}(2020{\natexlab{a}})\citenamefont {Sirunyan} \emph
  {et~al.}}]{Sirunyan:2020cao}%
  \BibitemOpen
  \bibfield  {author} {\bibinfo {author} {\bibfnamefont {A.~M.}\ \bibnamefont
  {Sirunyan}} \emph {et~al.} (\bibinfo {collaboration} {CMS}),\ }\href@noop {}
  {\  (\bibinfo {year} {2020}{\natexlab{a}})},\ \Eprint
  {https://arxiv.org/abs/2012.01581} {arXiv:2012.01581 [hep-ex]} \BibitemShut
  {NoStop}%
\bibitem [{\citenamefont {Sirunyan}\ \emph {et~al.}(2021)\citenamefont
  {Sirunyan} \emph {et~al.}}]{Sirunyan:2021kty}%
  \BibitemOpen
  \bibfield  {author} {\bibinfo {author} {\bibfnamefont {A.~M.}\ \bibnamefont
  {Sirunyan}} \emph {et~al.} (\bibinfo {collaboration} {CMS}),\ }\href@noop {}
  {\  (\bibinfo {year} {2021})},\ \Eprint {https://arxiv.org/abs/2104.13474}
  {arXiv:2104.13474 [hep-ex]} \BibitemShut {NoStop}%
\bibitem [{\citenamefont {Sirunyan}\ \emph
  {et~al.}(2019{\natexlab{b}})\citenamefont {Sirunyan} \emph
  {et~al.}}]{Sirunyan:2018vlw}%
  \BibitemOpen
  \bibfield  {author} {\bibinfo {author} {\bibfnamefont {A.~M.}\ \bibnamefont
  {Sirunyan}} \emph {et~al.} (\bibinfo {collaboration} {CMS}),\ }\href
  {https://doi.org/10.1103/PhysRevD.99.032011} {\bibfield  {journal} {\bibinfo
  {journal} {Phys. Rev. D}\ }\textbf {\bibinfo {volume} {99}},\ \bibinfo
  {pages} {032011} (\bibinfo {year} {2019}{\natexlab{b}})},\ \Eprint
  {https://arxiv.org/abs/1811.07991} {arXiv:1811.07991 [hep-ex]} \BibitemShut
  {NoStop}%
\bibitem [{\citenamefont {Sirunyan}\ \emph
  {et~al.}(2018{\natexlab{a}})\citenamefont {Sirunyan} \emph
  {et~al.}}]{Sirunyan:2018pwn}%
  \BibitemOpen
  \bibfield  {author} {\bibinfo {author} {\bibfnamefont {A.~M.}\ \bibnamefont
  {Sirunyan}} \emph {et~al.} (\bibinfo {collaboration} {CMS}),\ }\href
  {https://doi.org/10.1103/PhysRevD.98.092011} {\bibfield  {journal} {\bibinfo
  {journal} {Phys. Rev. D}\ }\textbf {\bibinfo {volume} {98}},\ \bibinfo
  {pages} {092011} (\bibinfo {year} {2018}{\natexlab{a}})},\ \Eprint
  {https://arxiv.org/abs/1808.03078} {arXiv:1808.03078 [hep-ex]} \BibitemShut
  {NoStop}%
\bibitem [{\citenamefont {Sirunyan}\ \emph
  {et~al.}(2020{\natexlab{b}})\citenamefont {Sirunyan} \emph
  {et~al.}}]{Sirunyan:2019nfw}%
  \BibitemOpen
  \bibfield  {author} {\bibinfo {author} {\bibfnamefont {A.~M.}\ \bibnamefont
  {Sirunyan}} \emph {et~al.} (\bibinfo {collaboration} {CMS}),\ }\href
  {https://doi.org/10.1088/2632-2153/ab9023} {\bibfield  {journal} {\bibinfo
  {journal} {Mach. Learn. Sci. Tech.}\ }\textbf {\bibinfo {volume} {1}},\
  \bibinfo {pages} {035012} (\bibinfo {year} {2020}{\natexlab{b}})},\ \Eprint
  {https://arxiv.org/abs/1912.12238} {arXiv:1912.12238 [hep-ex]} \BibitemShut
  {NoStop}%
\bibitem [{\citenamefont {Aaboud}\ \emph
  {et~al.}(2018{\natexlab{a}})\citenamefont {Aaboud} \emph
  {et~al.}}]{Aaboud:2017iio}%
  \BibitemOpen
  \bibfield  {author} {\bibinfo {author} {\bibfnamefont {M.}~\bibnamefont
  {Aaboud}} \emph {et~al.} (\bibinfo {collaboration} {ATLAS}),\ }\href
  {https://doi.org/10.1103/PhysRevD.97.052012} {\bibfield  {journal} {\bibinfo
  {journal} {Phys. Rev. D}\ }\textbf {\bibinfo {volume} {97}},\ \bibinfo
  {pages} {052012} (\bibinfo {year} {2018}{\natexlab{a}})},\ \Eprint
  {https://arxiv.org/abs/1710.04901} {arXiv:1710.04901 [hep-ex]} \BibitemShut
  {NoStop}%
\bibitem [{\citenamefont {Aaboud}\ \emph
  {et~al.}(2019{\natexlab{a}})\citenamefont {Aaboud} \emph
  {et~al.}}]{Aaboud:2019opc}%
  \BibitemOpen
  \bibfield  {author} {\bibinfo {author} {\bibfnamefont {M.}~\bibnamefont
  {Aaboud}} \emph {et~al.} (\bibinfo {collaboration} {ATLAS}),\ }\href
  {https://doi.org/10.1140/epjc/s10052-019-6962-6} {\bibfield  {journal}
  {\bibinfo  {journal} {Eur. Phys. J. C}\ }\textbf {\bibinfo {volume} {79}},\
  \bibinfo {pages} {481} (\bibinfo {year} {2019}{\natexlab{a}})},\ \Eprint
  {https://arxiv.org/abs/1902.03094} {arXiv:1902.03094 [hep-ex]} \BibitemShut
  {NoStop}%
\bibitem [{\citenamefont {Aaboud}\ \emph
  {et~al.}(2019{\natexlab{b}})\citenamefont {Aaboud} \emph
  {et~al.}}]{Aaboud:2018aqj}%
  \BibitemOpen
  \bibfield  {author} {\bibinfo {author} {\bibfnamefont {M.}~\bibnamefont
  {Aaboud}} \emph {et~al.} (\bibinfo {collaboration} {ATLAS}),\ }\href
  {https://doi.org/10.1103/PhysRevD.99.052005} {\bibfield  {journal} {\bibinfo
  {journal} {Phys. Rev. D}\ }\textbf {\bibinfo {volume} {99}},\ \bibinfo
  {pages} {052005} (\bibinfo {year} {2019}{\natexlab{b}})},\ \Eprint
  {https://arxiv.org/abs/1811.07370} {arXiv:1811.07370 [hep-ex]} \BibitemShut
  {NoStop}%
\bibitem [{\citenamefont {Aad}\ \emph {et~al.}(2020{\natexlab{a}})\citenamefont
  {Aad} \emph {et~al.}}]{Aad:2019xav}%
  \BibitemOpen
  \bibfield  {author} {\bibinfo {author} {\bibfnamefont {G.}~\bibnamefont
  {Aad}} \emph {et~al.} (\bibinfo {collaboration} {ATLAS}),\ }\href
  {https://doi.org/10.1103/PhysRevD.101.052013} {\bibfield  {journal} {\bibinfo
   {journal} {Phys. Rev. D}\ }\textbf {\bibinfo {volume} {101}},\ \bibinfo
  {pages} {052013} (\bibinfo {year} {2020}{\natexlab{a}})},\ \Eprint
  {https://arxiv.org/abs/1911.12575} {arXiv:1911.12575 [hep-ex]} \BibitemShut
  {NoStop}%
\bibitem [{\citenamefont {Sirunyan}\ \emph
  {et~al.}(2019{\natexlab{c}})\citenamefont {Sirunyan} \emph
  {et~al.}}]{Sirunyan:2019ctn}%
  \BibitemOpen
  \bibfield  {author} {\bibinfo {author} {\bibfnamefont {A.~M.}\ \bibnamefont
  {Sirunyan}} \emph {et~al.} (\bibinfo {collaboration} {CMS}),\ }\href
  {https://doi.org/10.1007/JHEP10(2019)244} {\bibfield  {journal} {\bibinfo
  {journal} {JHEP}\ }\textbf {\bibinfo {volume} {10}},\ \bibinfo {pages}
  {244}},\ \Eprint {https://arxiv.org/abs/1908.04722} {arXiv:1908.04722
  [hep-ex]} \BibitemShut {NoStop}%
\bibitem [{\citenamefont {Sirunyan}\ \emph
  {et~al.}(2020{\natexlab{c}})\citenamefont {Sirunyan} \emph
  {et~al.}}]{Sirunyan:2019xwh}%
  \BibitemOpen
  \bibfield  {author} {\bibinfo {author} {\bibfnamefont {A.~M.}\ \bibnamefont
  {Sirunyan}} \emph {et~al.} (\bibinfo {collaboration} {CMS}),\ }\href
  {https://doi.org/10.1140/epjc/s10052-019-7493-x} {\bibfield  {journal}
  {\bibinfo  {journal} {Eur. Phys. J. C}\ }\textbf {\bibinfo {volume} {80}},\
  \bibinfo {pages} {3} (\bibinfo {year} {2020}{\natexlab{c}})},\ \Eprint
  {https://arxiv.org/abs/1909.03460} {arXiv:1909.03460 [hep-ex]} \BibitemShut
  {NoStop}%
\bibitem [{\citenamefont {Sirunyan}\ \emph
  {et~al.}(2018{\natexlab{b}})\citenamefont {Sirunyan} \emph
  {et~al.}}]{Sirunyan:2018vjp}%
  \BibitemOpen
  \bibfield  {author} {\bibinfo {author} {\bibfnamefont {A.~M.}\ \bibnamefont
  {Sirunyan}} \emph {et~al.} (\bibinfo {collaboration} {CMS}),\ }\href
  {https://doi.org/10.1007/JHEP05(2018)025} {\bibfield  {journal} {\bibinfo
  {journal} {JHEP}\ }\textbf {\bibinfo {volume} {05}},\ \bibinfo {pages}
  {025}},\ \Eprint {https://arxiv.org/abs/1802.02110} {arXiv:1802.02110
  [hep-ex]} \BibitemShut {NoStop}%
\bibitem [{\citenamefont {Aad}\ \emph {et~al.}(2020{\natexlab{b}})\citenamefont
  {Aad} \emph {et~al.}}]{Aad:2020aze}%
  \BibitemOpen
  \bibfield  {author} {\bibinfo {author} {\bibfnamefont {G.}~\bibnamefont
  {Aad}} \emph {et~al.} (\bibinfo {collaboration} {ATLAS}),\ }\href
  {https://doi.org/10.22323/1.364.0605} {\bibfield  {journal} {\bibinfo
  {journal} {PoS}\ }\textbf {\bibinfo {volume} {EPS-HEP2019}},\ \bibinfo
  {pages} {605} (\bibinfo {year} {2020}{\natexlab{b}})},\ \Eprint
  {https://arxiv.org/abs/2010.14293} {arXiv:2010.14293 [hep-ex]} \BibitemShut
  {NoStop}%
\bibitem [{\citenamefont {Aaboud}\ \emph
  {et~al.}(2018{\natexlab{b}})\citenamefont {Aaboud} \emph
  {et~al.}}]{Aaboud:2017vwy}%
  \BibitemOpen
  \bibfield  {author} {\bibinfo {author} {\bibfnamefont {M.}~\bibnamefont
  {Aaboud}} \emph {et~al.} (\bibinfo {collaboration} {ATLAS}),\ }\href
  {https://doi.org/10.1103/PhysRevD.97.112001} {\bibfield  {journal} {\bibinfo
  {journal} {Phys. Rev. D}\ }\textbf {\bibinfo {volume} {97}},\ \bibinfo
  {pages} {112001} (\bibinfo {year} {2018}{\natexlab{b}})},\ \Eprint
  {https://arxiv.org/abs/1712.02332} {arXiv:1712.02332 [hep-ex]} \BibitemShut
  {NoStop}%
\bibitem [{\citenamefont {Aad}\ \emph {et~al.}(2021)\citenamefont {Aad} \emph
  {et~al.}}]{Aad:2021egl}%
  \BibitemOpen
  \bibfield  {author} {\bibinfo {author} {\bibfnamefont {G.}~\bibnamefont
  {Aad}} \emph {et~al.} (\bibinfo {collaboration} {ATLAS}),\ }\href
  {https://doi.org/10.1103/PhysRevD.103.112006} {\bibfield  {journal} {\bibinfo
   {journal} {Phys. Rev. D}\ }\textbf {\bibinfo {volume} {103}},\ \bibinfo
  {pages} {112006} (\bibinfo {year} {2021})},\ \Eprint
  {https://arxiv.org/abs/2102.10874} {arXiv:2102.10874 [hep-ex]} \BibitemShut
  {NoStop}%
\bibitem [{\citenamefont {Aaboud}\ \emph
  {et~al.}(2018{\natexlab{c}})\citenamefont {Aaboud} \emph
  {et~al.}}]{Aaboud:2017phn}%
  \BibitemOpen
  \bibfield  {author} {\bibinfo {author} {\bibfnamefont {M.}~\bibnamefont
  {Aaboud}} \emph {et~al.} (\bibinfo {collaboration} {ATLAS}),\ }\href
  {https://doi.org/10.1007/JHEP01(2018)126} {\bibfield  {journal} {\bibinfo
  {journal} {JHEP}\ }\textbf {\bibinfo {volume} {01}},\ \bibinfo {pages}
  {126}},\ \Eprint {https://arxiv.org/abs/1711.03301} {arXiv:1711.03301
  [hep-ex]} \BibitemShut {NoStop}%
\bibitem [{\citenamefont {Alwall}\ \emph {et~al.}(2014)\citenamefont {Alwall},
  \citenamefont {Frederix}, \citenamefont {Frixione}, \citenamefont {Hirschi},
  \citenamefont {Maltoni}, \citenamefont {Mattelaer}, \citenamefont {Shao},
  \citenamefont {Stelzer}, \citenamefont {Torrielli},\ and\ \citenamefont
  {Zaro}}]{Alwall:2014hca}%
  \BibitemOpen
  \bibfield  {author} {\bibinfo {author} {\bibfnamefont {J.}~\bibnamefont
  {Alwall}}, \bibinfo {author} {\bibfnamefont {R.}~\bibnamefont {Frederix}},
  \bibinfo {author} {\bibfnamefont {S.}~\bibnamefont {Frixione}}, \bibinfo
  {author} {\bibfnamefont {V.}~\bibnamefont {Hirschi}}, \bibinfo {author}
  {\bibfnamefont {F.}~\bibnamefont {Maltoni}}, \bibinfo {author} {\bibfnamefont
  {O.}~\bibnamefont {Mattelaer}}, \bibinfo {author} {\bibfnamefont {H.~S.}\
  \bibnamefont {Shao}}, \bibinfo {author} {\bibfnamefont {T.}~\bibnamefont
  {Stelzer}}, \bibinfo {author} {\bibfnamefont {P.}~\bibnamefont {Torrielli}},\
  and\ \bibinfo {author} {\bibfnamefont {M.}~\bibnamefont {Zaro}},\ }\href
  {https://doi.org/10.1007/JHEP07(2014)079} {\bibfield  {journal} {\bibinfo
  {journal} {JHEP}\ }\textbf {\bibinfo {volume} {07}},\ \bibinfo {pages}
  {079}},\ \Eprint {https://arxiv.org/abs/1405.0301} {arXiv:1405.0301 [hep-ph]}
  \BibitemShut {NoStop}%
\bibitem [{\citenamefont {Conte}\ \emph {et~al.}(2013)\citenamefont {Conte},
  \citenamefont {Fuks},\ and\ \citenamefont {Serret}}]{Conte:2012fm}%
  \BibitemOpen
  \bibfield  {author} {\bibinfo {author} {\bibfnamefont {E.}~\bibnamefont
  {Conte}}, \bibinfo {author} {\bibfnamefont {B.}~\bibnamefont {Fuks}},\ and\
  \bibinfo {author} {\bibfnamefont {G.}~\bibnamefont {Serret}},\ }\href
  {https://doi.org/10.1016/j.cpc.2012.09.009} {\bibfield  {journal} {\bibinfo
  {journal} {Comput. Phys. Commun.}\ }\textbf {\bibinfo {volume} {184}},\
  \bibinfo {pages} {222} (\bibinfo {year} {2013})},\ \Eprint
  {https://arxiv.org/abs/1206.1599} {arXiv:1206.1599 [hep-ph]} \BibitemShut
  {NoStop}%
\bibitem [{\citenamefont {Araz}\ \emph {et~al.}(2022)\citenamefont {Araz},
  \citenamefont {Fuks}, \citenamefont {Goodsell},\ and\ \citenamefont
  {Utsch}}]{Araz:2021akd}%
  \BibitemOpen
  \bibfield  {author} {\bibinfo {author} {\bibfnamefont {J.~Y.}\ \bibnamefont
  {Araz}}, \bibinfo {author} {\bibfnamefont {B.}~\bibnamefont {Fuks}}, \bibinfo
  {author} {\bibfnamefont {M.~D.}\ \bibnamefont {Goodsell}},\ and\ \bibinfo
  {author} {\bibfnamefont {M.}~\bibnamefont {Utsch}},\ }\href
  {https://doi.org/10.1140/epjc/s10052-022-10511-w} {\bibfield  {journal}
  {\bibinfo  {journal} {Eur. Phys. J. C}\ }\textbf {\bibinfo {volume} {82}},\
  \bibinfo {pages} {597} (\bibinfo {year} {2022})},\ \Eprint
  {https://arxiv.org/abs/2112.05163} {arXiv:2112.05163 [hep-ph]} \BibitemShut
  {NoStop}%
\bibitem [{\citenamefont {Cho}\ \emph {et~al.}(2008)\citenamefont {Cho},
  \citenamefont {Choi}, \citenamefont {Kim},\ and\ \citenamefont
  {Park}}]{Cho:2007qv}%
  \BibitemOpen
  \bibfield  {author} {\bibinfo {author} {\bibfnamefont {W.~S.}\ \bibnamefont
  {Cho}}, \bibinfo {author} {\bibfnamefont {K.}~\bibnamefont {Choi}}, \bibinfo
  {author} {\bibfnamefont {Y.~G.}\ \bibnamefont {Kim}},\ and\ \bibinfo {author}
  {\bibfnamefont {C.~B.}\ \bibnamefont {Park}},\ }\href
  {https://doi.org/10.1103/PhysRevLett.100.171801} {\bibfield  {journal}
  {\bibinfo  {journal} {Phys. Rev. Lett.}\ }\textbf {\bibinfo {volume} {100}},\
  \bibinfo {pages} {171801} (\bibinfo {year} {2008})},\ \Eprint
  {https://arxiv.org/abs/0709.0288} {arXiv:0709.0288 [hep-ph]} \BibitemShut
  {NoStop}%
\bibitem [{\citenamefont {Han}\ \emph {et~al.}(2010)\citenamefont {Han},
  \citenamefont {Kim},\ and\ \citenamefont {Song}}]{Han:2009ss}%
  \BibitemOpen
  \bibfield  {author} {\bibinfo {author} {\bibfnamefont {T.}~\bibnamefont
  {Han}}, \bibinfo {author} {\bibfnamefont {I.-W.}\ \bibnamefont {Kim}},\ and\
  \bibinfo {author} {\bibfnamefont {J.}~\bibnamefont {Song}},\ }\href
  {https://doi.org/10.1016/j.physletb.2010.09.010} {\bibfield  {journal}
  {\bibinfo  {journal} {Phys. Lett. B}\ }\textbf {\bibinfo {volume} {693}},\
  \bibinfo {pages} {575} (\bibinfo {year} {2010})},\ \Eprint
  {https://arxiv.org/abs/0906.5009} {arXiv:0906.5009 [hep-ph]} \BibitemShut
  {NoStop}%
\bibitem [{\citenamefont {Agashe}\ \emph {et~al.}(2010)\citenamefont {Agashe},
  \citenamefont {Kim}, \citenamefont {Toharia},\ and\ \citenamefont
  {Walker}}]{Agashe:2010gt}%
  \BibitemOpen
  \bibfield  {author} {\bibinfo {author} {\bibfnamefont {K.}~\bibnamefont
  {Agashe}}, \bibinfo {author} {\bibfnamefont {D.}~\bibnamefont {Kim}},
  \bibinfo {author} {\bibfnamefont {M.}~\bibnamefont {Toharia}},\ and\ \bibinfo
  {author} {\bibfnamefont {D.~G.~E.}\ \bibnamefont {Walker}},\ }\href
  {https://doi.org/10.1103/PhysRevD.82.015007} {\bibfield  {journal} {\bibinfo
  {journal} {Phys. Rev. D}\ }\textbf {\bibinfo {volume} {82}},\ \bibinfo
  {pages} {015007} (\bibinfo {year} {2010})},\ \Eprint
  {https://arxiv.org/abs/1003.0899} {arXiv:1003.0899 [hep-ph]} \BibitemShut
  {NoStop}%
\bibitem [{\citenamefont {Cho}\ \emph {et~al.}(2014{\natexlab{a}})\citenamefont
  {Cho}, \citenamefont {Gainer}, \citenamefont {Kim}, \citenamefont {Matchev},
  \citenamefont {Moortgat}, \citenamefont {Pape},\ and\ \citenamefont
  {Park}}]{Cho:2014naa}%
  \BibitemOpen
  \bibfield  {author} {\bibinfo {author} {\bibfnamefont {W.~S.}\ \bibnamefont
  {Cho}}, \bibinfo {author} {\bibfnamefont {J.~S.}\ \bibnamefont {Gainer}},
  \bibinfo {author} {\bibfnamefont {D.}~\bibnamefont {Kim}}, \bibinfo {author}
  {\bibfnamefont {K.~T.}\ \bibnamefont {Matchev}}, \bibinfo {author}
  {\bibfnamefont {F.}~\bibnamefont {Moortgat}}, \bibinfo {author}
  {\bibfnamefont {L.}~\bibnamefont {Pape}},\ and\ \bibinfo {author}
  {\bibfnamefont {M.}~\bibnamefont {Park}},\ }\href
  {https://doi.org/10.1007/JHEP08(2014)070} {\bibfield  {journal} {\bibinfo
  {journal} {JHEP}\ }\textbf {\bibinfo {volume} {08}},\ \bibinfo {pages}
  {070}},\ \Eprint {https://arxiv.org/abs/1401.1449} {arXiv:1401.1449 [hep-ph]}
  \BibitemShut {NoStop}%
\bibitem [{\citenamefont {Hinchliffe}\ \emph {et~al.}(1997)\citenamefont
  {Hinchliffe}, \citenamefont {Paige}, \citenamefont {Shapiro}, \citenamefont
  {Soderqvist},\ and\ \citenamefont {Yao}}]{Hinchliffe:1996iu}%
  \BibitemOpen
  \bibfield  {author} {\bibinfo {author} {\bibfnamefont {I.}~\bibnamefont
  {Hinchliffe}}, \bibinfo {author} {\bibfnamefont {F.~E.}\ \bibnamefont
  {Paige}}, \bibinfo {author} {\bibfnamefont {M.~D.}\ \bibnamefont {Shapiro}},
  \bibinfo {author} {\bibfnamefont {J.}~\bibnamefont {Soderqvist}},\ and\
  \bibinfo {author} {\bibfnamefont {W.}~\bibnamefont {Yao}},\ }\href
  {https://doi.org/10.1103/PhysRevD.55.5520} {\bibfield  {journal} {\bibinfo
  {journal} {Phys. Rev. D}\ }\textbf {\bibinfo {volume} {55}},\ \bibinfo
  {pages} {5520} (\bibinfo {year} {1997})},\ \Eprint
  {https://arxiv.org/abs/hep-ph/9610544} {arXiv:hep-ph/9610544} \BibitemShut
  {NoStop}%
\bibitem [{\citenamefont {Lester}\ and\ \citenamefont
  {Summers}(1999)}]{Lester:1999tx}%
  \BibitemOpen
  \bibfield  {author} {\bibinfo {author} {\bibfnamefont {C.~G.}\ \bibnamefont
  {Lester}}\ and\ \bibinfo {author} {\bibfnamefont {D.~J.}\ \bibnamefont
  {Summers}},\ }\href {https://doi.org/10.1016/S0370-2693(99)00945-4}
  {\bibfield  {journal} {\bibinfo  {journal} {Phys. Lett. B}\ }\textbf
  {\bibinfo {volume} {463}},\ \bibinfo {pages} {99} (\bibinfo {year} {1999})},\
  \Eprint {https://arxiv.org/abs/hep-ph/9906349} {arXiv:hep-ph/9906349}
  \BibitemShut {NoStop}%
\bibitem [{\citenamefont {Allanach}\ \emph {et~al.}(2000)\citenamefont
  {Allanach}, \citenamefont {Lester}, \citenamefont {Parker},\ and\
  \citenamefont {Webber}}]{Allanach:2000kt}%
  \BibitemOpen
  \bibfield  {author} {\bibinfo {author} {\bibfnamefont {B.~C.}\ \bibnamefont
  {Allanach}}, \bibinfo {author} {\bibfnamefont {C.~G.}\ \bibnamefont
  {Lester}}, \bibinfo {author} {\bibfnamefont {M.~A.}\ \bibnamefont {Parker}},\
  and\ \bibinfo {author} {\bibfnamefont {B.~R.}\ \bibnamefont {Webber}},\
  }\href {https://doi.org/10.1088/1126-6708/2000/09/004} {\bibfield  {journal}
  {\bibinfo  {journal} {JHEP}\ }\textbf {\bibinfo {volume} {09}},\ \bibinfo
  {pages} {004}},\ \Eprint {https://arxiv.org/abs/hep-ph/0007009}
  {arXiv:hep-ph/0007009} \BibitemShut {NoStop}%
\bibitem [{\citenamefont {Barr}\ \emph {et~al.}(2003)\citenamefont {Barr},
  \citenamefont {Lester},\ and\ \citenamefont {Stephens}}]{Barr:2003rg}%
  \BibitemOpen
  \bibfield  {author} {\bibinfo {author} {\bibfnamefont {A.}~\bibnamefont
  {Barr}}, \bibinfo {author} {\bibfnamefont {C.}~\bibnamefont {Lester}},\ and\
  \bibinfo {author} {\bibfnamefont {P.}~\bibnamefont {Stephens}},\ }\href
  {https://doi.org/10.1088/0954-3899/29/10/304} {\bibfield  {journal} {\bibinfo
   {journal} {J. Phys. G}\ }\textbf {\bibinfo {volume} {29}},\ \bibinfo {pages}
  {2343} (\bibinfo {year} {2003})},\ \Eprint
  {https://arxiv.org/abs/hep-ph/0304226} {arXiv:hep-ph/0304226} \BibitemShut
  {NoStop}%
\bibitem [{\citenamefont {Miller}\ \emph {et~al.}(2006)\citenamefont {Miller},
  \citenamefont {Osland},\ and\ \citenamefont {Raklev}}]{Miller:2005zp}%
  \BibitemOpen
  \bibfield  {author} {\bibinfo {author} {\bibfnamefont {D.~J.}\ \bibnamefont
  {Miller}}, \bibinfo {author} {\bibfnamefont {P.}~\bibnamefont {Osland}},\
  and\ \bibinfo {author} {\bibfnamefont {A.~R.}\ \bibnamefont {Raklev}},\
  }\href {https://doi.org/10.1088/1126-6708/2006/03/034} {\bibfield  {journal}
  {\bibinfo  {journal} {JHEP}\ }\textbf {\bibinfo {volume} {03}},\ \bibinfo
  {pages} {034}},\ \Eprint {https://arxiv.org/abs/hep-ph/0510356}
  {arXiv:hep-ph/0510356} \BibitemShut {NoStop}%
\bibitem [{\citenamefont {Konar}\ \emph {et~al.}(2009)\citenamefont {Konar},
  \citenamefont {Kong},\ and\ \citenamefont {Matchev}}]{Konar:2008ei}%
  \BibitemOpen
  \bibfield  {author} {\bibinfo {author} {\bibfnamefont {P.}~\bibnamefont
  {Konar}}, \bibinfo {author} {\bibfnamefont {K.}~\bibnamefont {Kong}},\ and\
  \bibinfo {author} {\bibfnamefont {K.~T.}\ \bibnamefont {Matchev}},\ }\href
  {https://doi.org/10.1088/1126-6708/2009/03/085} {\bibfield  {journal}
  {\bibinfo  {journal} {JHEP}\ }\textbf {\bibinfo {volume} {03}},\ \bibinfo
  {pages} {085}},\ \Eprint {https://arxiv.org/abs/0812.1042} {arXiv:0812.1042
  [hep-ph]} \BibitemShut {NoStop}%
\bibitem [{\citenamefont {Burns}\ \emph {et~al.}(2009)\citenamefont {Burns},
  \citenamefont {Matchev},\ and\ \citenamefont {Park}}]{Burns:2009zi}%
  \BibitemOpen
  \bibfield  {author} {\bibinfo {author} {\bibfnamefont {M.}~\bibnamefont
  {Burns}}, \bibinfo {author} {\bibfnamefont {K.~T.}\ \bibnamefont {Matchev}},\
  and\ \bibinfo {author} {\bibfnamefont {M.}~\bibnamefont {Park}},\ }\href
  {https://doi.org/10.1088/1126-6708/2009/05/094} {\bibfield  {journal}
  {\bibinfo  {journal} {JHEP}\ }\textbf {\bibinfo {volume} {05}},\ \bibinfo
  {pages} {094}},\ \Eprint {https://arxiv.org/abs/0903.4371} {arXiv:0903.4371
  [hep-ph]} \BibitemShut {NoStop}%
\bibitem [{\citenamefont {Matchev}\ \emph {et~al.}(2009)\citenamefont
  {Matchev}, \citenamefont {Moortgat}, \citenamefont {Pape},\ and\
  \citenamefont {Park}}]{Matchev:2009iw}%
  \BibitemOpen
  \bibfield  {author} {\bibinfo {author} {\bibfnamefont {K.~T.}\ \bibnamefont
  {Matchev}}, \bibinfo {author} {\bibfnamefont {F.}~\bibnamefont {Moortgat}},
  \bibinfo {author} {\bibfnamefont {L.}~\bibnamefont {Pape}},\ and\ \bibinfo
  {author} {\bibfnamefont {M.}~\bibnamefont {Park}},\ }\href
  {https://doi.org/10.1088/1126-6708/2009/08/104} {\bibfield  {journal}
  {\bibinfo  {journal} {JHEP}\ }\textbf {\bibinfo {volume} {08}},\ \bibinfo
  {pages} {104}},\ \Eprint {https://arxiv.org/abs/0906.2417} {arXiv:0906.2417
  [hep-ph]} \BibitemShut {NoStop}%
\bibitem [{\citenamefont {Matchev}\ and\ \citenamefont
  {Park}(2011)}]{Matchev:2009ad}%
  \BibitemOpen
  \bibfield  {author} {\bibinfo {author} {\bibfnamefont {K.~T.}\ \bibnamefont
  {Matchev}}\ and\ \bibinfo {author} {\bibfnamefont {M.}~\bibnamefont {Park}},\
  }\href {https://doi.org/10.1103/PhysRevLett.107.061801} {\bibfield  {journal}
  {\bibinfo  {journal} {Phys. Rev. Lett.}\ }\textbf {\bibinfo {volume} {107}},\
  \bibinfo {pages} {061801} (\bibinfo {year} {2011})},\ \Eprint
  {https://arxiv.org/abs/0910.1584} {arXiv:0910.1584 [hep-ph]} \BibitemShut
  {NoStop}%
\bibitem [{\citenamefont {Cho}\ \emph {et~al.}(2014{\natexlab{b}})\citenamefont
  {Cho}, \citenamefont {Kim}, \citenamefont {Matchev},\ and\ \citenamefont
  {Park}}]{Cho:2012er}%
  \BibitemOpen
  \bibfield  {author} {\bibinfo {author} {\bibfnamefont {W.~S.}\ \bibnamefont
  {Cho}}, \bibinfo {author} {\bibfnamefont {D.}~\bibnamefont {Kim}}, \bibinfo
  {author} {\bibfnamefont {K.~T.}\ \bibnamefont {Matchev}},\ and\ \bibinfo
  {author} {\bibfnamefont {M.}~\bibnamefont {Park}},\ }\href
  {https://doi.org/10.1103/PhysRevLett.112.211801} {\bibfield  {journal}
  {\bibinfo  {journal} {Phys. Rev. Lett.}\ }\textbf {\bibinfo {volume} {112}},\
  \bibinfo {pages} {211801} (\bibinfo {year} {2014}{\natexlab{b}})},\ \Eprint
  {https://arxiv.org/abs/1206.1546} {arXiv:1206.1546 [hep-ph]} \BibitemShut
  {NoStop}%
\bibitem [{\citenamefont {Kim}\ \emph {et~al.}(2016)\citenamefont {Kim},
  \citenamefont {Matchev},\ and\ \citenamefont {Park}}]{Kim:2015bnd}%
  \BibitemOpen
  \bibfield  {author} {\bibinfo {author} {\bibfnamefont {D.}~\bibnamefont
  {Kim}}, \bibinfo {author} {\bibfnamefont {K.~T.}\ \bibnamefont {Matchev}},\
  and\ \bibinfo {author} {\bibfnamefont {M.}~\bibnamefont {Park}},\ }\href
  {https://doi.org/10.1007/JHEP02(2016)129} {\bibfield  {journal} {\bibinfo
  {journal} {JHEP}\ }\textbf {\bibinfo {volume} {02}},\ \bibinfo {pages}
  {129}},\ \Eprint {https://arxiv.org/abs/1512.02222} {arXiv:1512.02222
  [hep-ph]} \BibitemShut {NoStop}%
\bibitem [{\citenamefont {Debnath}\ \emph {et~al.}(2016)\citenamefont
  {Debnath}, \citenamefont {Gainer}, \citenamefont {Kim},\ and\ \citenamefont
  {Matchev}}]{Debnath:2015wra}%
  \BibitemOpen
  \bibfield  {author} {\bibinfo {author} {\bibfnamefont {D.}~\bibnamefont
  {Debnath}}, \bibinfo {author} {\bibfnamefont {J.~S.}\ \bibnamefont {Gainer}},
  \bibinfo {author} {\bibfnamefont {D.}~\bibnamefont {Kim}},\ and\ \bibinfo
  {author} {\bibfnamefont {K.~T.}\ \bibnamefont {Matchev}},\ }\href
  {https://doi.org/10.1209/0295-5075/114/41001} {\bibfield  {journal} {\bibinfo
   {journal} {EPL}\ }\textbf {\bibinfo {volume} {114}},\ \bibinfo {pages}
  {41001} (\bibinfo {year} {2016})},\ \Eprint
  {https://arxiv.org/abs/1506.04141} {arXiv:1506.04141 [hep-ph]} \BibitemShut
  {NoStop}%
\bibitem [{\citenamefont {Debnath}\ \emph {et~al.}(2017)\citenamefont
  {Debnath}, \citenamefont {Gainer}, \citenamefont {Kilic}, \citenamefont
  {Kim}, \citenamefont {Matchev},\ and\ \citenamefont
  {Yang}}]{Debnath:2016gwz}%
  \BibitemOpen
  \bibfield  {author} {\bibinfo {author} {\bibfnamefont {D.}~\bibnamefont
  {Debnath}}, \bibinfo {author} {\bibfnamefont {J.~S.}\ \bibnamefont {Gainer}},
  \bibinfo {author} {\bibfnamefont {C.}~\bibnamefont {Kilic}}, \bibinfo
  {author} {\bibfnamefont {D.}~\bibnamefont {Kim}}, \bibinfo {author}
  {\bibfnamefont {K.~T.}\ \bibnamefont {Matchev}},\ and\ \bibinfo {author}
  {\bibfnamefont {Y.-P.}\ \bibnamefont {Yang}},\ }\href
  {https://doi.org/10.1007/JHEP06(2017)092} {\bibfield  {journal} {\bibinfo
  {journal} {JHEP}\ }\textbf {\bibinfo {volume} {06}},\ \bibinfo {pages}
  {092}},\ \Eprint {https://arxiv.org/abs/1611.04487} {arXiv:1611.04487
  [hep-ph]} \BibitemShut {NoStop}%
\bibitem [{\citenamefont {Debnath}\ \emph {et~al.}(2019)\citenamefont
  {Debnath}, \citenamefont {Gainer}, \citenamefont {Kilic}, \citenamefont
  {Kim}, \citenamefont {Matchev},\ and\ \citenamefont
  {Yang}}]{Debnath:2018azt}%
  \BibitemOpen
  \bibfield  {author} {\bibinfo {author} {\bibfnamefont {D.}~\bibnamefont
  {Debnath}}, \bibinfo {author} {\bibfnamefont {J.~S.}\ \bibnamefont {Gainer}},
  \bibinfo {author} {\bibfnamefont {C.}~\bibnamefont {Kilic}}, \bibinfo
  {author} {\bibfnamefont {D.}~\bibnamefont {Kim}}, \bibinfo {author}
  {\bibfnamefont {K.~T.}\ \bibnamefont {Matchev}},\ and\ \bibinfo {author}
  {\bibfnamefont {Y.-P.}\ \bibnamefont {Yang}},\ }\href
  {https://doi.org/10.1007/JHEP05(2019)008} {\bibfield  {journal} {\bibinfo
  {journal} {JHEP}\ }\textbf {\bibinfo {volume} {05}},\ \bibinfo {pages}
  {008}},\ \Eprint {https://arxiv.org/abs/1809.04517} {arXiv:1809.04517
  [hep-ph]} \BibitemShut {NoStop}%
\bibitem [{\citenamefont {Agashe}\ \emph {et~al.}(2013)\citenamefont {Agashe},
  \citenamefont {Franceschini},\ and\ \citenamefont {Kim}}]{Agashe:2012bn}%
  \BibitemOpen
  \bibfield  {author} {\bibinfo {author} {\bibfnamefont {K.}~\bibnamefont
  {Agashe}}, \bibinfo {author} {\bibfnamefont {R.}~\bibnamefont
  {Franceschini}},\ and\ \bibinfo {author} {\bibfnamefont {D.}~\bibnamefont
  {Kim}},\ }\href {https://doi.org/10.1103/PhysRevD.88.057701} {\bibfield
  {journal} {\bibinfo  {journal} {Phys. Rev. D}\ }\textbf {\bibinfo {volume}
  {88}},\ \bibinfo {pages} {057701} (\bibinfo {year} {2013})},\ \Eprint
  {https://arxiv.org/abs/1209.0772} {arXiv:1209.0772 [hep-ph]} \BibitemShut
  {NoStop}%
\bibitem [{\citenamefont {Agashe}\ \emph {et~al.}(2014)\citenamefont {Agashe},
  \citenamefont {Franceschini},\ and\ \citenamefont {Kim}}]{Agashe:2013eba}%
  \BibitemOpen
  \bibfield  {author} {\bibinfo {author} {\bibfnamefont {K.}~\bibnamefont
  {Agashe}}, \bibinfo {author} {\bibfnamefont {R.}~\bibnamefont
  {Franceschini}},\ and\ \bibinfo {author} {\bibfnamefont {D.}~\bibnamefont
  {Kim}},\ }\href {https://doi.org/10.1007/JHEP11(2014)059} {\bibfield
  {journal} {\bibinfo  {journal} {JHEP}\ }\textbf {\bibinfo {volume} {11}},\
  \bibinfo {pages} {059}},\ \Eprint {https://arxiv.org/abs/1309.4776}
  {arXiv:1309.4776 [hep-ph]} \BibitemShut {NoStop}%
\bibitem [{\citenamefont {Bayatian}\ \emph {et~al.}()\citenamefont {Bayatian}
  \emph {et~al.}}]{Bayatian:2006nff}%
  \BibitemOpen
  \bibfield  {author} {\bibinfo {author} {\bibfnamefont {G.~L.}\ \bibnamefont
  {Bayatian}} \emph {et~al.} (\bibinfo {collaboration} {CMS}),\ }\href@noop {}
  {\ }\bibinfo {note} {CERN-LHCC-2006-001, CMS-TDR-8-1}\BibitemShut {NoStop}%
\bibitem [{\citenamefont {Contardo}\ \emph {et~al.}()\citenamefont {Contardo},
  \citenamefont {Klute}, \citenamefont {Mans}, \citenamefont {Silvestris},\
  and\ \citenamefont {Butler}}]{Contardo:2015bmq}%
  \BibitemOpen
  \bibfield  {author} {\bibinfo {author} {\bibfnamefont {D.}~\bibnamefont
  {Contardo}}, \bibinfo {author} {\bibfnamefont {M.}~\bibnamefont {Klute}},
  \bibinfo {author} {\bibfnamefont {J.}~\bibnamefont {Mans}}, \bibinfo {author}
  {\bibfnamefont {L.}~\bibnamefont {Silvestris}},\ and\ \bibinfo {author}
  {\bibfnamefont {J.}~\bibnamefont {Butler}},\ }\href@noop {} {\ }\bibinfo
  {note} {CERN-LHCC-2015-010, LHCC-P-008, CMS-TDR-15-02}\BibitemShut {NoStop}%
\bibitem [{\citenamefont {Dienes}\ \emph {et~al.}()\citenamefont {Dienes},
  \citenamefont {Kim}, \citenamefont {Leininger}, \citenamefont {Thomas},\ and\
  \citenamefont {Wilhelm}}]{TumblersVsAntlers}%
  \BibitemOpen
  \bibfield  {author} {\bibinfo {author} {\bibfnamefont {K.~R.}\ \bibnamefont
  {Dienes}}, \bibinfo {author} {\bibfnamefont {D.}~\bibnamefont {Kim}},
  \bibinfo {author} {\bibfnamefont {T.}~\bibnamefont {Leininger}}, \bibinfo
  {author} {\bibfnamefont {B.}~\bibnamefont {Thomas}},\ and\ \bibinfo {author}
  {\bibfnamefont {J.}~\bibnamefont {Wilhelm}},\ }\href@noop {} {\ }\bibinfo
  {note} {$\text{in preparation}$}\BibitemShut {NoStop}%
\bibitem [{\citenamefont {Ambrogi}\ \emph {et~al.}(2020)\citenamefont {Ambrogi}
  \emph {et~al.}}]{Ambrogi:2018ujg}%
  \BibitemOpen
  \bibfield  {author} {\bibinfo {author} {\bibfnamefont {F.}~\bibnamefont
  {Ambrogi}} \emph {et~al.},\ }\href
  {https://doi.org/10.1016/j.cpc.2019.07.013} {\bibfield  {journal} {\bibinfo
  {journal} {Comput. Phys. Commun.}\ }\textbf {\bibinfo {volume} {251}},\
  \bibinfo {pages} {106848} (\bibinfo {year} {2020})},\ \Eprint
  {https://arxiv.org/abs/1811.10624} {arXiv:1811.10624 [hep-ph]} \BibitemShut
  {NoStop}%
\bibitem [{\citenamefont {Desai}\ \emph {et~al.}(2021)\citenamefont {Desai},
  \citenamefont {Domingo}, \citenamefont {Kim}, \citenamefont {Bazan},
  \citenamefont {Rolbiecki}, \citenamefont {Sonawane},\ and\ \citenamefont
  {Wang}}]{Desai:2021jsa}%
  \BibitemOpen
  \bibfield  {author} {\bibinfo {author} {\bibfnamefont {N.}~\bibnamefont
  {Desai}}, \bibinfo {author} {\bibfnamefont {F.}~\bibnamefont {Domingo}},
  \bibinfo {author} {\bibfnamefont {J.~S.}\ \bibnamefont {Kim}}, \bibinfo
  {author} {\bibfnamefont {R.~R. d.~A.}\ \bibnamefont {Bazan}}, \bibinfo
  {author} {\bibfnamefont {K.}~\bibnamefont {Rolbiecki}}, \bibinfo {author}
  {\bibfnamefont {M.}~\bibnamefont {Sonawane}},\ and\ \bibinfo {author}
  {\bibfnamefont {Z.~S.}\ \bibnamefont {Wang}},\ }\href
  {https://doi.org/10.1140/epjc/s10052-021-09727-z} {\bibfield  {journal}
  {\bibinfo  {journal} {Eur. Phys. J. C}\ }\textbf {\bibinfo {volume} {81}},\
  \bibinfo {pages} {968} (\bibinfo {year} {2021})},\ \Eprint
  {https://arxiv.org/abs/2104.04542} {arXiv:2104.04542 [hep-ph]} \BibitemShut
  {NoStop}%
\bibitem [{\citenamefont {Cranmer}\ and\ \citenamefont
  {Yavin}(2011)}]{Cranmer:2010hk}%
  \BibitemOpen
  \bibfield  {author} {\bibinfo {author} {\bibfnamefont {K.}~\bibnamefont
  {Cranmer}}\ and\ \bibinfo {author} {\bibfnamefont {I.}~\bibnamefont
  {Yavin}},\ }\href {https://doi.org/10.1007/JHEP04(2011)038} {\bibfield
  {journal} {\bibinfo  {journal} {JHEP}\ }\textbf {\bibinfo {volume} {04}},\
  \bibinfo {pages} {038}},\ \Eprint {https://arxiv.org/abs/1010.2506}
  {arXiv:1010.2506 [hep-ex]} \BibitemShut {NoStop}%
\bibitem [{\citenamefont {{ATLAS Collaboration}}()}]{ATLAS:2020viz}%
  \BibitemOpen
  \bibfield  {author} {\bibinfo {author} {\bibnamefont {{ATLAS
  Collaboration}}},\ }\href@noop {} {\ }\bibinfo {note}
  {ATL-PHYS-PUB-2020-007}\BibitemShut {NoStop}%
\bibitem [{\citenamefont {Chou}\ \emph {et~al.}(2017)\citenamefont {Chou},
  \citenamefont {Curtin},\ and\ \citenamefont {Lubatti}}]{Chou:2016lxi}%
  \BibitemOpen
  \bibfield  {author} {\bibinfo {author} {\bibfnamefont {J.~P.}\ \bibnamefont
  {Chou}}, \bibinfo {author} {\bibfnamefont {D.}~\bibnamefont {Curtin}},\ and\
  \bibinfo {author} {\bibfnamefont {H.~J.}\ \bibnamefont {Lubatti}},\ }\href
  {https://doi.org/10.1016/j.physletb.2017.01.043} {\bibfield  {journal}
  {\bibinfo  {journal} {Phys. Lett. B}\ }\textbf {\bibinfo {volume} {767}},\
  \bibinfo {pages} {29} (\bibinfo {year} {2017})},\ \Eprint
  {https://arxiv.org/abs/1606.06298} {arXiv:1606.06298 [hep-ph]} \BibitemShut
  {NoStop}%
\bibitem [{\citenamefont {Feng}\ \emph {et~al.}(2018)\citenamefont {Feng},
  \citenamefont {Galon}, \citenamefont {Kling},\ and\ \citenamefont
  {Trojanowski}}]{Feng:2017uoz}%
  \BibitemOpen
  \bibfield  {author} {\bibinfo {author} {\bibfnamefont {J.~L.}\ \bibnamefont
  {Feng}}, \bibinfo {author} {\bibfnamefont {I.}~\bibnamefont {Galon}},
  \bibinfo {author} {\bibfnamefont {F.}~\bibnamefont {Kling}},\ and\ \bibinfo
  {author} {\bibfnamefont {S.}~\bibnamefont {Trojanowski}},\ }\href
  {https://doi.org/10.1103/PhysRevD.97.035001} {\bibfield  {journal} {\bibinfo
  {journal} {Phys. Rev. D}\ }\textbf {\bibinfo {volume} {97}},\ \bibinfo
  {pages} {035001} (\bibinfo {year} {2018})},\ \Eprint
  {https://arxiv.org/abs/1708.09389} {arXiv:1708.09389 [hep-ph]} \BibitemShut
  {NoStop}%
\bibitem [{\citenamefont {Curtin}\ \emph {et~al.}(2018)\citenamefont {Curtin},
  \citenamefont {Dienes},\ and\ \citenamefont {Thomas}}]{Curtin:2018ees}%
  \BibitemOpen
  \bibfield  {author} {\bibinfo {author} {\bibfnamefont {D.}~\bibnamefont
  {Curtin}}, \bibinfo {author} {\bibfnamefont {K.~R.}\ \bibnamefont {Dienes}},\
  and\ \bibinfo {author} {\bibfnamefont {B.}~\bibnamefont {Thomas}},\ }\href
  {https://doi.org/10.1103/PhysRevD.98.115005} {\bibfield  {journal} {\bibinfo
  {journal} {Phys. Rev. D}\ }\textbf {\bibinfo {volume} {98}},\ \bibinfo
  {pages} {115005} (\bibinfo {year} {2018})},\ \Eprint
  {https://arxiv.org/abs/1809.11021} {arXiv:1809.11021 [hep-ph]} \BibitemShut
  {NoStop}%
\end{thebibliography}%

\end{document}